%% file: manuscript.tex
\documentclass[aip,jcp,reprint,amsmath,amssymb]{revtex4-2}
\input{sections/preamble}
\begin{document}
\input{sections/frontmatter}
\input{sections/body}

\end{document}

%% file: sections/preamble.tex
\usepackage[T1]{fontenc}
\usepackage{lmodern}
\usepackage{amsthm}
\usepackage{bm}
\usepackage{hyperref}
\hypersetup{colorlinks=true,citecolor=blue,linkcolor=blue,urlcolor=blue}
\newcommand{\Tr}{\operatorname{Tr}}
\newcommand{\M}{\mathcal M}
\newcommand{\F}{\mathcal F}
\newcommand{\Rset}{\mathcal R}
\newcommand{\SM}[1]{Sec.~S#1 of the supplementary material}
\newtheorem{theorem}{Theorem}
\newtheorem{lemma}[theorem]{Lemma}
\newtheorem{proposition}[theorem]{Proposition}
\newtheorem{corollary}[theorem]{Corollary}
\theoremstyle{remark}

\newtheorem{conjecture}[theorem]{Conjecture}
\newenvironment{sketch}%
 {\par\smallskip\noindent\textit{Proof sketch.}\ \ignorespaces}%
 {\unskip\nobreak\hfill$\square$\par\smallskip}
\IfFileExists{tables/numbers.tex}{\input{tables/numbers.tex}}{}
\usepackage{pgfplots}
\pgfplotsset{compat=1.17}

%% file: tables/numbers.tex

\newcommand{\HubToneStepKthree}{5.3\times10^{-5}}
\newcommand{\HubToneRatio}{60}
\newcommand{\HubTtwoDiffMax}{1\times10^{-13}}
\newcommand{\MolChfourNumTtwoP}{6.5\times10^{-5}}
\newcommand{\MolNtwoSpinTtwoP}{6.1\times10^{-6}}
\newcommand{\MolNtwoSpinTtwo}{9\times10^{-4}}

\newcommand{\HubLone}{-2.00903218}
\newcommand{\HubLoneWidth}{1.1\times10^{-9}}
\newcommand{\HubEoneU}{-2.0087310}
\newcommand{\HubEoneS}{-2.0017921}

\newcommand{\HubDefSpinPct}{95.8}

\newcommand{\HubNonaddPerS}{-7.24\times10^{-3}}
\newcommand{\HubDevKtwoS}{2\times10^{-10}}
\newcommand{\HubDevKtwoU}{-4.1\times10^{-8}}

\newcommand{\HubKliftSing}{3}
\newcommand{\HubEta}{0.10}
\newcommand{\HubVarF}{8.994\times10^{-4}}
\newcommand{\HubSpinSq}{2.395\times10^{-2}}
\newcommand{\MolEtaMin}{8.9\times10^{-7}}
\newcommand{\MolEtaMax}{1.8\times10^{-5}}
\newcommand{\HubCovF}{-8.994\times10^{-4}}

\newcommand{\BeDef}{-5.7\times10^{-7}}

\newcommand{\ChFourDef}{-4.02\times10^{-5}}

\newcommand{\NtwoDef}{-1.25\times10^{-3}}

\newcommand{\HtwoODef}{-8.07\times10^{-5}}

\newcommand{\HtwoOKtwoDevU}{1.9\times10^{-6}}
\newcommand{\HtwoOKtwoDevS}{5\times10^{-6}}
\newcommand{\NtwoMthirtytwoDevU}{7.4\times10^{-7}}
\newcommand{\NtwoMthirtytwoDevS}{1.6\times10^{-6}}

\newcommand{\NtwoNYFit}{-108.71553}
\newcommand{\ChFourNYDef}{3\times10^{-4}}

\newcommand{\HubGmpThreeTwo}{-1.1\times10^{-14}}
\newcommand{\HubGmpGapPerFrag}{2.6\times10^{-13}}
\newcommand{\HubSharpSDevMax}{3.9\times10^{-8}}
\newcommand{\HubSharpSWidthMax}{1.1\times10^{-7}}
\newcommand{\HubSharpSpin}{2.45\times10^{-2}}
\newcommand{\HubCutChangeMax}{1.3\times10^{-7}}
\newcommand{\BeWidthSum}{4.9\times10^{-9}}
\newcommand{\NtwoNYLoneDiff}{2.6\times10^{-5}}
\newcommand{\NtwoMthirtytwoWidthSumU}{8.7\times10^{-7}}

\newcommand{\HtwoOSatDevMax}{1.7\times10^{-7}}
\newcommand{\HtwoOSatWidthMax}{5.9\times10^{-7}}
\newcommand{\NtwoMonoMinStepU}{4.8\times10^{-6}}
\newcommand{\NtwoMonoMinStepS}{7.8\times10^{-6}}
\newcommand{\HubGmpDevMax}{5.9\times10^{-8}}
\newcommand{\HubGmpGapKone}{2.1\times10^{-8}}
\newcommand{\HubGmpGapMaxKtwoThree}{7.8\times10^{-13}}

%% file: sections/frontmatter.tex
\title[The PQG replication limit from a single-fragment SDP]{The replication-energy limit of PQG reduced-density-matrix theory from a single-fragment semidefinite program}
\author{Maho Nakata}
\email{maho@riken.jp}
\affiliation{RIKEN Cluster for Pioneering Research, Wako, Saitama 351-0198, Japan}
\date{\today}

\begin{abstract}
Variational two-electron reduced-density-matrix calculations are not size
extensive: the energy per fragment can decrease when identical, strictly
noninteracting fragments are replicated [M. Nakata and K. Yasuda, Phys.\ Rev.\ A \textbf{80}, 042109 (2009)]. For the $P$, $Q$, and $G$ conditions (PQG) we prove the following.
(i) The energy per fragment converges to the minimum of a semidefinite program (SDP) on one fragment. The limit is reached through negative interfragment covariances of order $1/(k-1)$ for $k$ replicas, which cancel the local electron number and spin fluctuations consistently with a global singlet. (ii) The energy per fragment
is nonincreasing in the replica count. (iii) It reaches the limit exactly at a finite replica count. (iv) The defect is tied to the
local electron-number variance: a sharp local electron number restores additivity without a spin constraint. (v) In chains of identical units the energy per unit has a limit that
tends to the same minimum as the coupling vanishes.
The key tool is that the $G$ condition [C. Garrod and J. K. Percus, J. Math.\ Phys.\ \textbf{5}, 1756 (1964)] keeps the variances and covariances
of one-body operators positive semidefinite without a representing state,
so that the Cauchy--Schwarz inequality applies. PQG therefore admits explicitly constructed replicated points whose compensated local fluctuations lower the energy per fragment below the isolated value. Be, CH$_4$, N$_2$, and H$_2$O (STO-6G) and a
four-site Hubbard fragment illustrate the results; in these examples the gap to the isolated value persists with three-index conditions $T_1$ and $T_2$.
\end{abstract}
\maketitle

%% file: sections/body.tex
\input{sections/introduction}
\input{sections/problems}
\input{sections/limit}
\input{sections/monotone}
\input{sections/reduced}
\input{sections/cuts}
\input{sections/stronger}
\input{sections/chains}
\input{sections/numerics}

\input{sections/conclusions}
\input{sections/backmatter}

%% file: sections/introduction.tex
\section{Introduction}
\label{sec:intro}
The variational two-electron reduced-density-matrix (v2RDM) method
minimizes the energy over reduced density matrices that satisfy selected
necessary $N$-representability conditions.\cite{Coleman1963} Necessary
conditions such as the $P$, $Q$, and $G$ conditions (the PQG
approximation) were formulated in the 1960s and used in early
calculations in the
1970s.\cite{GarrodPercus1964,GarrodMihailovicRosina1975,RosinaGarrod1975,MihailovicRosina1975,GarrodFusco1976}
The method was later
formulated as a semidefinite program (SDP),\cite{Nakata2001} extended by
the three-index $T_1$ and $T_2$
conditions,\cite{Zhao2004,Erdahl1978,Fukuda2007,Nakata2008} and developed
with efficient algorithms for strongly correlated and larger molecules.\cite{Mazziotti2004,GidofalviMazziotti2008} Reviews are given in
Refs.~\onlinecite{Mazziotti2007Book} and~\onlinecite{DePrince2024}.

Size consistency of PQG was examined in early work. Nakata et
al.\cite{Nakata2002} showed that the $G$ condition implies a necessary
condition for size consistency and found the energy at dissociation
equal, within numerical accuracy, to the sum of the atomic energies for
some molecules but not for others (deviations of up to $0.033\,E_h$, for
C$_2$), and Gidofalvi and Mazziotti\cite{GidofalviMazziotti2005} found
correlation energies of hydrocarbon chains that grow linearly in magnitude
with the number of electrons, a property of connected chains that does not
imply additivity for disconnected fragments.

Van Aggelen, Bultinck, Ayers, Verstichel, and coworkers then identified a
central failure of the method at dissociation: with PQG, diatomic molecules dissociate into fractionally charged atoms
with too low an energy (Refs.~\onlinecite{VanAggelen2009}
and~\onlinecite{VanAggelen2010}; see also Ref.~\onlinecite{DePrince2024}).
They traced it to the description of fractionally charged atoms and cured
it for diatomic molecules by subsystem constraints built from fragment
energies.\cite{Verstichel2010,VanAggelen2011}

For size consistency, Nakata and Anderson\cite{NakataAnderson2012} gave a
sufficient condition that uses subsystem RDMs of sharp electron number, an
integer number with zero variance. Integer electron-count constraints on
regions, used in Parts~I and~II of a companion study\cite{PartI,PartII}
and in Ref.~\onlinecite{Boyn2026}, largely remove spurious fractional
charges in molecules and clusters.

The PQG approximation also fails to be additive under replication.
Nakata and Yasuda\cite{NakataYasuda2009} computed disconnected replicas of
atoms and molecules, with the replica count entering only as a parameter,
and observed energies per fragment that decreased and appeared to approach
finite limits.

These replica results raise a specific question: which single-fragment
optimization problem determines the limit of repeated disconnected
replication, and which local constraints decide whether the defect is
present?

We answer this question for PQG. The limit is the minimum $E_1^{\mathsf c}$ of a
single-fragment SDP with the same variables as the isolated fixed-number
calculation but different constraints: it
fixes only the \emph{mean} electron number, omits the fixed-number
contraction and pair trace, and keeps the $P$ and $Q$ conditions and the
particle--hole covariance condition of Garrod and
Percus\cite{GarrodPercus1964} in an augmented semidefinite form. This
single-fragment problem is close to the Fock-space relaxation of
Part~II,\cite{PartII} which in addition fixes the mean numbers of the two
spins separately; here it arises as the replication limit.
Restricting a replicated solution to one fragment shows that these
constraints are necessary. Conversely, an explicit construction lifts any
local point with interior occupations to the replicated problem, for all
sufficiently large $k$, with
negative interfragment covariances of order $1/(k-1)$, which cancel the local
number and spin fluctuations while leaving every local energy unchanged.
Natural orbitals with occupation exactly $0$ or $1$ do not fluctuate and
can be split off, so every minimizer is lifted at a finite replica count;
for the global singlet the local point must be invariant under spin
rotations, and averaging any minimizer over spin rotations gives such a
point with the same energy.
The same limit holds with and without a global singlet constraint.
The $G$ condition of Garrod and Percus, in its original covariance form,
plays a double role: restriction produces it, and the Cauchy--Schwarz
inequality for the covariance form it defines, valid without a
representing state, is what keeps the lift positive.

The finite limit itself is not new: subadditivity and a lower bound give
its existence by Fekete's lemma, and its numerical approach was observed
in Ref.~\onlinecite{NakataYasuda2009}. We use \emph{replication
additivity} for $E_k=kE_1$ at every $k$, the disconnected-replica test of
size extensivity used in Ref.~\onlinecite{NakataYasuda2009}. All results concern a fixed finite
fragment basis and exactly zero interfragment integrals; they do not
establish size consistency or $N$-representability.

Our contributions are the following:
\begin{enumerate}
 \item The limit theorem:
 $\lim_{k\to\infty}E_k^{\mathsf x}/k=E_1^{\mathsf c}$ for unrestricted spin and for
 the global singlet, where $E_1^{\mathsf c}$ is the minimum of the
 single-fragment SDP.
 \item Monotonicity: the energy per fragment is nonincreasing in the
 replica count, because interfragment covariances can be diluted.
 \item Finite saturation: in a fixed finite basis there is always a finite
 replica count beyond which $E_k=kE_1^{\mathsf c}$ holds exactly, without
 assumptions on the local minimizer; moreover, equality at one count
 persists at all larger counts.
 \item A proof that the block structure used by Nakata and
 Yasuda,\cite{NakataYasuda2009} which makes every finite replica problem an
 SDP whose size does not depend on the replica count, loses no
 generality.
 \item The relation of the defect to the local number variance: integer
 cuts that use only the mean of a fragment's electron count do not remove
 it, whereas the equality that makes the count sharp restores additivity
 in the unrestricted setting.
 \item For full $p$-positivity, a proof that products of feasible points
 remain feasible, so that the replication limit exists, and the conjecture
 that it is again the single-fragment relaxation; for $T_1$ and $T_2$, a
 list of what carries over, and numerical evidence that the gap between
 the single-fragment SDP and the isolated fragment persists with
 three-index conditions: the $T_2'$ form removes most of the spin part of
 the gap but in general not the part due to the local electron number.
 \item For chains of identical units with a weak coupling $\lambda$, the
 existence of the energy per unit and its distance at most of order
 $\lambda$ from $E_1^{\mathsf c}$, so that the defect persists in weakly
 coupled polymers.
\end{enumerate}

Section~\ref{sec:problems} defines the replicated problems and the
single-fragment SDP. Section~\ref{sec:limit} proves the limit theorem
through a restriction bound and an explicit lift.
Section~\ref{sec:monotone} proves monotonicity and the finite-saturation
criteria and relates the results to de Finetti theorems, Sec.~\ref{sec:reduced} gives the symmetry reduction, and
Sec.~\ref{sec:cuts} discusses integer electron-count cuts, and
Sec.~\ref{sec:stronger} discusses what carries over to the three-index
conditions and to full $p$-positivity, and Sec.~\ref{sec:chains} extends
the limit to weakly coupled chains.
Section~\ref{sec:numerics} illustrates the results for Be, CH$_4$,
N$_2$, and H$_2$O and for a four-site Hubbard fragment, and
Sec.~\ref{sec:conclusions} concludes. Statements are proved in the main
text at the level of sketches that identify which quantities cancel and
which constraints matter; complete proofs are in the supplementary
material, whose Sec.~S1 summarizes the conventions and the logical
dependencies of the proofs.

%% file: sections/problems.tex
\section{Replicated and single-fragment problems}
\label{sec:problems}

\subsection{Disconnected replicas}
\label{sec:replicas}
For a set of spin orbitals, one fragment or the whole system of
fragments introduced below, we use the reduced density matrices
\begin{equation}
 \gamma_{ij}=\langle a_j^\dagger a_i\rangle,\qquad
 \Gamma_{ij;lm}=\tfrac12\langle a_l^\dagger a_m^\dagger a_j a_i\rangle.
 \label{eq:conventions}
\end{equation}
A monomial is a product of creation and annihilation operators, and its
degree $\deg\hat x$ is the number of factors; an operator has degree at
most $q$ if it is a linear combination of monomials of degree at most $q$.
For an ensemble of states the brackets are expectation values. For a
trial pair $\Rset=(\gamma,\Gamma)$ they denote the linear functional
obtained by reducing a number-conserving operator of degree at most four
with the canonical anticommutation relations (CAR); no representing
many-electron state is assumed. With $\hat N=\sum_ia_i^\dagger a_i$, CAR
gives
\begin{equation}
 \begin{aligned}
 \Tr\gamma&=\langle\hat N\rangle,\\
 2\sum_{i,j}\Gamma_{ij;ij}&=\langle\hat N(\hat N-1)\rangle,\\
 2\sum_p\Gamma_{ip;jp}&=\langle a_j^\dagger\hat Na_i\rangle
 =\langle a_j^\dagger a_i(\hat N-1)\rangle ,
 \end{aligned}
 \label{eq:numberidentities}
\end{equation}
which hold for ensembles and, by definition of the brackets, for trial
pairs. If the electron number is sharp, that is, equal to $N$ without
fluctuation, the right-hand sides are $N$, $N(N-1)$, and
$(N-1)\gamma_{ij}$. If only the mean $\langle\hat N\rangle=N$ is fixed,
the second right-hand side is $N(N-1)$ plus the number variance
$\langle(\hat N-N)^2\rangle$, and the third is $(N-1)\gamma_{ij}$ plus the
covariance $\langle a_j^\dagger a_i(\hat N-N)\rangle$.

The $N$-representability conditions used in this work are of
positive semidefinite (PSD) type: a matrix of moments
$\langle\hat O_I^\dagger\hat O_J\rangle$ over a fixed set of operators
$\hat O_J$ (for $T_1$ and $T_2$, of anticommutators) must be PSD. With
pairs $i<j$ in $P$ and $Q$ (all ordered pairs $(i,j)$, including $i=j$, in
$G$), triples $i<j<k$ for $T_1$, and triples with $j<k$ for $T_2$,
\begin{align}
 P_{ij,i'j'}&=\langle a_i^\dagger a_j^\dagger a_{j'}a_{i'}\rangle,
 \qquad
 Q_{ij,i'j'}=\langle a_ja_i\,a_{i'}^\dagger a_{j'}^\dagger\rangle,
 \nonumber\\
 G_{ij,i'j'}&=\langle a_j^\dagger a_i\,a_{i'}^\dagger a_{j'}\rangle,
 \nonumber\\
 (T_1)_{ijk,i'j'k'}&=\langle a_i^\dagger a_j^\dagger a_k^\dagger\,
 a_{k'}a_{j'}a_{i'}+a_{k'}a_{j'}a_{i'}\,
 a_i^\dagger a_j^\dagger a_k^\dagger\rangle,
 \nonumber\\
 (T_2)_{ijk,i'j'k'}&=\langle a_j^\dagger a_k^\dagger a_i\,
 a_{i'}^\dagger a_{k'}a_{j'}+a_{i'}^\dagger a_{k'}a_{j'}\,
 a_j^\dagger a_k^\dagger a_i\rangle,
 \nonumber\\
 T_2'&=\begin{pmatrix}T_2&X\\X^\dagger&\gamma^{\rm T}\end{pmatrix},
 \qquad X_{ijk,p}=\langle a_j^\dagger a_k^\dagger a_i\,a_p\rangle .
 \label{eq:psdconditions}
\end{align}
The operators are $\hat O_{i'j'}=a_{j'}a_{i'}$ ($P$),
$a_{i'}^\dagger a_{j'}^\dagger$ ($Q$), and $a_{i'}^\dagger a_{j'}$ ($G$);
$\hat O_{i'j'k'}=a_{k'}a_{j'}a_{i'}$ ($T_1$) and
$a_{i'}^\dagger a_{k'}a_{j'}$ ($T_2$); and for $T_2'$ the $T_2$ operators
together with the annihilators $a_p$, whose block is
$\langle a_p^\dagger a_q\rangle=\gamma_{qp}$. The operators of $P$, $Q$, and
$G$ are the monomials of degree two that change the electron number by
$-2$, $+2$, and $0$, and those of $T_1$ and $T_2$ have degree three; the PQG
conditions are thus the moment matrices indexed by monomials of degree two,
with entries of degree four. In terms of an arbitrary
combination $\hat B$ of the operators, the conditions read
$\langle\hat B^\dagger\hat B\rangle\ge0$ ($P$, $Q$, $G$),
$\langle\hat B^\dagger\hat B+\hat B\hat B^\dagger\rangle\ge0$ ($T_1$, $T_2$),
and $\langle\hat B^\dagger\hat B+\hat B_3\hat B_3^\dagger\rangle\ge0$ with
$\hat B=\hat B_3+\hat B_1$ split into its $T_2$ and one-particle parts
($T_2'$). For a state each is a sum of expectation values
$\langle\hat C^\dagger\hat C\rangle$ and is nonnegative. For a trial pair
the brackets are reduced by CAR; in $T_1$, $T_2$, and $T_2'$ the
three-body terms cancel, so all six are linear matrix inequalities in
$(\gamma,\Gamma)$; in particular $P_{ij,i'j'}=2\Gamma_{i'j';ij}$. The
conditions $P$ and $Q$ are due to Coleman,\cite{Coleman1963} $G$ to Garrod
and Percus,\cite{GarrodPercus1964} $T_1$ and $T_2$ to
Erdahl\cite{Erdahl1978} (see also Ref.~\onlinecite{Zhao2004}), and $T_2'$
to Braams, Percus, and Zhao\cite{BraamsPercusZhao2007} (see also
Ref.~\onlinecite{Mazziotti2006}). The bordered matrix $\widetilde G$ of
Sec.~\ref{sec:local} adds the identity operator to the operators of $G$.

Let a fragment $F$ have $r$ orthonormal spin orbitals and an integer
reference electron number $0<N_F<r$, with Hamiltonian
\begin{equation}
 \hat H_F=\sum_{p,q}h_{pq}a_p^\dagger a_q
 +\frac12\sum_{p,q,r,s}v_{pq,rs}\,a_p^\dagger a_q^\dagger a_s a_r+c_F,
 \label{eq:H}
\end{equation}
with $h_{pq}=\overline{h_{qp}}$, $v_{pq,rs}=\overline{v_{rs,pq}}$, and a
real constant $c_F$; the indices run over the spin orbitals of $F$, and
$v$ may be taken antisymmetric in $p,q$ and in $r,s$. The energy is
linear in the reduced density matrices,
$\langle \hat H_F\rangle=\sum_{pq}h_{pq}\gamma_{qp}+\sum_{pqrs}v_{pq,rs}
\Gamma_{rs;pq}+c_F$. On $k$ disjoint copies of $F$ we put
\begin{equation}
 \hat H_{\rm tot}=\sum_{a=1}^k\hat H_F^{(a)},
 \label{eq:replicas}
\end{equation}
so that the fragments are identical and strictly noninteracting. The
subscript ``tot'' marks quantities of the whole $k$-fragment system, for
example $\Rset_{\rm tot}=(\gamma_{\rm tot},\Gamma_{\rm tot})$ and
$\hat N_{\rm tot}=\sum_Ia_I^\dagger a_I$, and a superscript $(a)$ marks the copy
on fragment $a$. Spin orbitals of the whole system are labeled by capital
letters $I,J,P,\dots$, or by $i_a$ for spin orbital $i$ of fragment $a$
when the fragment matters; lower-case indices without a fragment label,
and quantities without ``tot'', refer to one fragment.

The global set $\F_k^{\mathsf u}$ consists of pairs
$\Rset_{\rm tot}$ on the $kr$ spin orbitals that satisfy the $P$, $Q$, and
$G$ conditions, $\Tr\gamma_{\rm tot}=kN_F$, and the fixed-number contraction
\begin{equation}
 2\sum_P(\Gamma_{\rm tot})_{IP;JP}=(kN_F-1)(\gamma_{\rm tot})_{IJ}.
 \label{eq:contraction}
\end{equation}
By the last identity of Eq.~\eqref{eq:numberidentities}, this is the form
that a sharp electron number $kN_F$ gives; its trace makes the pair trace
$kN_F(kN_F-1)/2$, so that the total number has no fluctuation,
$\langle(\hat N_{\rm tot}-kN_F)^2\rangle=0$.
The label $\mathsf u$ refers to unrestricted total spin. When $\hat H_F$ is
invariant under spin rotations, with $d$ spatial orbitals ($r=2d$) and
even $N_F$, the set $\F_k^{\mathsf s}$ is obtained by adding the global singlet
condition (every statement about $\mathsf x={\mathsf s}$ below assumes this
setting)
\begin{equation}
 \langle\hat{\bm S}_{\rm tot}^2\rangle=0 .
 \label{eq:singlet}
\end{equation}
We write
$E_k^{\mathsf x}=\min_{\Rset_{\rm tot}\in\F_k^{\mathsf x}}\langle \hat H_{\rm tot}\rangle_{\Rset_{\rm tot}}$ for
$\mathsf x\in\{{\mathsf u},{\mathsf s}\}$. Here $\mathsf x$ labels the spin setting,
unrestricted (${\mathsf u}$) or global singlet (${\mathsf s}$); it is not a
spin index. For $k=1$, $E_1^{\mathsf u}$ and $E_1^{\mathsf s}$ are the
energies of the isolated fixed-number PQG calculation on one fragment,
without and with the singlet condition. Both sets are nonempty (a Slater determinant is
feasible, and a closed-shell determinant for $\mathsf x={\mathsf s}$) and compact
(\SM{2}), so the minima exist.

We say that the PQG energy is \emph{replication additive} when
$E_k^{\mathsf x}=kE_1^{\mathsf x}$ for every $k$. This is the disconnected-replica
test of size extensivity used by Nakata and Yasuda.\cite{NakataYasuda2009}
All statements concern a fixed finite fragment basis, and no basis-set
limit is taken. Sections~\ref{sec:problems}--\ref{sec:stronger} concern
exactly zero interfragment integrals; Sec.~\ref{sec:chains} treats chains
with a weak coupling between neighboring units.

\subsection{The single-fragment problem}
\label{sec:local}
For one-body operators on $F$ define the particle--hole matrix and its
centered form,
\begin{align}
 G_{ij;kl}&=\langle a_j^\dagger a_i\,a_k^\dagger a_l\rangle,\nonumber\\
 C_{ij;kl}&=G_{ij;kl}-\gamma_{ij}\gamma_{lk}
 =\delta_{ik}\gamma_{lj}-2\Gamma_{li;jk}-\gamma_{ij}\gamma_{lk}.
 \label{eq:C}
\end{align}
The matrix $C$ is the covariance table of the local one-body operators.
Each index pair is a single label: the row $(ij)$ belongs to the operator
$a_j^\dagger a_i$, the column $(kl)$ to $a_k^\dagger a_l$ (the adjoint of
the row operator with the same pair), and the entry
$C_{ij;kl}=\langle a_j^\dagger a_i\,a_k^\dagger a_l\rangle
-\langle a_j^\dagger a_i\rangle\langle a_k^\dagger a_l\rangle$ is their
covariance; the diagonal entries are variances. Positivity of this
table is the particle--hole condition of Garrod and
Percus.\cite{GarrodPercus1964} (Their matrix called $G$ is our $C$, and
their two-particle normalization accounts for the factor $2$ multiplying
our $\Gamma$.) We impose it in the augmented form $\widetilde G\succeq0$
below and use $C$ only where covariances are computed explicitly.

For operators $\hat A$ and $\hat B$ we write
$\mathrm{cov}(\hat A,\hat B)=\langle\hat A\hat B\rangle-\langle\hat A\rangle
\langle\hat B\rangle$ and
$\mathrm{var}(\hat A)=\mathrm{cov}(\hat A^\dagger,\hat A)
=\langle\hat A^\dagger\hat A\rangle-|\langle\hat A\rangle|^2$, evaluated
from $(\gamma,\Gamma)$ when the operators are one-body. For
$\hat Z=\sum_{pq}z_{pq}a_p^\dagger a_q$ one has $\mathrm{var}(\hat Z)=
z^\dagger Cz$, so $\widetilde G\succeq0$ makes these variances
nonnegative and gives the Cauchy--Schwarz inequality
$|\mathrm{cov}(\hat A^\dagger,\hat B)|^2\le\mathrm{var}(\hat A)\,
\mathrm{var}(\hat B)$, although $(\gamma,\Gamma)$ need not come from a
state. By linearity of the brackets, for a number $\beta$ and a Hermitian
$\hat B$,
\begin{equation}
 \begin{aligned}
 \mathrm{var}(\hat A-\beta\hat B)
 &=\mathrm{var}(\hat A)-\beta\,\mathrm{cov}(\hat A^\dagger,\hat B)\\
 &\quad-\bar\beta\,\mathrm{cov}(\hat B,\hat A)+|\beta|^2\mathrm{var}(\hat B),
 \end{aligned}
 \label{eq:varexpand}
\end{equation}
and since $\mathrm{cov}(\hat A^\dagger,\hat B)$ is the complex conjugate
of $\mathrm{cov}(\hat B,\hat A)$, the choice
$\beta=\mathrm{cov}(\hat B,\hat A)/\mathrm{var}(\hat B)$ gives
\begin{equation}
 \mathrm{var}(\hat A-\beta\hat B)
 =\mathrm{var}(\hat A)-\frac{|\mathrm{cov}(\hat B,\hat A)|^2}{\mathrm{var}(\hat B)}\ge0 .
 \label{eq:regression}
\end{equation}
This is linear regression of $\hat A$ on $\hat B$: $\beta\hat B$ is the
part of $\hat A$ that follows $\hat B$, and the explained variance cannot
exceed the total variance.

Let $\M^{\mathsf c}$ be the set of local pairs on the $r$ spin
orbitals of $F$ satisfying (the definition applies to any mean
$0\le N_F\le r$; the end points occur after frozen orbitals are removed
in Sec.~\ref{sec:monotone})
\begin{align}
 &0\preceq\gamma\preceq I,\qquad \Tr\gamma=N_F,\qquad
  P\succeq0,\qquad Q\succeq0,\nonumber\\
 &\widetilde G=
 \begin{pmatrix}
 1 & (\operatorname{vec}\gamma)^\dagger\\
 \operatorname{vec}\gamma & G
 \end{pmatrix}\succeq0 .
 \label{eq:M}
\end{align}
The augmented (bordered) matrix $\widetilde G$ collects the moments
$\langle x^\dagger y\rangle$ of the identity and all one-body operators
$a_i^\dagger a_j$. Its quadratic form on
$c+\sum_{ij}z_{ij}a_i^\dagger a_j$ equals
$|c+(\operatorname{vec}\gamma)^\dagger z|^2+z^\dagger Cz$; that is, $C$ is
the Schur complement of the corner entry $1$, and
$\widetilde G\succeq0$ is equivalent to $C\succeq0$. Every entry of
$\widetilde G$ is affine in $(\gamma,\Gamma)$; the augmented form is thus a
semidefinite expression of the classical covariance condition and not a
new representability condition. Erdahl and Garrod wrote the nonlinear $G$
condition in exactly this form, as positivity of
$\langle g_0^\dagger g_0\rangle$ for $g_0=c+\sum_{ij}z_{ij}a_i^\dagger
a_j$, and showed that at fixed electron number it is equivalent to the
linear $G$ condition.\cite{ErdahlGarrod1974} The single-fragment value is
\begin{equation}
 \boxed{E_1^{\mathsf c}=\min_{\Rset\in\M^{\mathsf c}}\langle \hat H_F\rangle_{\Rset}.}
 \label{eq:L1}
\end{equation}
The label $\mathsf c$ indicates that the covariance condition is kept while
the fixed-number contraction is dropped; like $E_1^{\mathsf u}$ and
$E_1^{\mathsf s}$, $E_1^{\mathsf c}$ is a minimum over the RDMs of one
fragment, and it depends on $N_F$ through $\Tr\gamma=N_F$. Without the
contraction, $\gamma$ is not obtained from $\Gamma$ as in the isolated
problem; it is the border of $\widetilde G$, and it is tied to $\Gamma$ only
through $Q\succeq0$ and $\widetilde G\succeq0$, the latter equivalent to
$C\succeq0$ (Table~\ref{tab:constraints}). It is a moment relaxation at fixed \emph{mean} electron number, not an
optimization over physical grand-canonical states. The same augmented
particle--hole condition defines the Fock-space relaxation used in Part~II
of a companion study for atoms at fractional electron number, there with
the traces of the two spin blocks of $\gamma$ fixed
separately;\cite{PartII} here it arises as the replication limit.
Mazziotti formulated a hierarchy of representability conditions for
Fock-space states without particle-number conservation, and uses the
condition of zero number variance to recover the fixed-number
sector.\cite{Mazziotti2026} In that language, $\M^{\mathsf c}$ is a Fock-space
relaxation at the level of PQG for number-conserving moments, with the
mean electron number fixed and the variance condition dropped;
Theorem~\ref{thm:limit} identifies its minimum as the replication limit.

Let $\hat N_F=\sum_ia_i^\dagger a_i$. Its mean is
$\langle\hat N_F\rangle=\Tr\gamma=N_F$ by Eq.~\eqref{eq:numberidentities}, and
we define the local number variance by
\begin{equation}
 v=\langle\hat N_F^2\rangle-\langle\hat N_F\rangle^2
  =\langle\hat N_F^2\rangle-N_F^2 .
 \label{eq:vdef}
\end{equation}
By the linearity of the brackets this equals $\langle(\hat N_F-N_F)^2\rangle$.
With $u$ the coefficient vector of $\hat N_F$, one has $v=u^\dagger Cu$.
Thus $v$ is the quadratic form of $\widetilde G$ on $\hat N_F-N_F$, the
deviation of $\hat N_F$ from a multiple of the identity; for $N_F>0$,
\begin{equation}
 \Big\langle\Big(\frac{\hat N_F}{N_F}-1\Big)^2\Big\rangle=\frac{v}{N_F^2}.
 \label{eq:identitygap}
\end{equation}
In the isolated problem $v$ vanishes, so in the entries of $\widetilde G$
the identity may be replaced by $\hat N_F/N_F$, and the border of
$\widetilde G$, that is, $\gamma$, carries no information beyond $\Gamma$:
$\gamma$ follows from $\Gamma$ by the contraction. In $\M^{\mathsf c}$, $v$
is not fixed and $\gamma$ is an independent variable; in the replicated
system the local contributions $kv$ to the total number variance are
cancelled by negative covariances between fragments
(Sec.~\ref{sec:cancel}).

\begin{lemma}[Local compactness and fixed-number reduction]
\label{lem:local}
The set $\M^{\mathsf c}$ is nonempty and compact. Within $\M^{\mathsf c}$ the following
conditions are equivalent: (i) $v=0$; (ii) $Cu=0$; (iii) the local
fixed-number contraction $2\sum_p\Gamma_{ip;jp}=(N_F-1)\gamma_{ij}$ holds
for all $i,j$. Adding any of them to Eq.~\eqref{eq:M} recovers the
isolated unrestricted fixed-$N_F$ PQG problem $\F_1^{\mathsf u}$.
\end{lemma}

\begin{sketch}
Because $\widetilde G\succeq0$, its Schur complement $C$ is PSD, and a zero
quadratic form $v=u^\dagger Cu$ forces the whole
row $Cu$ to vanish, and $(Cu)_{ij}=\langle a_j^\dagger a_i(\hat N_F-N_F)\rangle$
is, after CAR, the residual of the contraction. Conversely, the
contraction is the trace relation $\langle\hat\Omega \hat N_F\rangle=N_F\langle\hat\Omega
\rangle$ of Erdahl and Garrod\cite{ErdahlGarrod1974} for one-body
$\hat\Omega$; together with $\langle\hat N_F^2\rangle=N_F^2$ it allows the
identity to be replaced by $\hat N_F/N_F$ in the entries of
$\widetilde G$. A constant plus a
one-body operator, $c+\hat O$, then acts as the one-body operator $\hat O+c\hat N_F/N_F$, and
the linear condition $G\succeq0$ gives $\langle(c+\hat O)^\dagger(c+\hat O)\rangle\ge0$,
that is, $\widetilde G\succeq0$. Without the contraction this replacement is
not available, which is why $\widetilde G\succeq0$ is imposed in $\M^{\mathsf c}$.
Compactness follows from $0\preceq\gamma\preceq I$ and the PSD bounds on
off-diagonal entries. The complete proof is in \SM{2}.
\end{sketch}

\subsection{Isolated and limiting problems compared}
\label{sec:compare}
The isolated value $E_1^{\mathsf x}$ and the limiting value $E_1^{\mathsf c}$ minimize
the same energy over the same variables, the RDMs of one fragment. They
differ only in the constraints (Table~\ref{tab:constraints}). The three
single-fragment values and the replica energies are ordered as
\begin{equation}
 E_1^{\mathsf c}\le\frac{E_k^{\mathsf x}}{k}\le E_1^{\mathsf x}
 \quad(k\ge1),\qquad
 E_1^{\mathsf c}\le E_1^{\mathsf u}\le E_1^{\mathsf s}.
 \label{eq:ordering}
\end{equation}
The first chain combines restriction (Proposition~\ref{prop:restriction})
with product extension [Eq.~\eqref{eq:productbounds}]; the second holds
because the isolated feasible sets are contained in $\M^{\mathsf c}$
(Lemma~\ref{lem:local}) and $\F_1^{\mathsf s}\subseteq\F_1^{\mathsf u}$. By
Theorem~\ref{thm:limit}, $E_k^{\mathsf x}/k$ moves from the upper to the
lower end of the first chain as $k\to\infty$. Four consequences are used
throughout.

\begin{table*}[t]
\caption{Constraints of the isolated fixed-number problem $E_1^{\mathsf x}$
($\F_1^{\mathsf x}$) and of the single-fragment problem $E_1^{\mathsf c}$
[$\M^{\mathsf c}$, Eq.~\eqref{eq:M}]. Both use one fragment's pair
$(\gamma,\Gamma)$ and the energy $\langle \hat H_F\rangle$. ``Implied'' means
that the condition follows from the imposed ones.}
\label{tab:constraints}
\begin{ruledtabular}
\begin{tabular}{p{0.24\linewidth}p{0.33\linewidth}p{0.33\linewidth}}
Condition & Isolated problem $E_1^{\mathsf x}$ & Single-fragment problem $E_1^{\mathsf c}$ \\
\hline
$P\succeq0$, $Q\succeq0$ & imposed & imposed \\
Particle--hole channel & $G\succeq0$ & augmented $\widetilde G\succeq0$ \\
1-RDM $\gamma$ & obtained from $\Gamma$ via the trace condition (fixed-number contraction) & extracted from $\widetilde G$ (its border $\operatorname{vec}\gamma$) \\
One-particle bounds & $0\preceq\gamma\preceq I$ (implied) & $0\preceq\gamma\preceq I$ (imposed) \\
Mean electron number & $\Tr\gamma=N_F$ & $\Tr\gamma=N_F$ \\
Fixed-number contraction & imposed & not imposed \\
Pair trace $\sum_{ij}\Gamma_{ij;ij}$ & $N_F(N_F-1)/2$ (implied) & $[N_F(N_F-1)+v]/2$, free through $v$ \\
Number variance $v$ & $0$ (implied) & $v\ge0$, free \\
Local spin & none (u) or $\langle\hat{\bm S}_F^2\rangle=0$ (s) & none \\
\end{tabular}
\end{ruledtabular}
\end{table*}

First, the pair trace must not be imposed in $\M^{\mathsf c}$. By
Eq.~\eqref{eq:numberidentities}, $\langle\hat N_F\rangle=N_F$, and by the
definition~\eqref{eq:vdef}, $\langle\hat N_F^2\rangle=N_F^2+v$. Hence
\begin{equation}
 \begin{aligned}
 \sum_{ij}\Gamma_{ij;ij}
 &=\tfrac12\langle\hat N_F(\hat N_F-1)\rangle
  =\tfrac12\bigl(\langle\hat N_F^2\rangle-\langle\hat N_F\rangle\bigr)\\
 &=\tfrac12\bigl(N_F^2+v-N_F\bigr)
  =\tfrac12\bigl[N_F(N_F-1)+v\bigr].
 \end{aligned}
\end{equation}
For a physical state, $v\ge0$ is obvious, because $v$ is a variance, and
$v=0$ holds exactly when the electron number is sharp. For a trial pair
neither statement is obvious, since the brackets are only a linear
functional. Both follow from $\widetilde G\succeq0$: $v=u^\dagger Cu$ is a
quadratic form of the PSD matrix $C$, so $v\ge0$, and $v=0$ holds exactly
when $Cu=0$, that is, when the local fixed-number contraction holds
(Lemma~\ref{lem:local}). Hence the pair trace in $\M^{\mathsf c}$ is at
least $N_F(N_F-1)/2$, and fixing it at this value gives $v=0$ and restores
the contraction.
The minimum then becomes $E_1^{\mathsf u}$, which is not a lower bound on
$E_k^{\mathsf x}/k$ in general.

Second, the spin setting affects $E_1^{\mathsf x}$ but, as shown in
Sec.~\ref{sec:limit}, not the limit: $E_1^{\mathsf c}$ carries no spin constraint,
whereas $E_1^{\mathsf s}\ge E_1^{\mathsf u}$. The defect $E_1^{\mathsf x}-E_1^{\mathsf c}$ can
therefore depend strongly on $\mathsf x$. In the singlet setting it splits
as $E_1^{\mathsf s}-E_1^{\mathsf c}=(E_1^{\mathsf s}-E_1^{\mathsf u})
+(E_1^{\mathsf u}-E_1^{\mathsf c})$, a spin part and a number part.

Third, part of the defect need not be an artifact of PQG. Fragments that
do not interact may still be correlated, and for a general Hamiltonian
even the exact energy per fragment can decrease under replication: if the
exact energies $E(M)$ of $M$ electrons are not convex at $N_F$, the
replicated system lowers its energy by distributing the electrons
unevenly among the fragments (Sec.~\ref{sec:ppos}). For example,
$\hat H_F=-(\hat N_F-2)^2$ on four spin orbitals with $N_F=2$ has exact
energy $0$, whereas two fragments with $0$ and $4$ electrons have energy
$-4$ per fragment. In the singlet setting a second exact mechanism exists:
if the exact ground state of the fragment has nonzero spin, local spins
can couple to a global singlet. For $\hat H_F=-\hat{\bm S}_F^2$ on two
spatial orbitals with $N_F=2$, the exact energies $E(M)$ are convex at
$N_F$, yet the isolated singlet has energy $0$, whereas two local triplets
coupled to a global singlet have energy $-2$ per fragment. The number part
of the defect is attributable to PQG when the exact problem is stable
against charge disproportionation, that is, when $E(N_F)$ equals the lower
convex envelope of the exact energies at $N_F$; the spin part, in
addition, when the exact isolated singlet minimum equals the exact
unrestricted one.

Fourth, fixing $N_\alpha$ and $N_\beta$ separately does not change the
singlet sets. If $\langle\hat{\bm S}_{\rm tot}^2\rangle=0$ and $G\succeq0$, the
$G$ quadratic form vanishes on $\hat S^z_{\rm tot}$, hence
$\langle \hat Z \hat S^z_{\rm tot}\rangle=0$ for every one-body $\hat Z$; together with
Eq.~\eqref{eq:contraction} this gives the spin-resolved contractions
(\SM{2}).

%% file: sections/limit.tex
\section{The replication limit}
\label{sec:limit}
The main theorem combines two inequalities. Averaging a global feasible
point over permutations of the fragments and restricting it to one
fragment gives a point of $\M^{\mathsf c}$, hence a lower bound
(Sec.~\ref{sec:restriction}). Conversely, a local point is lifted to a
global feasible point with the same RDMs on every fragment
(Secs.~\ref{sec:cancel}--\ref{sec:lift}). Because $\hat H_{\rm tot}$ has no
interfragment terms, such a lift has energy exactly
$k\langle \hat H_F\rangle$; the only question is feasibility.

\subsection{Restriction: a lower bound at every replica count}
\label{sec:restriction}
\begin{proposition}[Restriction]
\label{prop:restriction}
For $\mathsf x\in\{{\mathsf u},{\mathsf s}\}$ and every $k\ge1$,
$E_k^{\mathsf x}/k\ge E_1^{\mathsf c}$.
\end{proposition}

\begin{sketch}
Average a global minimizer over permutations of the fragments. The average
is feasible, has the same energy, and gives every fragment the mean
electron number $N_F$ and the energy $E_k^{\mathsf x}/k$. The global contraction
and global $G\succeq0$ imply the global augmented condition
(Lemma~\ref{lem:local} on the whole orbital space), and the local
$\widetilde G$, $P$, $Q$, and one-particle blocks are principal
submatrices of global ones. The restriction therefore lies in $\M^{\mathsf c}$.
Nothing forces it to have a sharp local electron number or a local
singlet. Details are in \SM{2}.
\end{sketch}

\subsection{What an extension must cancel}
\label{sec:cancel}
With unrestricted spin, an exact ground state of noninteracting fragments
can be chosen as a product of fragment eigenstates, each with an integer
electron number that does not fluctuate, equal to $N_F$ when the fragment
is stable against charge disproportionation (Sec.~\ref{sec:compare}). In
the singlet setting this can fail even exactly, because local spins may
have to couple to a global singlet (Sec.~\ref{sec:compare}). PQG does not
enforce integer local numbers. A local point of
$\M^{\mathsf c}$ may have a nonzero electron-number variance (and a nonzero
local spin), whereas in the replicated system the total electron
number must not fluctuate (and, for the singlet, the total spin must
vanish). If the fragments
were uncorrelated, their local fluctuations would add up instead of
vanishing. They cancel only if correlations appear between fragments that
are physically independent, namely negative interfragment covariances. We
now compute how large these must be.

Let $\hat X$ be a Hermitian one-body operator on $F$, let $\hat X^{(a)}$ be its copy
on fragment $a$, and let $\hat X_{\rm tot}=\sum_{a=1}^k\hat X^{(a)}$. Consider a
global point that is symmetric under permutations of the fragments, as
the averaged point of Proposition~\ref{prop:restriction} is. Every fragment
then has the same mean $\xi_X=\langle \hat X^{(a)}\rangle$ and the same local
variance $v_X=\langle(\hat X^{(a)}-\xi_X)^2\rangle$, and every pair $a\ne b$ has the
same interfragment covariance
$c_X=\langle(\hat X^{(a)}-\xi_X)(\hat X^{(b)}-\xi_X)\rangle$. Expanding the square of
$\hat X_{\rm tot}-k\xi_X=\sum_a(\hat X^{(a)}-\xi_X)$ gives $k$ diagonal terms and
$k(k-1)$ ordered off-diagonal terms,
\begin{align}
 \bigl\langle(\hat X_{\rm tot}-k\xi_X)^2\bigr\rangle
 &=\sum_{a}\bigl\langle(\hat X^{(a)}-\xi_X)^2\bigr\rangle\nonumber\\
 &\quad+\sum_{a\ne b}\bigl\langle(\hat X^{(a)}-\xi_X)(\hat X^{(b)}-\xi_X)\bigr\rangle\nonumber\\
 &=k\,v_X+k(k-1)\,c_X .
 \label{eq:totalfluctuation}
\end{align}

Take $\hat X=\hat N_F$ and label its moments by $N$, so that $\xi_N=N_F$ and $\hat X_{\rm tot}=\hat N_{\rm tot}$. The local
variance $v_N$ is then the number variance $v$ of Sec.~\ref{sec:local}, and both
moments are affine in the 2-RDM: with $\hat N_a=\sum_{i\in a}a_i^\dagger a_i$,
\begin{align}
 v_N&=\langle \hat N_a^2\rangle-N_F^2\nonumber\\
 &=2\sum_{i,j}(\Gamma_{\rm tot})_{i_aj_a;i_aj_a}+N_F-N_F^2,\nonumber\\
 c_N&=\langle \hat N_a\hat N_b\rangle-N_F^2\nonumber\\
 &=2\sum_{i,p}(\Gamma_{\rm tot})_{i_ap_b;i_ap_b}-N_F^2 .
 \label{eq:numbermoments}
\end{align}
A global fixed-number point has no total-number fluctuation: tracing the
contraction~\eqref{eq:contraction} over all spin orbitals gives
\begin{equation}
 \begin{aligned}
 2\sum_{I,J}(\Gamma_{\rm tot})_{IJ;IJ}&=kN_F(kN_F-1),\\
 \langle \hat N_{\rm tot}^2\rangle=kN_F+2\sum_{I,J}(\Gamma_{\rm tot})_{IJ;IJ}&=(kN_F)^2,
 \end{aligned}
\end{equation}
and $\langle(\hat N_{\rm tot}-kN_F)^2\rangle=0$. In Eq.~\eqref{eq:totalfluctuation}
with $\hat X=\hat N_F$ the left side is therefore zero, while the right side contains
the local variance $v_N$ and the covariance $c_N$:
\begin{equation}
 0=\bigl\langle(\hat N_{\rm tot}-kN_F)^2\bigr\rangle
 =\underbrace{k\,v_N}_{\text{local}}
 +\underbrace{k(k-1)\,c_N}_{\text{interfragment}} .
\end{equation}
Without interfragment correlation ($c_N=0$) the right side would be
$kv_N>0$ whenever $v_N>0$: the $k$ local fluctuations would add up. To make it
vanish, the $k(k-1)$ interfragment terms must cancel the $k$ local ones.
Dividing by $k(k-1)$ (for $k\ge2$) gives
\begin{equation}
 c_N=-\frac{v_N}{k-1}.
 \label{eq:requiredcov}
\end{equation}
Each pair of fragments thus carries a negative number covariance, whose size
decreases as $1/(k-1)$ because there are $k-1$ partners per fragment. For
example, in the global-singlet SDPA-GMP solution of the four-site Hubbard
fragment at $k=2$ (Sec.~\ref{sec:hubbard}), each fragment has $v_N=\HubVarF$,
and the two fragments have number covariance $\HubCovF$, as
Eq.~\eqref{eq:requiredcov} requires.
For the global singlet the same steps apply to the three spin components.
The condition $\langle\hat{\bm S}_{\rm tot}^2\rangle=0$ with $G\succeq0$ forces
$\langle \hat S^\alpha_{\rm tot}\rangle=0$, hence $\langle \hat S_F^\alpha\rangle=0$ in
a fragment-symmetric point, and summing Eq.~\eqref{eq:totalfluctuation}
over $\alpha=x,y,z$ gives
$0=\langle\hat{\bm S}_{\rm tot}^2\rangle
=k\langle\hat{\bm S}_F^2\rangle+k(k-1)\langle\hat{\bm S}_F^{(a)}\!\cdot\hat{\bm S}_F^{(b)}\rangle$,
that is,
$\langle\hat{\bm S}_F^{(a)}\!\cdot\hat{\bm S}_F^{(b)}\rangle=-\langle\hat{\bm S}_F^2\rangle/(k-1)$.
Like the number covariance, this spin correlation is affine in the
interfragment 2-RDM. Write the spin orbitals as $p\sigma$ (spatial
orbital $p$, spin $\sigma$), so that
$\hat{\bm S}_F^{(a)}=\tfrac12\sum_p\sum_{\sigma\tau}a_{p\sigma,a}^\dagger
\bm\sigma_{\sigma\tau}a_{p\tau,a}$ with the Pauli matrices $\bm\sigma$.
The identity $\sum_\alpha\sigma^\alpha_{\sigma\tau}\sigma^\alpha_{\sigma'\tau'}
=2\delta_{\sigma\tau'}\delta_{\tau\sigma'}-\delta_{\sigma\tau}\delta_{\sigma'\tau'}$
and normal ordering across the two fragments give, for $a\ne b$,
\begin{align}
 \hat{\bm S}_F^{(a)}\!\cdot\hat{\bm S}_F^{(b)}
 &=\tfrac12\sum_{p,q}\sum_{\sigma\tau}a_{p\sigma,a}^\dagger a_{p\tau,a}\,
  a_{q\tau,b}^\dagger a_{q\sigma,b}-\tfrac14\hat N_a\hat N_b,\nonumber\\
 \langle\hat{\bm S}_F^{(a)}\!\cdot\hat{\bm S}_F^{(b)}\rangle
 &=\sum_{p,q}\sum_{\sigma\tau}
  (\Gamma_{\rm tot})_{(p\tau)_a(q\sigma)_b;(p\sigma)_a(q\tau)_b}\nonumber\\
 &\quad-\tfrac12\sum_{i,j}(\Gamma_{\rm tot})_{i_aj_b;i_aj_b} .
 \label{eq:spincorr}
\end{align}
The first term exchanges the spins of the two fragments.
In the same Hubbard solution $\langle\hat{\bm S}_F^2\rangle=\HubSpinSq$, so the two
fragments must have $\langle\hat{\bm S}_F^{(a)}\!\cdot\hat{\bm S}_F^{(b)}\rangle=-\HubSpinSq$.

Only these quantities have to cancel. The fixed-number condition forbids any
fluctuation of the total electron number, and the singlet condition makes
the total spin vanish; no other one-body operator is constrained in this way. The total
occupation of a given orbital, summed over the fragments, may fluctuate
freely, for example. The lift therefore needs to prescribe the required
interfragment covariances only for $\hat N_F$ and, in the singlet case, for the
three spin components; these are the controlled operators of the next
subsection.
The product of local RDMs has $c_X=0$ and therefore violates the
fixed-number condition as soon as the fragment has a nonzero number
variance, and the singlet condition as soon as it has a nonzero local
spin. A lift must create \emph{negative interfragment covariances of order
$1/(k-1)$} that cancel the local fluctuations; the fluctuations themselves
remain. This is consistent with the observation of Nakata and
Yasuda\cite{NakataYasuda2009} that the interfragment 2-RDM elements of
replicated PQG solutions do not follow the cumulant expansion. In this
sense the lowering of the energy per fragment under replication is not
local: it is made possible by interfragment covariances, which let each
fragment carry the number and spin fluctuations that the isolated problem
forbids, whereas every intrafragment element of the lifted RDMs is that of
the single-fragment problem. Whether the resulting lowering is an artifact
of PQG depends on the exact problem: it is when the exact energies are
stable against charge disproportionation and, in the singlet setting, the
exact isolated ground state is a singlet (Sec.~\ref{sec:compare}).

\subsection{Controlled operators and the covariance kernel}
\label{sec:kernel}
The operators whose total fluctuation must vanish are
$\hat X_N=\hat N_F$ for $\mathsf x={\mathsf u}$, and $\hat X_N$ together with the three spin
components $\hat X_\alpha=\hat S_F^\alpha$ for $\mathsf x={\mathsf s}$. We write
$\mathcal A_{\mathsf u}=\{N\}$ and $\mathcal A_{\mathsf s}=\{N,x,y,z\}$, so
that the number of controlled operators is $|\mathcal A_{\mathsf u}|=1$ and
$|\mathcal A_{\mathsf s}|=4$ (here $|\cdot|$ counts elements), and
$\chi^\alpha$ for the coefficient matrix of $\hat X_\alpha$. For
$\alpha\in\mathcal A_{\mathsf x}$ define the covariance kernel and variance
\begin{align}
 (C_\alpha)_{ij}&=\langle a_j^\dagger a_i(\hat X_\alpha-\xi_\alpha)\rangle,
 \nonumber\\
 v_\alpha&=\langle(\hat X_\alpha-\xi_\alpha)^2\rangle=\Tr(\chi^\alpha C_\alpha),
 \label{eq:kappa}
\end{align}
which for fixed $\xi_\alpha$ are affine in $(\gamma,\Gamma)$ by CAR
($\xi_\alpha=N_F$ for the number operator, and $\xi_\alpha=0$ for the spin
components at the spin-rotation-invariant points used with
$\mathsf x={\mathsf s}$); $\xi_\alpha$ and $v_\alpha$ are
the mean $\xi_X$ and the local variance $v_X$ of Sec.~\ref{sec:cancel} for
$\hat X=\hat X_\alpha$. Since $\hat X_\alpha=\sum_{kl}\chi^\alpha_{kl}a_k^\dagger a_l$ is
a combination of the operators that label the columns of $C$, the
covariance of each one-body operator with $\hat X_\alpha$ is the same
combination of columns, $C_\alpha=C\chi^\alpha$. Thus $C_\alpha$ is the
column of the covariance table $C$ that belongs to $\hat X_\alpha$, with its
$r^2$ entries arranged as an $r\times r$ matrix: $(C_\alpha)_{ij}$ is the
covariance, within one fragment, of $a_j^\dagger a_i$ with $\hat X_\alpha$. For
$\hat X_N=\hat N_F$ one has $\chi^N=I$, and $C_N$ is the sum of the columns of the
occupation numbers $a_k^\dagger a_k$. $C_\alpha$ does not depend on the
replica count $k$. Define
\begin{equation}
 \Pi_{ij,pq}=\sum_{\alpha\in\mathcal A_{\mathsf x},\;v_\alpha>0}
 \frac{(C_\alpha)_{ij}\,(C_\alpha)_{qp}}{v_\alpha}.
 \label{eq:Pi}
\end{equation}
Section~\ref{sec:lift} builds replicated RDMs from a local point of
$\M^{\mathsf c}$; we call this construction the lift. In the lift, two
one-body operators on different fragments receive $-1/(k-1)$ times the
corresponding entry of $\Pi$ as their covariance. At a
fragment-symmetric point the covariances with a controlled operator are
forced: the centered total $\hat X_{\alpha,{\rm tot}}-k\xi_\alpha$ has zero
variance, so its covariance with any local one-body operator $\hat A^{(a)}$
vanishes, and $\mathrm{cov}(\hat A^{(a)},\hat X_\alpha^{(b)})=
-\mathrm{cov}(\hat A,\hat X_\alpha)/(k-1)$ for $b\ne a$, which for
$\hat A=\hat X_\alpha$ is Eq.~\eqref{eq:requiredcov}. The lift meets these
conditions; for all other pairs its prescription is a choice. It is not
the only one, and a numerical optimum need not have this form, but with
it the number-conserving covariance blocks remain PSD for every $k$
(Lemma~\ref{lem:projection} below), and the remaining blocks for $k$ above
an explicit threshold (Proposition~\ref{prop:lift}).

Since $\mathcal A_{\mathsf x}$ contains at most the electron number and the
three spin components, Eq.~\eqref{eq:Pi} can be written out. As a matrix
on index pairs, with the Hermitian $C_\alpha$ (shown after Lemma~\ref{lem:projection}) read as vectors,
\begin{equation}
 \begin{aligned}
 \Pi^{\mathsf u}&=\frac{C_NC_N^\dagger}{v_N},\\
 \Pi^{\mathsf s}&=\frac{C_NC_N^\dagger}{v_N}
   +\frac{1}{v_S}\sum_{\alpha=x,y,z}C_\alpha C_\alpha^\dagger,
 \qquad v_S=\tfrac13\langle\hat{\bm S}_F^2\rangle,
 \end{aligned}
 \label{eq:Piexplicit}
\end{equation}
where a term with zero variance is omitted. The four terms of
$\Pi^{\mathsf s}$ are independent, and the three spin components share the
variance $v_S$, because at a spin-rotation-invariant point
\begin{equation}
 \mathrm{cov}(\hat N_F,\hat S_F^\alpha)=0,\qquad
 \mathrm{cov}(\hat S_F^\alpha,\hat S_F^\beta)=\delta_{\alpha\beta}\,v_S .
 \label{eq:spinmoments}
\end{equation}
The first holds because $\langle\hat S_F^\alpha\rangle=0$ and
$\langle\hat N_F\hat S_F^\alpha\rangle$ is the expectation value of a vector
operator, which vanishes at an invariant point; the second because an
invariant symmetric tensor of rank two is proportional to
$\delta_{\alpha\beta}$, and its trace is $\langle\hat{\bm S}_F^2\rangle$.
The same invariance also fixes the spin structure of the kernels:
$C_N=L\otimes I_2$ and $C_\alpha=\tfrac12K\otimes\sigma^\alpha$ with spatial
matrices $L$ and $K$ and the Pauli matrices $\sigma^\alpha$, so that $\Pi^{\mathsf s}$ has
one spin-singlet direction tied to the number fluctuation (if $v_N>0$)
and three spin-triplet directions tied to the spin fluctuation (if
$v_S>0$), hence rank four when both variances are positive (\SM{3}).
Only spin-rotation invariance of the local point is used, not a local
singlet. A local point with $\langle\hat{\bm S}_F^2\rangle>0$, such as an equal
mixture of the two components of a doublet, has $v_S>0$, and the spin
terms of $\Pi^{\mathsf s}$ are what let the local spins couple to a global
singlet; for a local singlet $v_S=0$, the spin terms drop out, and
$\Pi^{\mathsf s}=\Pi^{\mathsf u}$. Any point of $\M^{\mathsf c}$ becomes
spin-rotation invariant, with the same energy for a spin-rotation-invariant
$\hat H_F$, when it is averaged over spin rotations
(Theorem~\ref{thm:limit}).

Consider first the unrestricted case, in which only the electron number
is controlled and $\Pi=\Pi^{\mathsf u}=C_NC_N^\dagger/v_N$ has rank one (zero if
$v_N=0$). Each
factor $(C_N)_{ij}$ in $\Pi_{ij,pq}=(C_N)_{ij}(C_N)_{qp}/v_N$ is
a covariance with $\hat N_F$: by the definition~\eqref{eq:kappa} with
$\xi_N=\langle \hat N_F\rangle=N_F$,
\begin{equation}
 \begin{aligned}
 (C_N)_{ij}&=\langle a_j^\dagger a_i(\hat N_F-N_F)\rangle\\
 &=\langle a_j^\dagger a_i\hat N_F\rangle-\langle a_j^\dagger a_i\rangle\langle \hat N_F\rangle\\
 &=\mathrm{cov}(a_j^\dagger a_i,\hat N_F),
 \end{aligned}
\end{equation}
and since $C_N$ is Hermitian,
$(C_N)_{qp}=\mathrm{cov}(\hat N_F,a_p^\dagger a_q)$. The interfragment covariance
that the lift assigns to $a_j^\dagger a_i$ on fragment $a$ and
$a_p^\dagger a_q$ on fragment $b\ne a$ is therefore
\begin{equation}
 -\frac{1}{k-1}\,\Pi_{ij,pq}
 =-\frac{1}{k-1}\,
 \frac{\mathrm{cov}(a_j^\dagger a_i,\hat N_F)\,
       \mathrm{cov}(\hat N_F,a_p^\dagger a_q)}{v_N},
\end{equation}
and, by linearity, any one-body operators $\hat A$ on fragment $a$ and $\hat B$ on
fragment $b$ receive
\begin{equation}
 -\frac{1}{k-1}\,\frac{\mathrm{cov}(\hat A,\hat N_F)\,\mathrm{cov}(\hat N_F,\hat B)}{v_N}.
\end{equation}
This is the required rule. For $\hat A=\hat B=\hat N_F$ both factors in the numerator
equal $\mathrm{cov}(\hat N_F,\hat N_F)=v_N$, and the covariance is
$-v_N^2/[(k-1)v_N]=-v_N/(k-1)$, which is Eq.~\eqref{eq:requiredcov}. An operator
uncorrelated with $\hat N_F$ receives no interfragment covariance, so the lift
disturbs only what is tied to the number fluctuation. When $v_N=0$, that is,
for the isolated fixed-number problem, $C_N=0$ and $\Pi=0$.

For the global singlet the three spin components are controlled in
addition to $\hat N_F$, and $\Pi=\Pi^{\mathsf s}$ has the four terms of
Eq.~\eqref{eq:Piexplicit}. The same steps show that the lift assigns
to one-body operators $\hat A$ and $\hat B$ on different fragments the covariance
\begin{equation}
 \begin{aligned}
 -\frac{1}{k-1}\Bigl[&
 \frac{\mathrm{cov}(\hat A,\hat N_F)\,\mathrm{cov}(\hat N_F,\hat B)}{v_N}\\
 &+\sum_{\alpha=x,y,z}
 \frac{\mathrm{cov}(\hat A,\hat S_F^\alpha)\,\mathrm{cov}(\hat S_F^\alpha,\hat B)}{v_S}\Bigr].
 \end{aligned}
\end{equation}
For $\hat A=\hat B=\hat N_F$ or $\hat A=\hat B=\hat S_F^\beta$ only one term survives by
Eq.~\eqref{eq:spinmoments}, and the covariance is $-v_N/(k-1)$ or
$-v_S/(k-1)$, as required. In statistical terms, $\Pi$ is the part of
the local covariance $C$ explained by linear regression on the controlled
operators, and $C-\Pi$ is the residual.

For the lift to be feasible, $\Pi$ must not exceed the local covariance
itself, since otherwise some combination of one-body operators would get a
negative variance in the replicated system. The next lemma shows that this
holds.

\begin{lemma}[Covariance kernel]
\label{lem:projection}
Let $\Rset\in\M^{\mathsf c}$, and for $\mathsf x={\mathsf s}$ let $\Rset$ be invariant
under spin rotations. Then $\Pi$ of Eq.~\eqref{eq:Piexplicit} satisfies $\Pi\succeq0$,
\begin{equation}
 \widetilde G-\begin{pmatrix}0&0\\0&\Pi\end{pmatrix}\succeq0,
 \quad\text{equivalently}\quad \Pi\preceq C,
\end{equation}
$(C-\Pi)\chi^\alpha=0$ for every $\alpha\in\mathcal A_{\mathsf x}$, and
$\|C_\alpha\|_{\rm op}^2\le v_\alpha/4$.
\end{lemma}
Under these assumptions each $C_\alpha$ is Hermitian, because
$(C_\alpha)_{ij}-\overline{(C_\alpha)_{ji}}=\langle[a_j^\dagger a_i,\hat X_\alpha]\rangle$
vanishes: the commutator is zero for the number operator, and its
expectation value is zero for the spin components at a
spin-rotation-invariant point.

Here $\|C_\alpha\|_{\rm op}$, with $C_\alpha$ arranged as an $r\times r$
matrix, is the largest absolute eigenvalue of the
Hermitian matrix $C_\alpha$, that is, the largest covariance
$|\mathrm{cov}(n_b,\hat X_\alpha)|$ of the occupation $n_b$ of a
normalized orbital $b$ with $\hat X_\alpha$; the bound is used for the
mixed blocks in Proposition~\ref{prop:lift}.

\begin{sketch}
For a one-body operator $\hat Z=\sum_{pq}z_{pq}a_p^\dagger a_q$ on the
fragment, $z^\dagger Cz=\mathrm{var}(\hat Z)$ is the variance of $\hat Z$ computed
from $(\gamma,\Gamma)$, and $\widetilde G\succeq0$, whose Schur complement is
$C$, makes these variances nonnegative. Covariances $\mathrm{cov}(\hat A,\hat B)$ of one-body
operators are formed in the same way, so the Cauchy--Schwarz inequality
$|\mathrm{cov}(\hat A^\dagger,\hat B)|^2\le\mathrm{var}(\hat A)\,\mathrm{var}(\hat B)$ holds even though
$(\gamma,\Gamma)$ need not come from a state. By the definition of $C_\alpha$,
\begin{equation}
 z^\dagger\Pi z=\sum_{\alpha:\,v_\alpha>0}
 \frac{|\mathrm{cov}(\hat Z,\hat X_\alpha)|^2}{v_\alpha};
\end{equation}
the two covariance factors are complex conjugates of each other because
$C_\alpha$ is Hermitian, which is automatic for $\hat N_F$ (it commutes with
every one-body operator) and follows from spin-rotation invariance for the
spin components.

In the unrestricted case only $\hat N_F$ is controlled, and
\begin{equation}
 \begin{aligned}
 z^\dagger Cz-z^\dagger\Pi z
 &=\mathrm{var}(\hat Z)-\frac{|\mathrm{cov}(\hat Z,\hat N_F)|^2}{v_N}\\
 &=\mathrm{var}(\hat Z-\beta \hat N_F)\ge0,
 \end{aligned}
\end{equation}
with $\beta=\mathrm{cov}(\hat N_F,\hat Z)/v_N$, by Eq.~\eqref{eq:regression}
with $\hat A=\hat Z$ and $\hat B=\hat N_F$.
This is linear regression: $\beta \hat N_F$ is the part of $\hat Z$ that follows the
number fluctuation, $\hat Z-\beta \hat N_F$ is the residual, and the explained
variance cannot exceed the total variance. Hence $\Pi\preceq C$, which
is the augmented form because $C-\Pi$ is the Schur complement of the
corner entry of $\widetilde G-(0\oplus\Pi)$, and
$\Pi\succeq0$ is clear from the sum of squares. For $\hat Z=\hat N_F$ (coefficient
matrix $\chi^N=I$), Eq.~\eqref{eq:regression} with $\hat A=\hat B=\hat N_F$
gives
\begin{equation}
 \begin{aligned}
 (\chi^N)^\dagger(C-\Pi)\chi^N
 &=\mathrm{var}(\hat N_F)-\frac{|\mathrm{cov}(\hat N_F,\hat N_F)|^2}{v_N}\\
 &=v_N-\frac{v_N^2}{v_N}=0 :
 \end{aligned}
\end{equation}
the residual of $\hat N_F$ has zero variance, and in the direction of the
controlled operator itself $\Pi$ removes all of the variance. Since
$C-\Pi\succeq0$, a vector with zero quadratic form lies in its kernel,
so $(C-\Pi)\chi^N=0$. In the singlet case one regresses on $\hat N_F$ and the three spin
components together. Because they are mutually uncorrelated and the spin
components share the variance $v_S$ [Eq.~\eqref{eq:spinmoments}], the
regression separates into the four terms of Eq.~\eqref{eq:Piexplicit},
\begin{equation}
 \mathrm{var}(\hat Z)-z^\dagger\Pi^{\mathsf s}z
 =\mathrm{var}\Bigl(\hat Z-\beta_N\hat N_F-\sum_{\alpha=x,y,z}\beta_\alpha\hat S_F^\alpha\Bigr)\ge0,
\end{equation}
with $\beta_N=\mathrm{cov}(\hat N_F,\hat Z)/v_N$ and
$\beta_\alpha=\mathrm{cov}(\hat S_F^\alpha,\hat Z)/v_S$; the cross terms between
the four regressors vanish by Eq.~\eqref{eq:spinmoments}. A regressor
with zero variance is dropped (its $\beta$ is set to zero): by
Cauchy--Schwarz it is uncorrelated with every $\hat Z$, so $C_\alpha=0$, and
it contributes neither to $\Pi^{\mathsf s}$ nor to the residual. The same
argument as above gives $(C-\Pi^{\mathsf s})\chi^\alpha=0$ for each
controlled operator; for one with zero variance this is
$C\chi^\alpha=C_\alpha=0$.

For the norm bound take $\hat Z=n_b=b^\dagger b$, the occupation of a single
normalized orbital $b$, with $g=\langle n_b\rangle$. The CAR give
$n_b^2=n_b$, so $\mathrm{var}(n_b)=g(1-g)\le1/4$, and Cauchy--Schwarz
gives $|\mathrm{cov}(n_b,\hat X_\alpha)|^2\le v_\alpha/4$. The left-hand side is
the square of a diagonal element of $C_\alpha$ in the orbital $b$; since
$C_\alpha$ is Hermitian, its largest such element bounds
$\|C_\alpha\|_{\rm op}$. In the cation--anion example of
Sec.~\ref{sec:lift}, each frontier occupation equals $(\hat N_F-N_F+1)/2$ in both
determinants and follows the number fluctuation completely, so $C-\Pi$
vanishes on it; an operator that moves an electron, for example from a
core to a virtual orbital, is uncorrelated with $\hat N_F$ and keeps its full
variance in $C-\Pi$. Details,
including the general hypotheses under which the lemma holds, are in
\SM{3}.
\end{sketch}

Both ingredients of the lift, $\Pi\preceq C$ and the bound on
$\|C_\alpha\|_{\rm op}$, rest on the Cauchy--Schwarz inequality for the
PSD form $C$, that is, on $\widetilde G\succeq0$, and hold although
$(\gamma,\Gamma)$ need not come from a state; the same condition is what
restriction produces (Proposition~\ref{prop:restriction}).

\subsection{The explicit lift}
\label{sec:lift}
The lift copies the same local RDMs $\Rset=(\gamma,\Gamma)\in\M^{\mathsf c}$
onto every fragment and fills in only the elements between fragments, so
as to supply the required covariances. The result is a pair
$\Rset_{\rm tot}=(\gamma_{\rm tot},\Gamma_{\rm tot})$ for the whole system
of $k$ fragments. Fix
$k\ge2$, put $\tau=1/(k-1)$, and write $i_a$ for spin orbital $i$ of
fragment $a$. As in Eq.~\eqref{eq:Pi}, every sum over $\alpha$ below runs
over $\alpha\in\mathcal A_{\mathsf x}$ with $v_\alpha>0$; a term with
$v_\alpha=0$ is omitted, and then $C_\alpha=0$. Then
\begin{align}
 (\gamma_{\rm tot})_{i_a j_b}&=\delta_{ab}\,\gamma_{ij},\nonumber\\
 (\Gamma_{\rm tot})_{i_aj_a;l_am_a}&=\Gamma_{ij;lm},\nonumber\\
 (\Gamma_{\rm tot})_{i_ap_b;j_aq_b}
 &=\tfrac12\Bigl[\gamma_{ij}\gamma_{pq}
  -\tau\sum_{\alpha}\frac{(C_\alpha)_{ij}(C_\alpha)_{pq}}{v_\alpha}\Bigr]
 \nonumber\\&\qquad(a\ne b),
 \label{eq:liftGamma}
\end{align}
together with the entries related to these by antisymmetry, and all other
entries, which would change some fragment's electron number, are zero. Every
fragment thus carries a copy of $(\gamma,\Gamma)$, and there are no 1-RDM
elements between fragments. If the fragments were independent, the
interfragment 2-RDM elements would have the product form
$\tfrac12\gamma_{ij}\gamma_{pq}$ alone, with no interfragment covariance;
the lift adds to it the correction prescribed by $\Pi$, which supplies the
negative covariances of Sec.~\ref{sec:cancel}. The global 2-RDM is
therefore not a product. Because $\hat H_{\rm tot}$ has no interfragment terms, the
energy is exactly $k\langle \hat H_F\rangle$, whatever the interfragment
elements are; the only question is whether $\Rset_{\rm tot}$ satisfies the PQG
conditions and the fixed-number contraction.

\emph{Example.} The following case shows every step explicitly. Let the
fragment have $r$ spin orbitals and $1\le N_F\le r-1$, and write a Slater
determinant by the occupations of the spin orbitals $1,\dots,r$. Take the
two determinants of the Hartree--Fock type
\begin{align}
 |\Phi_-\rangle&=|\underbrace{1\cdots1}_{N_F-1}\,\underbrace{00}_{2}\,
  \underbrace{0\cdots0}_{r-N_F-1}\rangle,\nonumber\\
 |\Phi_+\rangle&=|\underbrace{1\cdots1}_{N_F-1}\,\underbrace{11}_{2}\,
  \underbrace{0\cdots0}_{r-N_F-1}\rangle,
\end{align}
the cation with $N_F-1$ electrons and the anion with $N_F+1$ electrons. They
differ only in the spin orbitals $N_F$ and $N_F+1$. We call the first $N_F-1$
spin orbitals core, $N_F$ and $N_F+1$ frontier, and the last $r-N_F-1$ virtual,
and write $o_i^\pm\in\{0,1\}$ for the occupation of spin orbital $i$ in
$\Phi_\pm$, the $i$th digit above. For example, $r=4$ and $N_F=2$ give
$|\Phi_-\rangle=|1000\rangle$ and $|\Phi_+\rangle=|1110\rangle$, and $r=2$
and $N_F=1$ give $|00\rangle$ and $|11\rangle$.

\emph{(i) Local RDMs.} Take as local point the physical ensemble
$\tfrac12|\Phi_-\rangle\langle\Phi_-|+\tfrac12|\Phi_+\rangle\langle\Phi_+|$,
cation and anion with equal weights; it has mean electron number $N_F$ and
lies in $\M^{\mathsf c}$. Both determinants are built from the same
orbitals, so the RDMs are diagonal,
\begin{align}
 \gamma_{ij}&=\tfrac12\bigl(o_i^-+o_i^+\bigr)\delta_{ij},\nonumber\\
 \Gamma_{ij;lm}&=\tfrac14\bigl(o_i^-o_j^-+o_i^+o_j^+\bigr)
  \bigl(\delta_{il}\delta_{jm}-\delta_{im}\delta_{jl}\bigr),
\end{align}
for all spin orbitals $i,j,l,m$ of the fragment. That is,
$\langle n_i\rangle=1$, $\tfrac12$, and $0$ for core, frontier, and virtual
spin orbitals, and $\langle n_in_j\rangle=2\Gamma_{ij;ij}$ is $1$ for two
core spin orbitals, $\tfrac12$ for a core and a frontier or two frontier
spin orbitals, and $0$ if one of them is virtual. The smallest case,
$r=2$ and $N_F=1$, has no core or virtual orbitals: an empty and a doubly
occupied fragment, $\tfrac12|00\rangle\langle00|+\tfrac12|11\rangle\langle11|$.

\emph{(ii) Variance and covariance kernel.} From these RDMs we obtain the
three quantities that the lift uses. Since $\hat N_F-N_F=\mp1$ in $\Phi_\mp$, the
local number variance of Sec.~\ref{sec:local} is
$v_N=\langle(\hat N_F-N_F)^2\rangle=1$. The covariance kernel of
Eq.~\eqref{eq:kappa} is diagonal, because an off-diagonal operator
$a_j^\dagger a_i$ ($i\ne j$) changes the occupation pattern and has zero
expectation in both determinants, and
\begin{align}
 (C_N)_{ii}&=\langle n_i(\hat N_F-N_F)\rangle\nonumber\\
 &=\tfrac12\bigl[o_i^-(-1)+o_i^+(+1)\bigr]=\tfrac12\bigl(o_i^+-o_i^-\bigr),
\end{align}
which is $\tfrac12$ on the two frontier spin orbitals and zero on the core
and virtual ones: only the frontier occupations follow the number
fluctuation. Its trace is $\Tr C_N=1=v_N$. Finally, the matrix of
Eq.~\eqref{eq:Pi}, which has a single term because only $\hat N_F$ is
controlled, is $\Pi_{ij,pq}=(C_N)_{ij}(C_N)_{qp}/v_N=\tfrac14\delta_{ij}\delta_{pq}$
for frontier $i$ and $p$, and zero otherwise.

\emph{(iii) Lift.} For $k$ fragments the lift uses $\tau=1/(k-1)$, the
factor that spreads the required covariance over the $k-1$ partners of
each fragment (Sec.~\ref{sec:cancel}). Equation~\eqref{eq:liftGamma} with
these quantities gives, for fragments $a\ne b$,
\begin{align}
 (\gamma_{\rm tot})_{i_aj_b}&=0,\nonumber\\
 (\Gamma_{\rm tot})_{i_ap_b;j_aq_b}
 &=\tfrac12\bigl[\gamma_{ii}\gamma_{pp}-\tau (C_N)_{ii}(C_N)_{pp}\bigr]\delta_{ij}\delta_{pq},
\end{align}
while every fragment keeps the local RDMs of (i). The interfragment
elements therefore keep their product values $\tfrac12\gamma_{ii}\gamma_{pp}$, except
between two frontier spin orbitals, where
$(\Gamma_{\rm tot})_{i_ap_b;i_ap_b}=\tfrac12\bigl[\tfrac14-\tfrac{\tau}{4}\bigr]
=\tfrac{1-\tau}{8}$.

Taking traces recovers a fixed total electron number $kN_F$. First,
$\Tr\gamma_{\rm tot}=k\,(\Tr\gamma)=kN_F$. Second, the partial trace of the
2-RDM reproduces the 1-RDM through the fixed-number contraction. For spin
orbitals $i,j$ of fragment $a$ (fixed, not summed), split the sum over the
partner spin orbital $P$ into the $aaaa$ block ($P=p$ in fragment $a$) and
the $abab$ blocks ($P=p$ in a fragment $b\ne a$), and insert the elements:
inside a fragment $2\Gamma_{ip;jp}=\delta_{ij}\langle n_in_p\rangle$ for
$p\ne i$ and zero for $p=i$, and between fragments the expression above
with $q=p$. Then
\begin{align}
 &2\sum_P(\Gamma_{\rm tot})_{i_aP;j_aP}\nonumber\\
 &=\underbrace{2\sum_{p\in a}\Gamma_{ip;jp}}_{aaaa}
  +\underbrace{2\sum_{b\ne a}\sum_{p\in b}
   (\Gamma_{\rm tot})_{i_ap_b;j_ap_b}}_{abab}\nonumber\\
 &=\delta_{ij}\sum_{p\ne i}\langle n_in_p\rangle\nonumber\\
 &\quad+\delta_{ij}\sum_{b\ne a}\sum_{p}\bigl[\gamma_{ii}\gamma_{pp}-\tau (C_N)_{ii}(C_N)_{pp}\bigr]\nonumber\\
 &=\delta_{ij}\bigl[\langle n_i\hat N_F\rangle-\gamma_{ii}\bigr]\nonumber\\
 &\quad+\delta_{ij}(k-1)\bigl[N_F\gamma_{ii}-\tau (C_N)_{ii}\bigr]\nonumber\\
 &=\delta_{ij}\bigl[(C_N)_{ii}+N_F\gamma_{ii}-\gamma_{ii}\bigr]\nonumber\\
 &\quad+\delta_{ij}\bigl[(k-1)N_F\gamma_{ii}-(C_N)_{ii}\bigr]\nonumber\\
 &=(kN_F-1)\,\gamma_{ii}\,\delta_{ij}=(kN_F-1)(\gamma_{\rm tot})_{i_aj_a},
\end{align}
where $\sum_p\gamma_{pp}=N_F$, $\sum_p(C_N)_{pp}=\Tr C_N=1$, $\langle n_i\hat N_F\rangle=N_F\gamma_{ii}+(C_N)_{ii}$
(the definition of $C_N$), and $(k-1)\tau=1$ were used. The covariance
$(C_N)_{ii}$ in the $aaaa$ block is cancelled by the $abab$ blocks. Third, summing
the diagonal ($i=j$) over the $rk$ spin orbitals gives the pair
normalization $\sum_{I,J}(\Gamma_{\rm tot})_{IJ;IJ}=\tfrac12(kN_F-1)\Tr
\gamma_{\rm tot}=kN_F(kN_F-1)/2$.

The only nonzero interfragment elements are thus the pair occupations
(and those related by antisymmetry). Because orbital $i$ of fragment $a$
and orbital $p$ of fragment $b$ are different orbitals,
$n_{i,a}n_{p,b}=a_{i,a}^\dagger a_{p,b}^\dagger a_{p,b}a_{i,a}$, so that
\begin{equation}
 \langle n_{i,a}n_{p,b}\rangle=2(\Gamma_{\rm tot})_{i_ap_b;i_ap_b}
 =\gamma_{ii}\gamma_{pp}-\tau (C_N)_{ii}(C_N)_{pp} .
\end{equation}
For two frontier spin orbitals this is $\tfrac14-\tfrac{\tau}{4}
=\tfrac{1-\tau}{4}$; for every other pair it is the product $\gamma_{ii}\gamma_{pp}$ of
independent fragments. The product term $\tfrac14$ is the probability that
both frontier orbitals are occupied if the fragments are independent; the
correction $-\tau/4$ reduces it, and for $k=2$ ($\tau=1$) cancels it, so
that $\langle n_{i,a}n_{p,b}\rangle=0$. Table~\ref{tab:liftexample} lists
these elements. Summing over $i$ and $p$ gives
$\langle \hat N_a\hat N_b\rangle-N_F^2=-\tau\bigl(\sum_i(C_N)_{ii}\bigr)^2=-\tau=-v_N/(k-1)$,
as Eq.~\eqref{eq:requiredcov} requires, and the total number does not
fluctuate. For $k=2$ ($\tau=1$) this reads explicitly
\begin{align}
 \langle \hat N_a\rangle&=\textstyle\sum_i\gamma_{ii}=N_F,\nonumber\\
 \langle \hat N_a^2\rangle&=N_F^2+v_N=N_F^2+1,\nonumber\\
 \langle \hat N_a\hat N_b\rangle&=\textstyle\sum_{i,p}\bigl(\gamma_{ii}\gamma_{pp}-(C_N)_{ii}(C_N)_{pp}\bigr)=N_F^2-1,
\end{align}
so each fragment has variance $1=v_N$, the two fragments have covariance
$-1=-v_N/(k-1)$, and $\langle \hat N_{\rm tot}^2\rangle=2(N_F^2+1)+2(N_F^2-1)=4N_F^2
=\langle \hat N_{\rm tot}\rangle^2$: the total number $2N_F$ does not fluctuate.

\begin{table}[t]
\caption{Cation--anion example: RDM elements of the lift between two
frontier spin orbitals $i,p\in\{N_F,N_F+1\}$ of fragments $a\ne b$, compared
with two different frontier spin orbitals of one fragment ($i\ne p$;
for $i=p$ the 2-RDM element vanishes by antisymmetry). All other interfragment
elements keep their product values. $\tau=1/(k-1)$.}
\label{tab:liftexample}
\footnotesize
\begin{ruledtabular}
\begin{tabular}{lcccc}
 & $(\gamma_{\rm tot})_{i_aj_b}$ & $(\Gamma_{\rm tot})_{i_ap_b;i_ap_b}$
 & $\langle n_{i,a}n_{p,b}\rangle$ & covariance \\
\hline
same fragment & $\tfrac12\delta_{ij}$ & $\tfrac14$ & $\tfrac12$ & $\tfrac14$ \\
$k=2$ & $0$ & $0$ & $0$ & $-\tfrac14$ \\
$k=3$ & $0$ & $\tfrac1{16}$ & $\tfrac18$ & $-\tfrac18$ \\
$k=4$ & $0$ & $\tfrac1{12}$ & $\tfrac16$ & $-\tfrac1{12}$ \\
$k\to\infty$ & $0$ & $\tfrac18$ & $\tfrac14$ & $0$ \\
\end{tabular}
\end{ruledtabular}
\end{table}

\emph{(iv) What the lift represents.} For $k=2$ the frontier orbitals of
the two fragments are never occupied simultaneously, and the lift is the
physical ensemble
$\tfrac12|\Phi_+\Phi_-\rangle\langle\Phi_+\Phi_-|
+\tfrac12|\Phi_-\Phi_+\rangle\langle\Phi_-\Phi_+|$, in which one fragment is
the anion and the other the cation, with total number $2N_F$ (for $r=4$ and
$N_F=2$, the configurations $|1110,1000\rangle$ and $|1000,1110\rangle$). For $k=3$ the
lift has total number $3N_F$ without fluctuation. In a physical state with
these moments, the occupations $1$ and $0$ keep the core spin orbitals
occupied and the virtual ones empty, and
$\langle(n_{N_F}-n_{N_F+1})^2\rangle=\tfrac12+\tfrac12-2\times\tfrac12=0$ makes
the two frontier spin orbitals of each fragment both occupied or both
empty. Every fragment would then be a cation or an anion, $\hat N_a-N_F=\pm1$, and
the sum of three such numbers is odd and cannot vanish; no physical state
has these moments. The same parity argument applies to every odd $k$. The lift nevertheless satisfies the PQG conditions. For
$r=2$ and $N_F=1$, $\eta=\tfrac12$ and Eq.~\eqref{eq:Kexplicit} holds for
every $k\ge2$. With core and virtual orbitals the occupations $1$ and $0$
lie outside the hypothesis of Proposition~\ref{prop:lift}, but these
orbitals can be split off (Lemma~\ref{lem:frozen} of Sec.~\ref{sec:monotone}):
the lift is the product of the lift of the frontier pair with a fixed
determinant on the core, so the PQG conditions hold for every $k\ge2$, as
also confirmed numerically in \SM{4}. For odd $k$ the lift is
thus a PQG-feasible point that is not $N$-representable, and its
interfragment correlation is exactly what PQG admits between fragments
that are physically independent. As $k$ grows, the interfragment
covariance $-\tau/4$ is diluted and the elements approach the product
values of independent fragments.

To check the PQG
conditions one needs the blocks of the $P$, $Q$, and $G$ matrices that
involve two different fragments (the mixed blocks): pairs of annihilators $a_{p,a}a_{r,b}$ (the $P$
condition), pairs of creators (the $Q$ condition), and transfers
$a_{p,a}^\dagger a_{r,b}$ of one electron from fragment $b$ to fragment $a$
(the $G$ condition). With Eq.~\eqref{eq:liftGamma} these mixed blocks
are
\begin{align}
 P_{ab}&=\gamma^T\!\otimes\gamma^T
   -\tau\sum_\alpha\frac{C_\alpha^T\!\otimes C_\alpha^T}{v_\alpha},
   \nonumber\\
 Q_{ab}&=(I-\gamma)\otimes(I-\gamma)
   -\tau\sum_\alpha\frac{C_\alpha\otimes C_\alpha}{v_\alpha},
   \nonumber\\
 G_{a\leftarrow b}&=(I-\gamma)\otimes\gamma^T
   +\tau\sum_\alpha\frac{C_\alpha\otimes C_\alpha^T}{v_\alpha},
 \label{eq:mixedblocks}
\end{align}
which reduce to the product blocks at $\tau=0$. Unlike Gram matrices of a
quantum state, these blocks are not positive semidefinite automatically:
the correction is subtracted in $P_{ab}$ and $Q_{ab}$, and
Proposition~\ref{prop:lift} gives the condition under which they remain
PSD.

\begin{proposition}[Explicit lift]
\label{prop:lift}
Let $\Rset\in\M^{\mathsf c}$ have $0\prec\gamma\prec I$, and put
\begin{equation}
 \eta=\min\{\lambda_{\min}(\gamma),\lambda_{\min}(I-\gamma)\}.
\end{equation}
For
$\mathsf x={\mathsf s}$ let $\Rset$ be invariant under spin rotations. For every
$k\ge2$ with
\begin{equation}
 4(k-1)\,\eta^2\ge|\mathcal A_{\mathsf x}|,
 \label{eq:Kexplicit}
\end{equation}
the lift $\Rset_{\rm tot}$ lies in $\F_k^{\mathsf x}$, has the local RDMs $\Rset$ on
every fragment, and has energy $k\langle \hat H_F\rangle_{\Rset}$.
\end{proposition}

\begin{sketch}
Because no element changes a local electron number, every global $P$,
$Q$, and $G$ matrix splits into blocks. The number-conserving block of
$G$ has as rows and columns the operators $a_{i,a}^\dagger a_{j,a}$ of all
fragments, each acting within one fragment; its diagonal fragment blocks
carry the local covariance $C$ and its off-diagonal ones the
interfragment covariance $-\tau\Pi$. Because the lift is symmetric under
permutations of the fragments, this centered block has the form
$I_k\otimes A+(J_k-I_k)\otimes B$, with $J_k$ the $k\times k$ all-ones
matrix. Such a matrix has the eigenblock $A+(k-1)B$ on the
fragment-symmetric direction and $A-B$ on the $k-1$ directions orthogonal
to it. With $A=C$, $B=-\tau\Pi$, and $\tau=1/(k-1)$ these are
\begin{equation}
 C-\Pi\quad(\text{once}),\qquad C+\tau\Pi\quad(k-1\ \text{times}),
\end{equation}
for every $k\ge2$. The transverse blocks $C+\tau\Pi$ are PSD because $C$
and $\Pi$ are. The fragment-symmetric block $C-\Pi$ is PSD by
Lemma~\ref{lem:projection}; it is the direction of the total number (and
spin), where the negative covariances cancel the local fluctuations, and
the only place where positivity is not automatic. Each mixed block in Eq.~\eqref{eq:mixedblocks} is a
product block with smallest eigenvalue at least $\eta^2$, perturbed by a
term of norm at most $\tau|\mathcal A_{\mathsf x}|/4$; Eq.~\eqref{eq:Kexplicit}
keeps it PSD. The total-number variance vanishes by construction, which
gives the global contraction, and in the singlet case
$\langle\hat{\bm S}_{\rm tot}^2\rangle=0$. The energy involves only
intrafragment elements. All blocks are checked in \SM{4}, and the
contraction and spin conditions in \SM{5}.
\end{sketch}

Since $|\mathcal A_{\mathsf x}|\le4$, the condition $k\ge1+1/\eta^2$
suffices in both settings. Condition~\eqref{eq:Kexplicit} is uniform but conservative. Since each
block in Eq.~\eqref{eq:mixedblocks} is affine in $\tau$ with a positive
definite value at $\tau=0$, the set of admissible $\tau$ is an interval
$[0,\tau_c(\Rset)]$, where $\tau_c$ is the smallest of three
generalized-eigenvalue endpoints. The lift is feasible exactly when
$k\ge K_{\rm lift}=\max\{2,\lceil1+1/\tau_c\rceil\}$ (\SM{4}). This
threshold concerns this construction only; other extensions may be
feasible at smaller $k$, so $K_{\rm lift}$ is a sufficient replica count,
not the onset of saturation.

\subsection{The limit theorem}
\label{sec:limitproof}
\begin{theorem}[The replication limit is a single-fragment SDP minimum]
\label{thm:limit}
For the Hamiltonian and global sets of Sec.~\ref{sec:problems},
\begin{equation}
 \boxed{\lim_{k\to\infty}\frac{E_k^{\mathsf u}}{k}=E_1^{\mathsf c}>-\infty.}
 \label{eq:main}
\end{equation}
If $\hat H_F$ is invariant under spin rotations and $N_F$ is even, then also
\begin{equation}
 \boxed{\lim_{k\to\infty}\frac{E_k^{\mathsf s}}{k}=E_1^{\mathsf c}.}
 \label{eq:mainspin}
\end{equation}
The unrestricted-spin and global-singlet limits therefore coincide.
Moreover, for $k\ge1+|\mathcal A_{\mathsf x}|/(4m^2)$,
\begin{align}
 0&\le\frac{E_k^{\mathsf x}}{k}-E_1^{\mathsf c}\nonumber\\
 &\le\frac{e_{\rm ref}-E_1^{\mathsf c}}{2m}\sqrt{\frac{|\mathcal A_{\mathsf x}|}{k-1}},
 \label{eq:rate}
\end{align}
where $m=\min\{N_F/r,1-N_F/r\}$ and $e_{\rm ref}$ is the energy
$\langle \hat H_F\rangle$ of the Fock-space ensemble in which every spin
orbital is occupied independently with probability $N_F/r$.
\end{theorem}

\begin{sketch}
Take a minimizer $\Rset_*$ of $E_1^{\mathsf c}$; for the singlet case, average it
over spin rotations, which keeps feasibility and energy. A minimizer may
have occupations exactly $0$ or $1$, where Proposition~\ref{prop:lift}
does not apply. Mix it with the RDMs $\Rset_{\rm ref}$ of the physical
Fock-space ensemble in which every orbital is occupied independently with
probability $N_F/r$:
$\Rset_\epsilon=(1-\epsilon)\Rset_*+\epsilon\Rset_{\rm ref}$. The mixture
lies in the convex set $\M^{\mathsf c}$, has $\eta\ge\epsilon m$, and has energy
$(1-\epsilon)E_1^{\mathsf c}+\epsilon e_{\rm ref}$. Lifting it with
Proposition~\ref{prop:lift} and combining with
Proposition~\ref{prop:restriction} gives
$E_1^{\mathsf c}\le E_k^{\mathsf x}/k\le(1-\epsilon)E_1^{\mathsf c}+\epsilon e_{\rm ref}$
whenever $4(k-1)\epsilon^2m^2\ge|\mathcal A_{\mathsf x}|$. Choosing the
smallest admissible $\epsilon$ gives Eq.~\eqref{eq:rate}. Compactness of
$\M^{\mathsf c}$ makes $E_1^{\mathsf c}$ finite for any Hermitian two-body
interaction. The complete proof is in \SM{5}.
\end{sketch}

The rate in Eq.~\eqref{eq:rate} comes from the regularization and is not
expected to be sharp; Sec.~\ref{sec:monotone} gives exact equality at
finite $k$ in a fixed finite basis. The theorem concerns exact SDP
values. It does not assert that a globally feasible pair represents a
physical singlet, nor does it fix local number or spin quantum numbers;
global singlet feasibility means Eq.~\eqref{eq:singlet} within the PQG
relaxation.

%% file: sections/monotone.tex
\section{Monotonicity, additivity, and finite saturation}
\label{sec:monotone}
The product of local RDMs on disjoint orbital spaces (product closure:
the local RDMs are kept, the interfragment 1-RDM is zero, and the mixed
2-RDM elements are $\Gamma_{i_ap_b;j_aq_b}=\tfrac12(\gamma_a)_{ij}(\gamma_b)_{pq}$,
completed by antisymmetry) preserves
fixed-number PQG, and preserves a singlet when both factors are singlets
(\SM{6}). Hence
\begin{equation}
 E_{k+l}^{\mathsf x}\le E_k^{\mathsf x}+E_l^{\mathsf x},\qquad
 E_1^{\mathsf c}\le E_k^{\mathsf x}/k\le E_1^{\mathsf x} .
 \label{eq:productbounds}
\end{equation}
Subadditivity and the lower bound already give existence of
$\lim_kE_k^{\mathsf x}/k$ by Fekete's lemma.\cite{Fekete1923} They identify
neither the value of the limit nor monotonicity at consecutive replica
counts; the next result supplies the latter.

\begin{proposition}[Covariance dilution and monotonicity]
\label{prop:dilution}
For $k\ge M\ge2$ and $\mathsf x\in\{{\mathsf u},{\mathsf s}\}$, every point of
$\F_M^{\mathsf x}$ can be mapped to a point of $\F_k^{\mathsf x}$ with the same
permutation-averaged single-fragment RDMs. No assumption on the
occupations is needed. Consequently
\begin{equation}
 \frac{E_k^{\mathsf x}}{k}\le\frac{E_M^{\mathsf x}}{M},\qquad k\ge M\ge1 ,
 \label{eq:monotonicity}
\end{equation}
where the case $M=1$ is Eq.~\eqref{eq:productbounds}.
\end{proposition}

\begin{sketch}
Average the $M$-fragment point over fragment permutations and over
independent phase rotations of each fragment. Let $C$ be the local
covariance of one-body operators and $D$ the covariance between two
different fragments. Global feasibility at $M$ fragments means that
$C-D$ (transverse) and $C+(M-1)D$ (fragment-symmetric) are PSD. For $k$
fragments keep every local RDM and \emph{dilute} the interfragment
covariance to $D_k=\theta D$ with $\theta=(M-1)/(k-1)$. The
fragment-symmetric block is unchanged, $C+(k-1)D_k=C+(M-1)D$, so the
number (and spin) null vectors that encode the global contraction (and
singlet) are kept; the transverse block $C-D_k=(1-\theta)C+\theta(C-D)$ is
a convex combination of PSD matrices. The pair, hole, and transfer blocks
between two fragments are, likewise, convex combinations of product blocks
and of the parent blocks. The negative covariances that were needed among
$M$ fragments are thus spread over $k$ fragments. The complete proof is in
\SM{6}.
\end{sketch}

\begin{corollary}[A local criterion for replication additivity]
\label{cor:additivity}
In each setting of Theorem~\ref{thm:limit}, $E_k^{\mathsf x}=kE_1^{\mathsf x}$ for
all $k$ if and only if $E_1^{\mathsf c}=E_1^{\mathsf x}$. Moreover
$E_1^{\mathsf c}=\inf_{k\ge1}E_k^{\mathsf x}/k$, and if $E_m^{\mathsf x}/m<E_1^{\mathsf x}$ for
some $m$, then the defect per fragment persists:
\begin{equation}
 \lim_{k\to\infty}\frac{E_k^{\mathsf x}-kE_1^{\mathsf x}}{k}
 =E_1^{\mathsf c}-E_1^{\mathsf x}\le\frac{E_m^{\mathsf x}}{m}-E_1^{\mathsf x}<0 .
 \label{eq:persistent}
\end{equation}
\end{corollary}

\begin{sketch}
Combine Eq.~\eqref{eq:productbounds}, monotonicity, and
Theorem~\ref{thm:limit}: a nonincreasing sequence bounded below by its
limit has that limit as its infimum.
\end{sketch}

Natural orbitals of a local point with occupation exactly $0$ or $1$ do
not obstruct the lift: they can be split off.

\begin{lemma}[Frozen orbitals]
\label{lem:frozen}
Let $\Rset\in\M^{\mathsf c}$, and let $b$ be a natural orbital of $\gamma$
with occupation $0$ (empty) or $1$ (filled). Then every element of
$\Gamma$ that involves $b$ is that of the product of the remaining pair
with a vacuum ($b$ empty) or with a single filled orbital ($b$ filled),
and the remaining pair on the other $r-1$ spin orbitals lies in the set
$\M^{\mathsf c}$ for those orbitals with mean electron number $N_F$ or
$N_F-1$.
\end{lemma}

\begin{sketch}
By CAR $\hat n_b^2=\hat n_b$, so
$\mathrm{var}(\hat n_b)=\langle\hat n_b\rangle(1-\langle\hat n_b\rangle)=0$.
The Cauchy--Schwarz inequality for the PSD form $C$ then gives
$\mathrm{cov}(\hat n_b,\hat A)=0$ for every one-body $\hat A$, in particular
$\langle\hat n_b\hat n_j\rangle=\langle\hat n_b\rangle\langle\hat n_j\rangle$.
If $b$ is empty, the diagonal entries $\langle\hat n_b\hat n_j\rangle$ of
$P$ on the pairs containing $b$ vanish, so these rows and columns of the
PSD matrix $P$ vanish, and so do all elements of $\Gamma$ with index $b$.
If $b$ is filled, the same argument with the hole occupation
$1-\hat n_b$ and the rows of the PSD matrix $G$ on the operators
$a_b^\dagger a_j$ gives the product with a filled orbital. The
remaining blocks are principal blocks of those of $\Rset$. The complete
proof is in \SM{6}.
\end{sketch}

\begin{corollary}[Finite saturation]
\label{cor:saturation}
(i) In each applicable setting there is a finite $K$ with
$E_k^{\mathsf x}=kE_1^{\mathsf c}$ for every $k\ge K$.
(ii) Independently, if $E_M^{\mathsf x}=ME_1^{\mathsf c}$ holds exactly at one
positive integer $M$, then it holds for every $k\ge M$.
\end{corollary}

\begin{sketch}
For (i), take a minimizer of $E_1^{\mathsf c}$, averaged over spin rotations
for $\mathsf x={\mathsf s}$, and pass to its natural orbitals. By
Lemma~\ref{lem:frozen} it is the product of a closed set of empty and
filled orbitals with an active part whose natural occupations lie strictly
between $0$ and $1$; the active mean electron number is $N_F$ minus the
number of filled orbitals, an integer. In the singlet case the
natural orbitals come in spin pairs, so the filled part is closed shell.
If there is no active part, the minimizer is a determinant and its
$k$-fold product already attains $kE_1^{\mathsf c}$. Otherwise
Proposition~\ref{prop:lift} lifts the active part for $k\ge K_{\rm lift}$,
evaluated on the active orbitals, and the product closure (\SM{6})
attaches $k$ copies of the frozen part, a determinant with a fixed number
of electrons and, for $\mathsf x={\mathsf s}$, zero spin. The result lies in $\F_k^{\mathsf x}$, has the
local RDMs of the minimizer, and has energy $kE_1^{\mathsf c}$, so the
restriction bound gives $E_k^{\mathsf x}=kE_1^{\mathsf c}$ for $k\ge K_{\rm lift}$.
For (ii), monotonicity and the restriction bound give
$E_1^{\mathsf c}\le E_k^{\mathsf x}/k\le E_M^{\mathsf x}/M=E_1^{\mathsf c}$ for
$k\ge M$. Details are in \SM{6}.
\end{sketch}

Part (i) gives finite attainment, not merely convergence, in a fixed finite
basis and without any assumption on the minimizer. It does not identify
$K$: the bound $K_{\rm lift}$ depends on the smallest margin $\eta$ of the
active occupations, which may be small, and $K_{\rm lift}$ need not be the
first saturation count. Part (ii) shows that saturation, once reached, is
permanent, but neither part identifies a saturation count from rounded
numerical energies.

\subsection{Relation to de Finetti theorems}
\label{sec:definetti}
Theorem~\ref{thm:limit} and Corollary~\ref{cor:saturation} reduce a
problem on $k$ copies that is symmetric under permutations of the copies
to a problem on one copy. This is the structure of de Finetti theorems.
The proofs above use no de Finetti estimate, and the comparison shows
what the fixed total electron number changes.

\emph{Background.} A sequence of random variables is exchangeable if its
joint law is invariant under permutations. De Finetti's theorem states
that an infinite exchangeable sequence is a mixture of independent and
identically distributed (i.i.d.) sequences.\cite{deFinetti1937} For $k$
exchangeable variables this fails, but the joint law of $n$ of them lies
within $2cn/k$ of a mixture of i.i.d.\ laws in total variation when each
variable takes $c$ values, and within $n(n-1)/k$ in
general.\cite{DiaconisFreedman1980} Quantum versions bound the distance
of the $n$-copy marginal of a permutation-invariant state on $k$ copies
from mixtures of product states,\cite{KoenigRenner2005,CKMR2007} a
fermionic version exists,\cite{KZE2017} and a related argument derives
mean-field energies as the limit of the energy per particle of symmetric
many-body systems.\cite{LNR2014} In all of these the object of interest is
a marginal on $n\ge2$ copies, or a bipartite state: a state $\rho_{AB}$ is
called $k$-extendible if a state on $A$ and $k$ copies of $B$, symmetric
under permutations of the copies, has $\rho_{AB}$ as the marginal on $A$
and each copy, and the $k$-extendible states define a hierarchy of SDP
relaxations of the separable states.\cite{DPS2004} A marginal on a single
copy carries no information of this kind, because every state $\rho$ is
the single-copy marginal of the symmetric state $\rho^{\otimes k}$.

\emph{Correspondence.} The copies are the fragments. Let
$\M_k^{\mathsf x}$ be the set of single-fragment pairs $(\gamma,\Gamma)$
obtained from points of $\F_k^{\mathsf x}$ by averaging over fragment
permutations (and, for $\mathsf x={\mathsf s}$, over global spin rotations)
and restricting to one fragment; we call these the $k$-extendible
fragment RDMs. Unlike single-copy marginals of states, they are restricted
by the global constraints: the fixed total electron number and, for
$\mathsf x={\mathsf s}$, the global singlet. Proposition~\ref{prop:restriction} gives
$\M_k^{\mathsf x}\subseteq\M^{\mathsf c}$. For $k\ge M\ge2$,
Proposition~\ref{prop:dilution} gives $\M_M^{\mathsf x}\subseteq\M_k^{\mathsf x}$,
and for $M=1$ the $k$-fold product of a point of $\F_1^{\mathsf x}$ (\SM{6})
gives $\M_1^{\mathsf x}\subseteq\M_k^{\mathsf x}$. The proof of Corollary~\ref{cor:saturation}(i) uses the
minimizer only to evaluate the energy: every point of $\M^{\mathsf c}$
(invariant under spin rotations for $\mathsf x={\mathsf s}$) is a product of
a determinant and an active pair, and lifting the active pair places it in
$\M_k^{\mathsf x}$ for all sufficiently large $k$ (\SM{6}). Hence
\begin{equation}
 \M_1^{\mathsf x}\subseteq\M_2^{\mathsf x}\subseteq\cdots\subseteq\M^{\mathsf c},
 \qquad
 \bigcup_{k\ge1}\M_k^{\mathsf x}=\M^{\mathsf c}
 \label{eq:extendible}
\end{equation}
for $\mathsf x={\mathsf u}$, and the same with $\M^{\mathsf c}$ replaced by its
spin-rotation-invariant points for $\mathsf x={\mathsf s}$. Every point of
$\M^{\mathsf c}$ (every spin-rotation-invariant point for
$\mathsf x={\mathsf s}$) is thus exactly $k$-extendible at a finite replica
count, which depends on the point.

\emph{Differences.} Four features distinguish Eq.~\eqref{eq:extendible}
from the de Finetti setting.

(i) The direction of the inclusions is reversed. In the bipartite
setting a $(k+1)$-extendible state is $k$-extendible, because one copy of
$B$ can be traced out, so the extendible sets decrease with $k$. The
separable states lie in all of them, because in each product term of a
separable decomposition the factor on $B$ can be replaced by its tensor
power before mixing; the content of a de Finetti theorem is the converse,
that a highly extendible state is nearly separable. Here a fragment cannot
simply be removed, since the remaining fragments would no longer have a
fixed total electron number, and the sets increase with $k$. The
inclusion in $\M^{\mathsf c}$ is the easy step (principal submatrices),
and the content is the converse, that every point of $\M^{\mathsf c}$ is
extendible. The analog of the tensor power, the product of local RDMs, is
not available when $v>0$, because the local number fluctuations would add
up (Sec.~\ref{sec:cancel}); the lift has to be constructed.

(ii) The anticorrelation that is an error there is the mechanism here.
In a mixture of i.i.d.\ laws two different variables have nonnegative
covariance, the variance of the conditional mean. Drawing all $k$ balls,
one at a time without replacement, from an urn of $k$ balls labeled by
numbers gives an exchangeable law whose pairs have covariance
$-\sigma^2/(k-1)$, with $\sigma^2$ the single-draw variance, because the
sum of the labels is fixed. For $k\ge2$ and $\sigma^2>0$ this law is
therefore not a mixture of i.i.d.\ laws, and
finite de Finetti theorems bound its difference from sampling with
replacement.\cite{DiaconisFreedman1980}
Equation~\eqref{eq:requiredcov} is the same identity for the fragment
electron numbers: the fixed total number forces the covariance $-v_N/(k-1)$
on every pair of fragments. In the lift these covariances are not a
deviation to be bounded; they are what allows each fragment to keep its
number variance, the other $k-1$ fragments acting as a particle
reservoir. The limiting set is therefore not the fixed-$N_F$ set
$\F_1^{\mathsf x}$ but $\M^{\mathsf c}$, in which the local electron number
may fluctuate about $N_F$. For exact states the same construction relates
fractional electron numbers to ensembles. Perdew, Parr, Levy, and Balduz
treated a subsystem that exchanges electrons with its
surroundings.\cite{PPLB1982} Yang, Zhang, and Ayers placed $q$ identical,
infinitely separated systems with $qM+p$ electrons in total (integers
$0<p<q$) in a symmetrized ground state, the superposition over all
assignments of $M+1$ electrons to $p$ of the systems and $M$ to the
others. Each system then has the density and 1-RDM of the mixture with
weights $p/q$ and $(q-p)/q$ of the $(M+1)$- and $M$-electron ground
states, and, assuming convexity of the exact energies, they obtained the
energy at fractional electron number as the straight line between the
integers without invoking ensembles.\cite{YZA2000} The electron numbers of
the systems in this state follow the urn law above. The exact limit is then the lower convex
envelope of $E(M)$ at $N_F$ (Sec.~\ref{sec:ppos}), which equals $E(N_F)$
when the fragment is stable against charge disproportionation. Under
PQG no such stability is enforced, and when the exact problem is stable,
the number part of the defect $E_1^{\mathsf x}-E_1^{\mathsf c}$ is an artifact
of PQG (Sec.~\ref{sec:compare}).

(iii) The objects are moment relaxations, not states. Only one- and
two-body moments enter, and $\M_k^{\mathsf x}$ is not the set of marginals of
$k$-copy states. The lift at $k=3$ in Sec.~\ref{sec:lift} is PQG feasible
but not $N$-representable. Conversely, exact states obey conditions that
PQG does not see: for $k=2$ the fixed total number gives
$\hat N_b=2N_F-\hat N_a$ in every configuration, so the local number
distribution of a fragment-symmetric state is symmetric about $N_F$; in
particular its third central moment, a three-body quantity that PQG does
not constrain, vanishes. In the
lift, the interfragment covariance sits in the number-conserving block of
$G$, which is PSD for every $k\ge2$ by Lemma~\ref{lem:projection}. The
replica condition~\eqref{eq:Kexplicit} comes from the other blocks, the
mixed pair, hole, and transfer blocks of Eq.~\eqref{eq:mixedblocks}, in
which the correction must be absorbed by product blocks with smallest
eigenvalue at least $\eta^2$. When an occupation approaches $0$ or $1$
this margin vanishes and the bound on the replica count diverges;
occupations exactly $0$ or $1$ are removed by Lemma~\ref{lem:frozen}.

(iv) The finite-$k$ error vanishes. In a fixed finite basis,
Corollary~\ref{cor:saturation}(i) gives $E_k^{\mathsf x}=kE_1^{\mathsf c}$ for
$k\ge K$, whereas the urn example shows that a de Finetti error is
generally nonzero at every finite $k$. The rate in Eq.~\eqref{eq:rate}
is an explicit bound that does not require $K$, and it is not claimed to
be sharp.

%% file: sections/reduced.tex
\section{An exact finite-replica problem of fixed size}
\label{sec:reduced}
Finite-$k$ values are needed to observe monotonicity and saturation. A
direct $k$-replica SDP has $kr$ spin orbitals. Following the standard
symmetry reduction of SDPs,\cite{GatermannParrilo2004} the two averages
used in Proposition~\ref{prop:dilution} reduce it exactly to a problem
whose size does not depend on $k$. Besides one fragment's
$(\gamma,\Gamma)$, introduce the interfragment pair matrix
\begin{equation}
 \Xi_{ip,jq}=\langle a_{i,a}^\dagger a_{p,b}^\dagger a_{q,b}a_{j,a}\rangle,
 \qquad a\ne b ,
 \label{eq:Xi}
\end{equation}
which is Hermitian and, for a point symmetric under exchange of the two
fragments, invariant under the simultaneous exchange
$(i,j)\leftrightarrow(p,q)$.
By CAR the interfragment particle--hole matrix
$G^X_{ij;pq}=\langle(a_j^\dagger a_i)^{(a)}(a_p^\dagger a_q)^{(b)}\rangle$
equals $\Xi_{jp,iq}$, and the mixed blocks $P_{ab}$, $Q_{ab}$, and
$G_{a\leftarrow b}$ are affine in $(\gamma,\Xi)$, with $P_{ab}=\Xi$ (explicit
forms in \SM{7}).

\begin{proposition}[Exact reduction]
\label{prop:reduced}
For $k\ge2$, $E_k^{\mathsf x}/k$ equals the minimum of
$\langle \hat H_F\rangle_{(\gamma,\Gamma)}$ over $(\gamma,\Gamma,\Xi)$ subject to
\begin{itemize}
 \item[(R1)] local $P\succeq0$, $Q\succeq0$, and $\Tr\gamma=N_F$;
 \item[(R2)] $P_{ab}\succeq0$, $Q_{ab}\succeq0$, and $G_{a\leftarrow b}\succeq0$;
 \item[(R3)] $G-G^X\succeq0$ and $G+(k-1)G^X\succeq0$;
 \item[(R4)] $2\sum_p\Gamma_{ip;jp}+(k-1)\sum_p\Xi_{jp,ip}=(kN_F-1)\gamma_{ij}$;
 \item[(R5)] for $\mathsf x={\mathsf s}$,
 $\langle\hat{\bm S}_F^2\rangle+(k-1)\langle\hat{\bm S}_F^{(a)}\!\cdot\hat{\bm S}_F^{(b)}\rangle=0$.
\end{itemize}
Here $\Xi$ has the exchange symmetry stated above, so that the
reversed transfer block $G_{b\leftarrow a}$ coincides with
$G_{a\leftarrow b}$ after relabeling and one direction suffices. The
replica count enters only through the second condition of (R3), through
(R4), and through (R5).
\end{proposition}

\begin{sketch}
The $k$-replica problem is invariant under permutations of the fragments
and under a separate phase rotation of each fragment, because $\hat H_{\rm tot}$
conserves every local electron number. Averaging a minimizer over these
symmetries gives a minimizer with no number-changing interfragment
moments, identical local RDMs, and a single interfragment matrix $\Xi$.
Its $P$, $Q$, and $G$ matrices then split into the blocks (R1)--(R3); (R3) are the
fragment-symmetric and transverse eigenblocks, as in
Proposition~\ref{prop:lift}. The global contraction within one fragment
is (R4), and (R5) is Eq.~\eqref{eq:singlet} divided by $k$. Details are
in \SM{7}.
\end{sketch}

Nakata and Yasuda\cite{NakataYasuda2009} computed replica energies with
this block structure, obtained from two assumptions: all molecules carry
the same RDMs, and all 2-RDM elements that change a local electron number
vanish. Up to index and normalization conventions, their particle--hole
blocks and 1-RDM contraction coincide with (R3) and (R4).
Proposition~\ref{prop:reduced} shows that neither assumption loses
generality for the global PQG sets defined here.

Setting $k=1$ and omitting $\Xi$ recovers the isolated problem. In the
singlet case, $\langle\hat{\bm S}_{\rm tot}^2\rangle=0$ and $G\succeq0$ force the
three total-spin coefficient vectors into the kernel of the
fragment-symmetric block, so that block has no strictly feasible point.
We impose these kernel vectors as linear equalities and compress the block
to their orthogonal complement. This facial reduction\cite{PermenterParrilo2018}
is an equivalent reformulation (\SM{7}). All singlet values in this
work, including $E_1^{\mathsf s}$, are computed in this form.

%% file: sections/cuts.tex
\section{Integer cuts and sharp fragment numbers}
\label{sec:cuts}
The number part $E_1^{\mathsf u}-E_1^{\mathsf c}$ of the defect arises
because the restricted fragment may have a nonzero number variance $v$
(Lemma~\ref{lem:local}); in the singlet setting the spin part
$E_1^{\mathsf s}-E_1^{\mathsf u}$ comes from local spin fluctuations
(Sec.~\ref{sec:compare}). Integer
electron-count cuts, introduced in Parts~I and~II of a companion
study,\cite{PartI,PartII} constrain the distribution of a
region's electron count using only its first two moments. This section
asks whether such cuts remove the defect, and what does.

For a region $\Omega$ with electron-count operator $\hat N_\Omega$, mean
$\mu_\Omega$, and pair population
$\F_\Omega=\langle\hat N_\Omega(\hat N_\Omega-1)\rangle$, the integer cut at
level $\ell\in\mathbb Z$ (called $k$ in Part~I\cite{PartI}) is
\begin{equation}
 I_{\Omega,\ell}=\F_\Omega-2\ell\mu_\Omega+\ell(\ell+1)
 =\bigl\langle(\hat N_\Omega-\ell)(\hat N_\Omega-\ell-1)\bigr\rangle\ge0 ,
 \label{eq:intcut}
\end{equation}
valid for every physical state because $\hat N_\Omega$ has integer
spectrum.

\begin{proposition}[Cuts on a whole fragment]
\label{prop:T1}
For an integer mean $N_F$, every point of $\M^{\mathsf c}$ satisfies
$I_{F,\ell}=v+(N_F-\ell)(N_F-\ell-1)\ge v\ge0$ for every integer $\ell$. Since
averaging a global feasible point over fragment permutations keeps its
energy and restricts it to $\M^{\mathsf c}$ on every fragment, imposing these
cuts on every fragment changes neither $E_1^{\mathsf c}$ nor any
$E_k^{\mathsf x}$.
\end{proposition}

\begin{sketch}
Substitute $\mu_F=N_F$ and $\F_F=N_F(N_F-1)+v$ into Eq.~\eqref{eq:intcut}; the
product of two consecutive integers is nonnegative, and $v=u^\dagger
Cu\ge0$. Part~II\cite{PartII} made the same observation for the
covariance (bordered $G$) relaxation at integer mean; the proposition
applies it to the replica problem (\SM{8}).
\end{sketch}

Within a molecular fragment, a real-space atomic region $\Omega$ is
described in the fragment basis by the overlap matrix
$S^\Omega_{pq}=\int_\Omega\overline{\phi_p}\phi_q$, whose eigenvalues lie in $[0,1]$
but which is not a projection. Then
$\mu_\Omega=\sum_{pq}S^\Omega_{pq}\gamma_{qp}$ (summed over spin) and
\begin{equation}
 \F_\Omega=\int_\Omega\!\int_\Omega\rho_2
 =\sum_{pqrs}S^\Omega_{pq}S^\Omega_{rs}
 \langle a_p^\dagger a_r^\dagger a_s a_q\rangle .
 \label{eq:regionF}
\end{equation}
Only the pair term enters, because the one-particle operator
$\hat\chi_\Omega$, the indicator of $\Omega$, is a projection on the full
one-particle space. Replacing $\F_\Omega$ by
$\langle\hat N_\Omega'^2\rangle-\langle\hat N_\Omega'\rangle$ for the basis operator
$\hat N_\Omega'=\sum_{pq}S^\Omega_{pq}a_p^\dagger a_q$ would add
$\Tr\,[(S^\Omega)^2-S^\Omega]\gamma\le0$ and give an inequality that
physical states can violate.

\begin{proposition}[The lowest region cut]
\label{prop:T1region}
For every region, $I_{\Omega,0}=\F_\Omega\ge0$ is implied by $P\succeq0$.
\end{proposition}

\begin{sketch}
Equation~\eqref{eq:regionF} is the trace of the pair matrix against
$S^\Omega\otimes S^\Omega$, and both are PSD (\SM{8}).
\end{sketch}

Thus, as for whole fragments, the part of an integer cut that follows from
PQG adds nothing. Cuts with $\ell\ge1$ on regions with nonintegral mean are
not implied in general; Section~\ref{sec:numerics} shows that they are
nevertheless inactive for the molecules studied here.

What does remove the defect is a sharp local electron number. Add to the
$k$-replica problem the linear equalities (imposed on all fragments
together; with the fixed total electron number and $G\succeq0$ they fix
every local mean at $N_F$ and make every local variance zero)
\begin{equation}
 \begin{aligned}
 v_a=0\quad\Longleftrightarrow\quad
 &2\sum_{i,j}(\Gamma_{\rm tot})_{i_aj_a;i_aj_a}=N_F(N_F-1),\\
 &a=1,\dots,k,
 \end{aligned}
 \label{eq:sharp}
\end{equation}
and denote the minimum by $E_k^{\mathsf x,\sharp}$.

\begin{proposition}[Sharp fragment numbers restore additivity]
\label{prop:T2}
For every $k\ge1$,
\begin{equation}
 E_k^{{\mathsf u},\sharp}=kE_1^{\mathsf u},\qquad
 kE_1^{\mathsf u}\le E_k^{{\mathsf s},\sharp}\le kE_1^{\mathsf s}.
 \label{eq:T2}
\end{equation}
Moreover, there is a finite $K$ with $E_k^{{\mathsf s},\sharp}=kE_1^{\mathsf u}$
for every $k\ge K$.
\end{proposition}

\begin{sketch}
Permutation averaging preserves Eq.~\eqref{eq:sharp}, so the restricted
point lies in $\M^{\mathsf c}$ with $v=0$; by Lemma~\ref{lem:local} it satisfies
the local contraction and is feasible for the isolated problem, which gives
the lower bounds. The product of isolated minimizers satisfies
Eq.~\eqref{eq:sharp}, and the product of singlets is a singlet, which gives
the upper bounds. For the last statement, average a minimizer of
$E_1^{\mathsf u}$ over spin rotations; it has $v=0$, so $C_N=0$ and the lift
corrects only the spin covariances. After splitting off the frozen
orbitals (Lemma~\ref{lem:frozen}), Proposition~\ref{prop:lift} gives, for
$k\ge K_{\rm lift}$, a global singlet with the local RDMs of the minimizer,
hence with $v_a=0$ and energy $kE_1^{\mathsf u}$ (\SM{8}).
\end{sketch}

Equation~\eqref{eq:sharp} is the semidefinite form of the sharp-number
assumption in the sufficient condition for size consistency of Nakata and
Anderson,\cite{NakataAnderson2012} which in addition requires
unitary-invariant diagonal $N$-representability. For disconnected identical
replicas, Proposition~\ref{prop:T2} shows that PQG with sharp fragment
numbers already gives additivity in the unrestricted setting. Propositions~\ref{prop:T1}
and~\ref{prop:T2} together locate the defect: inequalities that use only
the mean of a fragment's electron count do not remove it, whereas the
equality that removes its variance restores additivity in the
unrestricted setting. In the singlet setting the lower end $kE_1^{\mathsf u}$ is attained for
all sufficiently large $k$, because local spin fluctuations can be
cancelled by interfragment spin correlation even at sharp local number;
for small $k$ Eq.~\eqref{eq:T2} leaves the interval between
$E_1^{\mathsf u}$ and $E_1^{\mathsf s}$ open. For the Hubbard fragment the
numerical results of Sec.~\ref{sec:numerics} show attainment already for
$k=2,3,4$, which is a numerical observation.

%% file: sections/stronger.tex
\section{Stronger conditions}
\label{sec:stronger}
The results above are proved for PQG. This section collects what carries
over to stronger conditions.

\subsection{Three-index conditions}
\label{sec:threeindex}
The conditions $T_1$ and $T_2$ are defined in
Eq.~\eqref{eq:psdconditions}.\cite{ErdahlJin2000,Zhao2004} We add either $T_1$ alone or $T_1$ and $T_2$ to every problem and write
$E_k^{\mathsf x,T}$ for the replicated minima, $\M^{\mathsf c,T}$ for
$\M^{\mathsf c}$ with the same local conditions, and $E_1^{\mathsf c,T}$ for
its minimum. Table~\ref{tab:threeindex} lists which statements carry over.
In both cases the inequalities
$E_1^{\mathsf c,T}\le E_1^{\mathsf u,T}\le E_1^{\mathsf s,T}$ and the
restriction bound $E_1^{\mathsf c,T}\le E_k^{\mathsf x,T}/k$ hold by the
arguments of Lemma~\ref{lem:local} and Proposition~\ref{prop:restriction}:
$T_1$ and $T_2$ are invariant under permutations of the orbitals, and
their local blocks are principal blocks of the global ones. Since adding
conditions can only raise a minimum, $E_k^{\mathsf x,T}/k\ge E_1^{\mathsf c}$
as well, so $E_k^{\mathsf x,T}/k$ is bounded below. For $T_1$ the products
also stay feasible.

\begin{proposition}[Products under $T_1$]
\label{prop:t1products}
Products, in the sense of the product closure, of points that satisfy the
PQG and $T_1$ conditions with sharp electron numbers on disjoint
fragments satisfy the $T_1$ condition. Consequently, with PQG and $T_1$,
$E_{k+l}^{\mathsf x,T}\le E_k^{\mathsf x,T}+E_l^{\mathsf x,T}$,
$E_k^{\mathsf x,T}/k\le E_1^{\mathsf x,T}$, and the limit of
$E_k^{\mathsf x,T}/k$ exists by Fekete's lemma.
\end{proposition}

\begin{sketch}
A $T_1$ operator $\hat X$ is a combination of products of three
annihilators and lowers the electron number by $3$. In the product,
moments of $\hat X^\dagger\hat X'$ and $\hat X'\hat X^\dagger$ vanish unless
$\hat X$ and $\hat X'$ remove the same numbers of electrons from the two
fragments, so the $T_1$ matrix is block diagonal in the splits $(3,0)$,
$(2,1)$, $(1,2)$, and $(0,3)$. The blocks $(3,0)$ and $(0,3)$ are local
$T_1$ matrices. In the block $(2,1)$, with $\hat X=a_{i,a}a_{j,a}a_{p,b}$,
the two terms factorize, and the block is
$P_a\otimes\rho_b+H_a\otimes(I-\rho_b)$, where $P_a$ is the local $P$
matrix, $H_a$ the transpose of the local $Q$ matrix on the same pairs, and
$\rho_b=\gamma_b^T$. Both terms are Kronecker products of PSD matrices. The
complete proof is in \SM{9}.
\end{sketch}

These steps can look as if they carried over at once to $T_2$, and it is
worth stating what is not obvious. Averaging over fragment permutations
and local phase rotations preserves $T_1$ and $T_2$; this gives the
restriction bound and a symmetric parent point, but not more. Products of
feasible points are the first gap. A $T_2$ operator lowers the electron
number by one, and in the split $(1,0)$ local operators such as
$a_{i,a}^\dagger a_{j,a}a_{l,a}$ and mixed ones such as
$a_{p,b}^\dagger a_{q,b}a_{l,a}$ remove the same numbers of electrons from
the two fragments. Their block therefore couples the local $T_2$ matrix
to products of a local 2-RDM and a 1-RDM, and it is not a sum of Kronecker
products of PSD matrices. For physical states the product is physical and
satisfies $T_2$, but for trial pairs this argument is not available. The
dilution of Proposition~\ref{prop:dilution} is the second gap. Its blocks
are affine in the dilution parameter $\theta$, so they are convex
combinations of the product blocks ($\theta=0$) and the parent blocks
($\theta=1$). The $\theta=0$ end needs the product property. At the
$\theta=1$ end, $T_1$ blocks with three annihilators on three different
fragments have no counterpart in a parent with two fragments. It is
therefore not obvious whether Proposition~\ref{prop:dilution} holds with
$T_1$, or with $T_1$ and $T_2$; monotonicity does not follow from symmetry
alone.

\begin{table*}[t]
\caption{Results of this paper with the three-index conditions added,
$T_1$ alone and $T_1$ with $T_2$. ``Holds'' means proved here;
``conjectured'' refers to Conjecture~\ref{conj:threeindex}; ``open'' means
neither proved nor refuted here.}
\label{tab:threeindex}
\footnotesize
\begin{ruledtabular}
\begin{tabular}{p{0.42\linewidth}p{0.11\linewidth}p{0.17\linewidth}p{0.22\linewidth}}
Result & PQG+$T_1$ & PQG+$T_1$+$T_2$ & Reason \\
\hline
Lemma~\ref{lem:local} (local reduction)\newline $v=0\Leftrightarrow$ local contraction & holds & holds & local statement \\
$E_1^{\mathsf c,T}\le E_1^{\mathsf u,T}\le E_1^{\mathsf s,T}$ & holds & holds & Lemma~\ref{lem:local}; added condition \\
Proposition~\ref{prop:restriction} (restriction bound)\newline $E_k^{\mathsf x,T}/k\ge E_1^{\mathsf c,T}$ & holds & holds & principal blocks \\
Eq.~\eqref{eq:productbounds} (products)\newline $E_{k+l}^{\mathsf x,T}\le E_k^{\mathsf x,T}+E_l^{\mathsf x,T}$, $E_k^{\mathsf x,T}/k\le E_1^{\mathsf x,T}$ & holds & conjectured & products (Proposition~\ref{prop:t1products}) \\
existence of the limit\newline $\lim_{k\to\infty}E_k^{\mathsf x,T}/k=\inf_kE_k^{\mathsf x,T}/k$ & holds & conjectured & Fekete's lemma \\
monotonicity along multiples\newline $E_{qM}^{\mathsf x,T}/(qM)\le E_M^{\mathsf x,T}/M$ & holds & conjectured & products \\
Proposition~\ref{prop:dilution} (monotonicity)\newline $E_k^{\mathsf x,T}/k\le E_M^{\mathsf x,T}/M$ for $k\ge M$ & open & open & products; $T_1$ blocks on three fragments \\
Corollary~\ref{cor:additivity}, ``if''\newline $E_1^{\mathsf c,T}=E_1^{\mathsf x,T}\Rightarrow E_k^{\mathsf x,T}=kE_1^{\mathsf x,T}$ & holds & conjectured & sandwich \\
Corollary~\ref{cor:additivity}, ``only if''\newline $E_k^{\mathsf x,T}=kE_1^{\mathsf x,T}\ \forall k\Rightarrow E_1^{\mathsf c,T}=E_1^{\mathsf x,T}$ & open & open & limit value \\
Proposition~\ref{prop:lift} and Theorem~\ref{thm:limit}\newline $\lim_{k\to\infty}E_k^{\mathsf x,T}/k=E_1^{\mathsf c,T}$ & open & open & lift; mixed $T$ blocks \\
Corollary~\ref{cor:saturation} (finite saturation)\newline $E_k^{\mathsf x,T}=kE_1^{\mathsf c,T}$ for $k\ge K$ & open & open & lift \\
Proposition~\ref{prop:reduced} (exact reduction)\newline $E_k^{\mathsf x}/k=\min\{\langle\hat H_F\rangle:\text{(R1)--(R5)}\}$ & open & open & $T$ blocks on three fragments \\
Propositions~\ref{prop:T1} and~\ref{prop:T1region} (implied cuts)\newline $I_{F,\ell}\ge0$ for all $\ell\in\mathbb Z$ if $\mu_F\in\mathbb Z$ & holds & holds & more conditions only \\
Proposition~\ref{prop:T2} (sharp numbers, unrestricted identity only)\newline $E_k^{{\mathsf u},\sharp,T}=kE_1^{{\mathsf u},T}$ & holds & lower bound holds; upper bound conjectured & restriction; products \\
\end{tabular}
\end{ruledtabular}
\end{table*}

\begin{conjecture}
\label{conj:threeindex}
Products of points satisfying the PQG, $T_1$, and $T_2$ conditions on
disjoint fragments, with interfragment 2-RDM elements
$\tfrac12\gamma_{ij}\gamma_{pq}$ as in the product closure, satisfy the
$T_2$ condition. Consequently the statements of
Proposition~\ref{prop:t1products} hold with $T_1$ and $T_2$.
\end{conjecture}

If the conjecture failed for a product with a physical state, the
condition that such products satisfy $T_2$ would be a new
necessary condition for a single fragment, since products of
$N$-representable states are $N$-representable. For the Hubbard fragment
the SDPA-GMP solutions agree with Proposition~\ref{prop:t1products} and
the conjecture
(Table~\ref{tab:energies}): adding $T_1$ reduces the defect at $k=2$ by
a factor of about $\HubToneRatio$, and the energy per fragment still
decreases by $\HubToneStepKthree\,t$ from $k=2$ to $k=3$; with $T_1$ and
$T_2$ the values for $k=1$ and $k=2$ agree to within $\HubTtwoDiffMax\,t$
per fragment. These are numerical observations for one fragment. Replica
calculations with three-index conditions at large $k$, such as the $k=32$
series for N$_2$ with PQG, are not attempted: the exact reduction of
Proposition~\ref{prop:reduced} is not established for these conditions,
whose blocks also couple three fragments, and the unreduced problems are
too large.

Table~\ref{tab:energies} also contains $T_2'$
[Eq.~\eqref{eq:psdconditions}], which borders the $T_2$ matrix by the
entries $\langle\hat X_i^\dagger a_p\rangle$ and
$\langle a_p^\dagger a_q\rangle$, where $\hat X_i$ are the operators of
$T_2$; the $T_2$ matrix is its principal block, so it implies $T_2$. Table~\ref{tab:threeindexdefect} compares, for each set
of conditions, the isolated singlet, the isolated unrestricted, and the
local SDP minima of the molecules and the Hubbard fragment, split into a
spin part $E_1^{\mathsf s,T}-E_1^{\mathsf u,T}$ and a local-number part
$E_1^{\mathsf u,T}-E_1^{\mathsf c,T}$. The two parts need not decrease as
conditions are added (for CH$_4$ the spin part is larger with $T_1$ and
$T_2$ than with PQG alone), and neither vanishes in general. With $T_1$
and $T_2$ the spin part of N$_2$ is still $\MolNtwoSpinTtwo\,E_h$. The border of $T_2'$
removes most of the spin part, which falls below $10^{-7}$ for H$_2$O,
CH$_4$, and the Hubbard fragment, but $\MolNtwoSpinTtwoP\,E_h$ remains for
N$_2$, and the local-number part of CH$_4$ is still
$\MolChfourNumTtwoP\,E_h$. In every system the local SDP solution returned
with $T_2$ violates $T_2'$. For PQG the local SDP minimum is the limit of the
energy per fragment (Theorem~\ref{thm:limit}). With $T_1$ the energy per
fragment lies between $E_1^{\mathsf c,T}$ and $E_1^{\mathsf x,T}$
(Proposition~\ref{prop:t1products}), and with $T_2$ or $T_2'$ only the
lower bound is established. With $T_1$ the total
$E_1^{\mathsf s,T}-E_1^{\mathsf c,T}$ is thus the width of an established
bracket for the singlet energy per fragment; with $T_2$ or $T_2'$ the
isolated value is not known to bound the replicated energy per fragment
from above, and the entries are differences between two single-fragment
minima. In neither case do they demonstrate a replication defect. These
are numerical observations.

\input{tables/three_index_defect}

\subsection{$p$-positivity}
\label{sec:ppos}
The obstacle for $T_2$ is that it combines moment matrices so that the
3-RDM cancels, and local and mixed operators then couple in a product.
Conditions that are themselves moment matrices behave better. Let
$p\ge2$ be an integer, with monomials and their degree as in
Sec.~\ref{sec:replicas}. The variables
are a moment family $(\Gamma^{(0)},\Gamma^{(1)},\dots,\Gamma^{(p)})$ with
$\Gamma^{(0)}=1$, $\Gamma^{(1)}=\gamma$, and $\Gamma^{(q)}$ the $q$-body
RDM; equivalently, a linear functional on the number-conserving operators
of degree at most $2p$, normalized and Hermitian, whose values are reduced
by CAR. When the electron number fluctuates, the lower members are
independent variables and are not fixed-number contractions of
$\Gamma^{(p)}$. The family satisfies full $p$-positivity if every matrix $\langle\hat x^\dagger\hat y\rangle$, with $\hat x$ and
$\hat y$ running over the monomials of degree at most $p$ that change the
electron number by the same amount, is PSD; positivity conditions of this
kind form the hierarchy of Erdahl and Jin\cite{ErdahlJin2000,ErdahlJin2000THEOCHEM} and of
Mazziotti and Erdahl,\cite{MazziottiErdahl2001} and Mazziotti's number-nonconserving
$(2,p)$-positivity conditions\cite{Mazziotti2026} constrain the 2-RDM alone
by cancelling the higher-body parts of $p$-positive operators, unlike the
full $p$-positivity used here, whose variable is the $p$-RDM. For $p=2$ this is the set of
$P$, $Q$, and augmented $G$ conditions together with
$0\preceq\gamma\preceq I$, that is, $\M^{\mathsf c}$ without its trace
condition; the number-conserving matrix contains the identity and is
therefore bordered. The conditions $T_1$ and $T_2$ follow from full
$3$-positivity. We write $E_k^{\mathsf x,p}$ for the replicated minima with
full $p$-positivity, the fixed total number, and the fixed-number
contractions between the members of the family, and $\M^{\mathsf c,p}$ and
$E_1^{\mathsf c,p}$ for the single-fragment set with full $p$-positivity and
$\Tr\gamma=N_F$ only, without fixed-number contractions.

\begin{proposition}[Products under $p$-positivity]
\label{prop:pproducts}
Let $\Rset_A$ and $\Rset_B$ satisfy full $p$-positivity with sharp
electron numbers (all fixed-number contractions up to order $p$) on disjoint sets of spin orbitals, and let their product
have all moments that change the electron number of $A$ or $B$ equal to
zero and all other moments factorized. Then the product satisfies full
$p$-positivity and the fixed-number contractions. Consequently
$E_{k+l}^{\mathsf x,p}\le E_k^{\mathsf x,p}+E_l^{\mathsf x,p}$ and
$E_1^{\mathsf c,p}\le E_k^{\mathsf x,p}/k\le E_1^{\mathsf x,p}$, and by
Fekete's lemma the limit of $E_k^{\mathsf x,p}/k$ exists.
\end{proposition}

\begin{sketch}
A monomial on $A\cup B$ is, up to sign, a product $\hat x_A\hat y_B$ with
$\deg\hat x_A+\deg\hat y_B\le p$. If $\hat x_A^\dagger\hat x'_A$ has odd
degree it changes the electron number of $A$, and its moment vanishes;
otherwise it is even and commutes with the operators on $B$, so that
$\langle(\hat x_A\hat y_B)^\dagger\hat x'_A\hat y'_B\rangle
=\langle\hat x_A^\dagger\hat x'_A\rangle_A\langle\hat y_B^\dagger\hat y'_B
\rangle_B$ without a sign. Each moment matrix of the product is therefore
block diagonal in the separate changes of the electron numbers of $A$ and
$B$, and each block is a principal submatrix of the Kronecker product of a
local moment matrix of $A$ and one of $B$, both of degree at most $p$ and
PSD. The contractions follow as in the product closure of \SM{6}, and the
lower bound $E_1^{\mathsf c,p}\le E_k^{\mathsf x,p}/k$ by restriction as in
Proposition~\ref{prop:restriction}. The complete proof is in \SM{9}.
\end{sketch}

For full $p$-positivity, then, subadditivity, the upper bound by the
isolated value, and the existence of the limit hold, as for $T_1$ alone
(Table~\ref{tab:ppos}). The sandwich
$E_1^{\mathsf c,p}\le E_k^{\mathsf x,p}/k\le E_1^{\mathsf x,p}$ also gives one
direction of Corollary~\ref{cor:additivity}: if
$E_1^{\mathsf c,p}=E_1^{\mathsf x,p}$, then $E_k^{\mathsf x,p}=kE_1^{\mathsf x,p}$
for every $k$. Monotonicity also holds along multiples: $q$ copies of an
optimal $M$-fragment point form a feasible $qM$-fragment point, so that
\begin{equation}
 E_{qM}^{\mathsf x,p}\le qE_M^{\mathsf x,p},\qquad
 \frac{E_{qM}^{\mathsf x,p}}{qM}\le\frac{E_M^{\mathsf x,p}}{M}
 \quad(q=1,2,\dots),
 \label{eq:multiples}
\end{equation}
and the energy per fragment does not increase along $M,2M,4M,\dots$. For
$k$ that is not a multiple of $M$, Proposition~\ref{prop:dilution} needs
the dilution, which for $p\ge3$ has to rescale moments that connect up to
$p$ fragments. Lemma~\ref{lem:local} also extends: for full $p$-positivity, zero local
number variance implies every fixed-number contraction of the moment
family, up to the highest order, because the moment matrices are PSD
separately (\SM{9}); with the products this gives
$E_k^{{\mathsf u},\sharp,p}=kE_1^{{\mathsf u},p}$ for every $k$. The argument
uses the individual moment matrices and does not apply to $T_1$ or $T_2$.
What remains open is the value of the limit, which for PQG
required the lift, and with it the converse.

\begin{table}[t]
\caption{Results of this paper for full $p$-positivity. ``Holds'' means
proved here (Proposition~\ref{prop:pproducts} and \SM{9}); ``open'' means
neither proved nor refuted here.}
\label{tab:ppos}
\footnotesize
\begin{ruledtabular}
\begin{tabular}{p{0.43\linewidth}p{0.19\linewidth}p{0.26\linewidth}}
Result & Full $p$-positivity & Reason \\
\hline
Proposition~\ref{prop:restriction} (restriction bound)\newline $E_k^{\mathsf x,p}/k\ge E_1^{\mathsf c,p}$ & holds & principal blocks \\
Eq.~\eqref{eq:productbounds} (products)\newline $E_{k+l}^{\mathsf x,p}\le E_k^{\mathsf x,p}+E_l^{\mathsf x,p}$, $E_k^{\mathsf x,p}/k\le E_1^{\mathsf x,p}$ & holds & Kronecker products \\
existence of the limit\newline $\lim_{k\to\infty}E_k^{\mathsf x,p}/k=\inf_kE_k^{\mathsf x,p}/k$ & holds & Fekete's lemma \\
Corollary~\ref{cor:additivity}, ``if''\newline $E_1^{\mathsf c,p}=E_1^{\mathsf x,p}\Rightarrow E_k^{\mathsf x,p}=kE_1^{\mathsf x,p}$ & holds & sandwich \\
Corollary~\ref{cor:additivity}, ``only if''\newline $E_k^{\mathsf x,p}=kE_1^{\mathsf x,p}\ \forall k\Rightarrow E_1^{\mathsf c,p}=E_1^{\mathsf x,p}$ & open & limit value \\
Propositions~\ref{prop:T1} and~\ref{prop:T1region} (implied cuts)\newline $I_{F,\ell}\ge0$ for all $\ell\in\mathbb Z$ if $\mu_F\in\mathbb Z$ & holds & more conditions only \\
Proposition~\ref{prop:T2} (sharp numbers)\newline $E_k^{{\mathsf u},\sharp,p}=kE_1^{{\mathsf u},p}$ & holds & products; zero variance gives all contractions (\SM{9}) \\
Proposition~\ref{prop:lift} and Theorem~\ref{thm:limit}\newline $\lim_kE_k^{\mathsf x,p}/k=E_1^{\mathsf c,p}$ & open (Conjecture~\ref{conj:ppos}) & lift \\
monotonicity along multiples, Eq.~\eqref{eq:multiples}\newline $E_{qM}^{\mathsf x,p}/(qM)\le E_M^{\mathsf x,p}/M$ & holds & products \\
Proposition~\ref{prop:dilution} (monotonicity)\newline $E_k^{\mathsf x,p}/k\le E_M^{\mathsf x,p}/M$ for $k\ge M$ & open & dilution of multi-fragment moments \\
Corollary~\ref{cor:saturation} (finite saturation)\newline $E_k^{\mathsf x,p}=kE_1^{\mathsf c,p}$ for $k\ge K$ & open & lift \\
Proposition~\ref{prop:reduced} (exact reduction)\newline $E_k^{\mathsf x,p}/k$ from an SDP of $k$-independent size & open & blocks on up to $p$ fragments \\
\end{tabular}
\end{ruledtabular}
\end{table}

\begin{conjecture}
\label{conj:ppos}
For every $p\ge2$ and both spin settings,
$\lim_{k\to\infty}E_k^{\mathsf x,p}/k=E_1^{\mathsf c,p}$: the replication
limit of full $p$-positivity is the minimum of the single-fragment
relaxation that fixes only the mean electron number.
\end{conjecture}

The conjecture holds at both ends of the hierarchy. For $p=2$ it is
Theorem~\ref{thm:limit}. At the other end, if positivity is imposed on all
operators of the fragment's Fock space, the single-fragment set consists
of the RDMs of Fock-space ensembles with mean number $N_F$, and its
minimum is the lower convex envelope of the exact energies $E(M)$ of $M$
electrons, evaluated at $N_F$. With unrestricted spin, the exact
replicated energy is the minimum of $\sum_aE(M_a)$ over distributions
with $\sum_aM_a=kN_F$, because noninteracting fragments may share the
electrons unevenly, and its value per fragment tends to the same convex
envelope. In the singlet setting the local states must also couple to
zero total spin; pairing each local state with an identical copy, which
can always be coupled to a singlet, gives the same envelope as
$k\to\infty$, although the finite-$k$ minima can differ. Between these ends, the limit could lie above
$E_1^{\mathsf c,p}$ if no construction analogous to the lift exists.

%% file: tables/three_index_defect.tex
\begin{table*}[t]
\caption{Isolated singlet, isolated unrestricted, and local SDP minima with three-index conditions: the spin part $E_1^{\mathsf s,T}-E_1^{\mathsf u,T}$ (S$-$U) and the local-number part $E_1^{\mathsf u,T}-E_1^{\mathsf c,T}$ (U$-$C) of $E_1^{\mathsf s,T}-E_1^{\mathsf c,T}$, from the midpoints of the primal-dual pairs (hartree for the STO-6G molecules, $t$ for the four-site Hubbard fragment). All values were computed in this work with double-precision SDPA 7.4.3 and accepted with the criteria of Sec.~\ref{sec:numerics}, except the Hubbard $E_1^{\mathsf s,T}$ with three-index conditions, which are the archived SDPA-GMP values ($k=1$ column of Table~\ref{tab:energies}). The singlet problems use the reduced formulation of $E_1^{\mathsf s}$ with the three-index blocks added. Differences below $10^{-7}$ are not distinguished from zero; the primal-dual widths are not certified bounds. For PQG and $T_1$ the restriction bound and the product construction bracket $E_k^{\mathsf s,T}/k$ between $E_1^{\mathsf c,T}$ and $E_1^{\mathsf s,T}$; for $T_2$ and $T_2'$ only the lower bound is established.}
\label{tab:threeindexdefect}
\footnotesize
\begin{ruledtabular}
\begin{tabular}{lrrrrrrrr}
 & \multicolumn{2}{c}{PQG} & \multicolumn{2}{c}{PQG+$T_1$} & \multicolumn{2}{c}{PQG+$T_1$+$T_2$} & \multicolumn{2}{c}{PQG+$T_1$+$T_2'$} \\
System & S$-$U & U$-$C & S$-$U & U$-$C & S$-$U & U$-$C & S$-$U & U$-$C \\
\hline
Be & $<10^{-7}$ & $5.7\times10^{-7}$ & $<10^{-7}$ & $4.1\times10^{-7}$ & $<10^{-7}$ & $2.6\times10^{-7}$ & $<10^{-7}$ & $2.3\times10^{-7}$ \\
H$_2$O & $2.7\times10^{-4}$ & $8.1\times10^{-5}$ & $2.7\times10^{-5}$ & $3.3\times10^{-4}$ & $2.2\times10^{-5}$ & $1.1\times10^{-5}$ & $<10^{-7}$ & $4.7\times10^{-6}$ \\
CH$_4$ & $2.7\times10^{-4}$ & $4\times10^{-5}$ & $1.2\times10^{-4}$ & $1.4\times10^{-4}$ & $9.2\times10^{-4}$ & $7\times10^{-5}$ & $<10^{-7}$ & $6.5\times10^{-5}$ \\
N$_2$ & $1.8\times10^{-3}$ & $1.2\times10^{-3}$ & $1.1\times10^{-3}$ & $1.9\times10^{-4}$ & $9\times10^{-4}$ & $1.8\times10^{-4}$ & $6.1\times10^{-6}$ & $8.7\times10^{-6}$ \\
Hubbard & $6.9\times10^{-3}$ & $3\times10^{-4}$ & $<10^{-7}$ & $2.2\times10^{-4}$ & $3.9\times10^{-5}$ & $8\times10^{-6}$ & $<10^{-7}$ & $<10^{-7}$ \\
\end{tabular}
\end{ruledtabular}
\end{table*}

%% file: sections/chains.tex
\section{Weakly coupled chains}
\label{sec:chains}
The replicated fragments above do not interact. The same tools describe
chains of identical units, such as polymers or molecular stacks, in which
neighboring units are coupled. Let
\begin{equation}
 \hat H_L(\lambda)=\sum_{a=1}^{L}\hat H_F^{(a)}
 +\lambda\sum_{a=1}^{L-1}\hat W^{(a,a+1)},
 \label{eq:chain}
\end{equation}
where $\hat W^{(a,a+1)}$ is a Hermitian, number-conserving operator of
degree at most four on the spin orbitals of units $a$ and $a+1$, with the
same coefficients for every $a$, and $\lambda\ge0$. With $LN_F$ electrons,
let $E_L^{\mathsf x}(\lambda)$ be the minimum of $\langle\hat H_L(\lambda)\rangle$
over $\F_L^{\mathsf x}$; for $\mathsf x={\mathsf s}$, $\hat W$ is assumed
invariant under spin rotations. At $\lambda=0$ this is the replicated
problem, $E_L^{\mathsf x}(0)=E_L^{\mathsf x}$. Let $B$ be the sum of the
absolute values of the coefficients of $\hat W^{(1,2)}$ in a fixed
normal-ordered monomial form, including its constant term (any such
form gives a valid bound).
Every PQG-feasible pair has $|\gamma_{ij}|\le1$ and
$|\Gamma_{ij;kl}|\le\tfrac12$, because $0\preceq\gamma\preceq I$ and the
diagonal entries $\langle\hat n_i\hat n_j\rangle\le\langle\hat n_i\rangle\le1$
of the PSD matrix $P$ bound its other entries. Hence
$|\langle\hat W^{(a,a+1)}\rangle|\le B$ at every feasible point.

\begin{theorem}[Energy per unit]
\label{thm:chainlimit}
For every $\lambda\ge0$ and both spin settings the limit
$e^{\mathsf x}(\lambda)=\lim_{L\to\infty}E_L^{\mathsf x}(\lambda)/L$ exists, and
\begin{equation}
 \bigl|e^{\mathsf x}(\lambda)-E_1^{\mathsf c}\bigr|\le\lambda B .
 \label{eq:chainbound}
\end{equation}
In particular $e^{\mathsf x}(\lambda)\to E_1^{\mathsf c}$ as $\lambda\to0$.
\end{theorem}

\begin{sketch}
Placing an optimal point of a chain of $L$ units next to one of $M$ units
gives, by the product closure, a feasible point of the chain of $L+M$
units; its energy differs from $E_L^{\mathsf x}+E_M^{\mathsf x}$ only by the
one coupling term across the junction, so
$E_{L+M}^{\mathsf x}(\lambda)\le E_L^{\mathsf x}(\lambda)+E_M^{\mathsf x}(\lambda)+\lambda B$.
The feasible sets do not depend on $\lambda$, so every feasible point has
energy at least its $\lambda=0$ energy minus $\lambda(L-1)B$, and the
restriction bound gives
$E_L^{\mathsf x}(\lambda)\ge LE_1^{\mathsf c}-\lambda(L-1)B$. Fekete's lemma
applied to $E_L^{\mathsf x}(\lambda)+\lambda B$ gives the limit. For the
upper bound in Eq.~\eqref{eq:chainbound}, take the point of
Corollary~\ref{cor:saturation} with energy $LE_1^{\mathsf c}$ at $\lambda=0$
for $L\ge K$; switching on the coupling raises its energy by at most
$\lambda(L-1)B$. The complete proof is in \SM{10}.
\end{sketch}

Theorem~\ref{thm:chainlimit} says that the replication defect persists in
weakly coupled chains. With exact $N$-representability and a fragment
that is stable against charge disproportionation and has a singlet
ground state, the energy per unit of a weakly coupled chain tends to the
exact $E_1$ as $\lambda\to0$. With PQG it tends to $E_1^{\mathsf c}$, which
lies below the isolated PQG value $E_1^{\mathsf x}$ by the defect
$E_1^{\mathsf x}-E_1^{\mathsf c}$. The existence of the limit uses only the
product closure and the bound on the moments, so it also holds with $T_1$
(Proposition~\ref{prop:t1products}) and with full $p$-positivity
(Proposition~\ref{prop:pproducts}); the lower bound
$e(\lambda)\ge E_1^{\mathsf c,T}-\lambda B$ holds there as well, but the
upper bound uses the lift and is proved only for PQG. The constant $B$ is
a crude bound, and the theorem does not describe the behavior at strong
coupling. The argument extends to couplings of any finite range; long-range
couplings, such as unscreened Coulomb interactions between distant units,
are not covered.

%% file: sections/numerics.tex
\section{Numerical examples}
\label{sec:numerics}
\subsection{Computational protocol}
\label{sec:protocol}
Every SDP was generated from the operator algebra: matrix entries were
obtained by normal ordering with CAR, linear equalities were eliminated by
exact rational Gaussian elimination (no equality was split into two
inequalities), and the problems were written in SDPA sparse format with 40
significant digits. All problems were solved in double precision with
SDPA~7.4.3.\cite{SDPA7} The SDPs use real, $S^z$-conserving RDMs with
zero local spin means, which does not change any minimum (\SM{7}), and
the singlet problems use the facially reduced form of
Sec.~\ref{sec:reduced}. Primal and dual objective values are reported as
pairs; they are not certified bounds. We call the interval between them
the bracket and its length the width. A solve was accepted when SDPA stopped in phase pdOPT, pdFEAS, or
pFEAS with primal and dual feasibility errors below $10^{-8}$ and
$10^{-6}$; comparisons below are made within the sum of the widths of the
solves involved. One-electron and two-electron integrals were computed
with PySCF\cite{PySCF} in canonical RHF orbitals. Details, solver diagnostics, and the comparison with SDPA-GMP solutions are given
in \SM{11}.

\subsection{Molecular fragments}
\label{sec:molecules}
Table~\ref{tab:molecules} gives $E_1^{\mathsf u}$, $E_1^{\mathsf s}$, and $E_1^{\mathsf c}$ for
Be, CH$_4$, N$_2$, and H$_2$O in the STO-6G basis.\cite{Hehre1969} CH$_4$
($T_d$, $r_{\rm CH}=1.091$~\AA) and N$_2$ ($r=1.0975$~\AA) use the
geometries of Nakata and Yasuda;\cite{NakataYasuda2009} H$_2$O has
$r_{\rm OH}=0.958$~\AA\ and $\angle{\rm HOH}=104.5^\circ$. The limiting
defect $E_1^{\mathsf c}-E_1^{\mathsf u}$ ranges from $\HtwoODef\,E_h$ for H$_2$O and
$\ChFourDef\,E_h$ for CH$_4$ to $\NtwoDef\,E_h$ for N$_2$. For Be it is
$\BeDef\,E_h$, larger than the sum of the widths of the two solves,
$\BeWidthSum\,E_h$. The widths are differences of objective values and do
not bound the error of an optimal value when the feasibility residuals are
nonzero, so we do not conclude from this that the Be defect is nonzero.

\input{tables/molecules}

\emph{Finite replica counts.} For H$_2$O (Table~\ref{tab:h2oreduced}) the
energy per fragment at $k=2$ is still $\HtwoOKtwoDevU\,E_h$ (u) and
$\HtwoOKtwoDevS\,E_h$ (s) above $E_1^{\mathsf c}$;
for every $k\ge3$ it agrees with $E_1^{\mathsf c}$ to within $\HtwoOSatDevMax\,E_h$,
inside the sum of the widths (at most $\HtwoOSatWidthMax\,E_h$), in both
spin settings. For N$_2$ (Table~\ref{tab:n2reduced}) the energy per
fragment decreases monotonically up to $k=32$ in both settings, as
Proposition~\ref{prop:dilution} requires (the smallest step between
consecutive computed counts is $\NtwoMonoMinStepU\,E_h$ for u and
$\NtwoMonoMinStepS\,E_h$ for s, larger than the widths), and approaches
$E_1^{\mathsf c}$ faster than $1/k$: whenever $k$ is doubled, $E_k/k-E_1^{\mathsf c}$ decreases by
more than a factor of two. At $k=32$ the unrestricted value lies $\NtwoMthirtytwoDevU\,E_h$
above $E_1^{\mathsf c}$, within the sum of the widths ($\NtwoMthirtytwoWidthSumU\,E_h$),
and the singlet value lies $\NtwoMthirtytwoDevS\,E_h$ above it, outside
that sum; the singlet solve at $k=32$ has its primal value below its dual
value, a sign of nonzero feasibility residuals, so whether either setting
has reached $E_1^{\mathsf c}$ at $k=32$ is not decided. N$_2$ thus needs many more replicas than H$_2$O ($k=3$)
before the energy per fragment reaches $E_1^{\mathsf c}$ within the
resolution.

Figure~\ref{fig:convergence} shows how the energy per fragment approaches
$E_1^{\mathsf c}$. For N$_2$ the curves are not straight lines on the
logarithmic scales: the local slope steepens from about $-1.2$ at small $k$
to about $-2$ near $k=12$--$16$ and to about $-3$ or steeper beyond $k=16$, so
the approach is close to $k^{-2}$ only in an intermediate range and becomes
faster at larger $k$. Such a steepening is what one expects when the
energy per fragment reaches $E_1^{\mathsf c}$ at a finite replica count
(Corollary~\ref{cor:saturation}), rather than approaching it as a power of
$k$; the data do not determine that count. Corollary~\ref{cor:saturation} guarantees that such a count exists, but
neither the data nor the construction of Proposition~\ref{prop:lift}
determines it. At the numerical minimizers of $E_1^{\mathsf c}$ for all four
molecules, some natural occupations, mainly those of core orbitals, lie
within $\eta=\MolEtaMin$--$\MolEtaMax$ of $0$ or $1$. Whether these are
exactly frozen, and can be split off by Lemma~\ref{lem:frozen}, or are
small but nonzero margins of the active part, is not decided by
double-precision solutions; in the second case the threshold $\tau_c$ of
the active part is tiny and not resolved. We therefore give no
$K_{\rm lift}$ for the molecules; for the Hubbard fragment it is
given in Sec.~\ref{sec:hubbard}. The bound of
Eq.~\eqref{eq:rate}, of order $k^{-1/2}$, is a worst-case estimate from the
regularization and lies far above these values.

\input{figures/convergence}

\input{tables/molecules_reduced_h2o}
\input{tables/molecules_reduced_n2}

\emph{Comparison with earlier replica calculations.} For CH$_4$ and N$_2$
the present $E_1^{\mathsf s}$ reproduces the values reported by Nakata and
Yasuda\cite{NakataYasuda2009} to the digits they quote: for CH$_4$ the
values differ by one unit in the seventh decimal, and for N$_2$, quoted to
six decimals, the difference of about $5\times10^{-7}\,E_h$ is consistent
with rounding (Table~\ref{tab:NYcompare}). For CH$_4$ they reported a decrease of about
$\ChFourNYDef\,E_h$ from $k=1$ to $2$ followed by a plateau, in agreement
with the present singlet defect. For N$_2$ they described the energies up to
$M=32$ replicas by the empirical fit $E_M/M=\NtwoNYFit+0.00302\,M^{-2}$
(in their notation, $M=k$). Its
constant term lies $\NtwoNYLoneDiff\,E_h$ above the present $E_1^{\mathsf c}$, and the
present singlet values for $k\ge24$ already lie below it. Since $E_1^{\mathsf c}$ is
both a lower bound at every $k$ and the limit (Theorem~\ref{thm:limit}),
the fit does not give the asymptotic value for the PQG problem. The
qualitative observations of that work are explained by the present
results: the monotone decrease by Proposition~\ref{prop:dilution}, and the
finite limit by Theorem~\ref{thm:limit}. That work also showed that
imposing the cumulant form of the interfragment 2-RDM in the limit leads
back to the isolated problem; the limit theorem does not need this
assumption, and the limit keeps the covariance condition while dropping
the contraction.

\input{tables/ny_compare}

\emph{Atomic-region integer cuts.} Imposing the integer cut
Eq.~\eqref{eq:intcut} with $\ell=\lfloor\mu_\Omega\rfloor$ on every atomic
region simultaneously (bond-critical-point planes for N$_2$ as in
Part~I,\cite{PartI} and Bragg-radius-weighted Voronoi cells for H$_2$O and
CH$_4$ as in Part~II\cite{PartII}) leaves both $E_1^{\mathsf u}$ and $E_1^{\mathsf c}$
unchanged within the widths (Table~\ref{tab:regioncuts}). All cuts are
inactive at the cutless solutions, including those on the heavy-atom
regions with $k\ge1$, which are not implied by
Proposition~\ref{prop:T1region}. Within this basis and these partitions,
inequalities on the first two moments of atomic electron counts therefore
do not reduce the replication defect.

\input{tables/molecules_region_cuts}

\subsection{Four-site Hubbard fragment}
\label{sec:hubbard}
The fragment is the open four-site Hubbard chain,\cite{Hubbard1963}
\begin{equation}
 \hat H_F=-t\sum_{i=1}^3\sum_\sigma
 (a_{i\sigma}^\dagger a_{i+1,\sigma}+\mathrm{h.c.})
 +U\sum_{i=1}^4n_{i\uparrow}n_{i\downarrow},
 \label{eq:Hubbard}
\end{equation}
with $U/t=4$, $r=8$, and $N_F=4$. This model fragment complements the
molecules: its full replica problems can be solved for several $k$, and
SDPA-GMP solutions are available. The single-fragment program gives
$E_1^{\mathsf c}/t=\HubLone$ with a width of $\HubLoneWidth$, compared with
$E_1^{\mathsf u}/t=\HubEoneU$ and $E_1^{\mathsf s}/t=\HubEoneS$
(Table~\ref{tab:energies}).

\input{tables/hubbard_archive}

\emph{Full and reduced replica problems.} We solved the full (unreduced)
$k$-replica problem for $k\le4$ and the reduced problem of
Proposition~\ref{prop:reduced} for $k$ up to $64$ ($k=1000$ for
$\mathsf x={\mathsf u}$; Tables~\ref{tab:reduced}
and~\ref{tab:saturation}). The two agree within their widths at every
$k$ where both were solved, as Proposition~\ref{prop:reduced} requires.
For $k=1,2,3$ the global-singlet values also agree with solutions
obtained with SDPA-GMP in 512-bit arithmetic\cite{Nakata2010SDPAGMP} to within $\HubGmpDevMax\,t$ per fragment, which is
inside the sum of the double-precision width and the SDPA-GMP primal--dual
gap (the latter is $\HubGmpGapKone\,t$ at $k=1$ and below $10^{-12}\,t$
at $k=2,3$).

\input{tables/hubbard_reduced}
\input{tables/hubbard_saturation}

\emph{Saturation within the numerical resolution.} In both spin settings
the energy per fragment drops by $E_1^{\mathsf x}-E_1^{\mathsf c}$ from $k=1$ to $k=2$ and
then stays at $E_1^{\mathsf c}$ within the widths up to the largest $k$ computed
(Table~\ref{tab:saturation}), earlier than for H$_2$O ($k=3$) and N$_2$: the replication error per fragment,
$E_k/k-E_1^{\mathsf x}$, is $\HubNonaddPerS\,t$ for $\mathsf x={\mathsf s}$ at every
$k\ge2$. The SDPA-GMP solutions resolve this further. Their values at
$k=2$ and $k=3$ differ by $E_3/3-E_2/2=\HubGmpThreeTwo\,t$, below their
primal--dual gaps of at most $\HubGmpGapPerFrag\,t$ per fragment, and both
lie inside the double-precision bracket of $E_1^{\mathsf c}$, whose width is
$\HubLoneWidth\,t$; for the SDPA-GMP value,
$E_2^{\mathsf s}/2-E_1^{\mathsf c}=\HubDevKtwoS\,t$ (with $E_1^{\mathsf c}$ at the midpoint of its
bracket). In double precision, $E_2^{\mathsf u}/2-E_1^{\mathsf c}=\HubDevKtwoU\,t$, inside
the widths. These are numerical
observations: the SDPA-GMP values at $k=2$ and $k=3$ agree with each
other within their gaps, and the computed values of $E_2^{\mathsf s}/2$ and
of $E_1^{\mathsf c}$ agree to about $10^{-9}\,t$. Because $E_1^{\mathsf c}$ is a
double-precision value and primal--dual differences are not error bounds
(\SM{11}), this is an agreement of computed values, not a bound on the
difference of the optimal values, and it does not establish exact
saturation at $k=2$. If
exact, Corollary~\ref{cor:saturation}(ii) would extend it to every $k\ge2$.

\emph{Lift thresholds.} Table~\ref{tab:liftthresholds} gives the thresholds,
evaluated numerically, of the construction of Proposition~\ref{prop:lift} at the
numerical minimizer of $E_1^{\mathsf c}$. With the number alone controlled, the lift is
feasible from $k=2$; with number and spin controlled, it is feasible from
$k=\HubKliftSing$. The singlet value nonetheless agrees with $E_1^{\mathsf c}$ at $k=2$
within the resolution; this numerical agreement does not show that an
exact extension exists at $k=2$.
$K_{\rm lift}$ is a property of the construction, not the onset of
saturation.

The contrast with the molecules explains why saturation is visible here
but not there. At the Hubbard minimizer the natural occupations stay at
least $\eta\approx\HubEta$ away from $0$ and $1$, so $K_{\rm lift}$ is small
and the energy per fragment reaches $E_1^{\mathsf c}$ within the computed
range. At the molecular minimizers some occupations lie within
$\MolEtaMin$--$\MolEtaMax$ of $0$ or $1$ (Sec.~\ref{sec:molecules}). If
these are small but nonzero margins of the active part rather than frozen
orbitals, the uniform bound of Eq.~\eqref{eq:Kexplicit} gives a $K_{\rm lift}$
of order $\eta^{-2}$, and over the accessible replica counts the energy per
fragment appears to approach $E_1^{\mathsf c}$ as a power of $k$, as for
N$_2$, although Corollary~\ref{cor:saturation} guarantees saturation at
some finite count.

\input{tables/hubbard_lift}

\emph{Spin and number parts of the defect.} The singlet defect splits as
$E_1^{\mathsf s}-E_1^{\mathsf c}=(E_1^{\mathsf s}-E_1^{\mathsf u})+(E_1^{\mathsf u}-E_1^{\mathsf c})$
(Table~\ref{tab:deficit}). The first (spin) part is
$\HubDefSpinPct\%$ of the total. The two parts behave differently at
sharp local number. With the local electron number made sharp by
Eq.~\eqref{eq:sharp}, the singlet replica values for $k=2,3,4$ equal
$E_1^{\mathsf u}$ to within $\HubSharpSDevMax\,t$ (widths up to
$\HubSharpSWidthMax\,t$), the value that Proposition~\ref{prop:T2}
gives for sufficiently large $k$. The lowering by the number part,
$E_1^{\mathsf u}-E_1^{\mathsf c}$, is thus blocked, whereas the lowering
from the isolated singlet by the spin part, $E_1^{\mathsf s}-E_1^{\mathsf u}$,
still occurs, here already at $k=2$. The corresponding solutions are not local singlets
($\langle\hat{\bm S}_F^2\rangle=\HubSharpSpin$); the global singlet is reached
through the interfragment spin correlation
$\langle\hat{\bm S}_F^{(a)}\!\cdot\hat{\bm S}_F^{(b)}\rangle=-\langle\hat{\bm S}_F^2\rangle/(k-1)$
of Sec.~\ref{sec:cancel}. The number part, $E_1^{\mathsf u}-E_1^{\mathsf c}$, requires
local number fluctuations and their negative interfragment covariance. In
contrast, integer cuts on the fragment, on single sites, and on half
chains, all of which have integer means, change $E_1^{\mathsf c}$ and the $k=2$ values
by at most $\HubCutChangeMax\,t$, in agreement with
Proposition~\ref{prop:T1}.

\input{tables/hubbard_deficit}

%% file: tables/molecules.tex
\begin{table*}[t]
\caption{PQG energies of the molecular fragments in the STO-6G basis (hartree; double-precision SDPA 7.4.3): primal (upper) and dual (lower) objective values, which are not certified bounds. The deficit $E_1^{\mathsf c}-E_1^{\mathsf u}$ is computed from the midpoints of the primal-dual pairs; the Be deficit ($\BeDef\,E_h$) exceeds the sum of the primal-dual widths of the two solves ($\BeWidthSum\,E_h$) but is close to the solver noise level.}
\label{tab:molecules}
\footnotesize
\begin{ruledtabular}
\begin{tabular}{llllll}
System & $E_{\rm FCI}$ & $E_1^{\mathsf u}$ & $E_1^{\mathsf s}$ & $E_1^{\mathsf c}$ & $E_1^{\mathsf c}-E_1^{\mathsf u}$ \\
\hline
Be/STO-6G & $-14.5560886$ & \begin{tabular}[t]{@{}l@{}}$-14.5560891$\\$-14.5560891$\end{tabular} & \begin{tabular}[t]{@{}l@{}}$-14.5560890$\\$-14.5560890$\end{tabular} & \begin{tabular}[t]{@{}l@{}}$-14.5560896$\\$-14.5560896$\end{tabular} & $-5.7\times10^{-7}$ \\
CH$_4$/STO-6G & $-40.1909382$ & \begin{tabular}[t]{@{}l@{}}$-40.2107929$\\$-40.2107930$\end{tabular} & \begin{tabular}[t]{@{}l@{}}$-40.2105266$\\$-40.2105266$\end{tabular} & \begin{tabular}[t]{@{}l@{}}$-40.2108331$\\$-40.2108332$\end{tabular} & $-4.02\times10^{-5}$ \\
N$_2$/STO-6G & $-108.7004158$ & \begin{tabular}[t]{@{}l@{}}$-108.7143086$\\$-108.7143086$\end{tabular} & \begin{tabular}[t]{@{}l@{}}$-108.7124885$\\$-108.7124886$\end{tabular} & \begin{tabular}[t]{@{}l@{}}$-108.7155558$\\$-108.7155559$\end{tabular} & $-1.25\times10^{-3}$ \\
H$_2$O/STO-6G & $-75.7288495$ & \begin{tabular}[t]{@{}l@{}}$-75.7310975$\\$-75.7310975$\end{tabular} & \begin{tabular}[t]{@{}l@{}}$-75.7308271$\\$-75.7308271$\end{tabular} & \begin{tabular}[t]{@{}l@{}}$-75.7311782$\\$-75.7311782$\end{tabular} & $-8.07\times10^{-5}$ \\
\end{tabular}
\end{ruledtabular}
\end{table*}

%% file: figures/convergence.tex
\begin{figure}[t]
\centering
\begin{tikzpicture}
\begin{loglogaxis}[
  width=0.95\columnwidth, height=0.72\columnwidth,
  xlabel={replica count $k$}, ylabel={$E_k/k-E_1^{\mathsf c}$ ($E_h$)},
  xmin=0.8, xmax=45, ymin=5e-7, ymax=6e-3,
  xtick={1,2,4,8,16,32}, xticklabels={1,2,4,8,16,32},
  legend pos=south west, legend cell align=left,
  legend style={font=\footnotesize, draw=none, fill=none},
  tick label style={font=\footnotesize}, label style={font=\small}]
\addplot[mark=*, mark size=1.6pt, black] table[x index=0, y index=1]
  {figures/convergence_n2_u.dat};
\addlegendentry{N$_2$, $\mathsf u$}
\addplot[mark=square*, mark size=1.5pt, black, dashed] table[x index=0, y index=1]
  {figures/convergence_n2_s.dat};
\addlegendentry{N$_2$, $\mathsf s$}
\addplot[mark=triangle*, mark size=2pt, gray] table[x index=0, y index=1]
  {figures/convergence_h2o_u.dat};
\addlegendentry{H$_2$O, $\mathsf u$}
\addplot[mark=triangle, mark size=2pt, gray, dashed] table[x index=0, y index=1]
  {figures/convergence_h2o_s.dat};
\addlegendentry{H$_2$O, $\mathsf s$}
\addplot[domain=1:40, samples=2, thin, dotted] {9.607062e-05*8/x};
\addlegendentry{$\propto k^{-1}$}
\addplot[domain=2:40, samples=30, thin, dashdotted] {9.607062e-05*64/x^2};
\addlegendentry{$\propto k^{-2}$}
\end{loglogaxis}
\end{tikzpicture}
\caption{Approach of the energy per fragment to $E_1^{\mathsf c}$ for the
reduced replica problems of N$_2$ and H$_2$O (STO-6G; midpoints of the
primal--dual pairs, double-precision SDPA~7.4.3). Points within the sum of
the widths of the two solves involved are not shown; for H$_2$O this
leaves only $k=1$ and $2$, and for N$_2$ with $\mathsf x={\mathsf u}$ it
removes $k=32$. The guide lines pass through the N$_2$ ($\mathsf u$) point
at $k=8$. The bound of Eq.~\eqref{eq:rate}, of order $k^{-1/2}$, lies
far above the plotted range.}
\label{fig:convergence}
\end{figure}
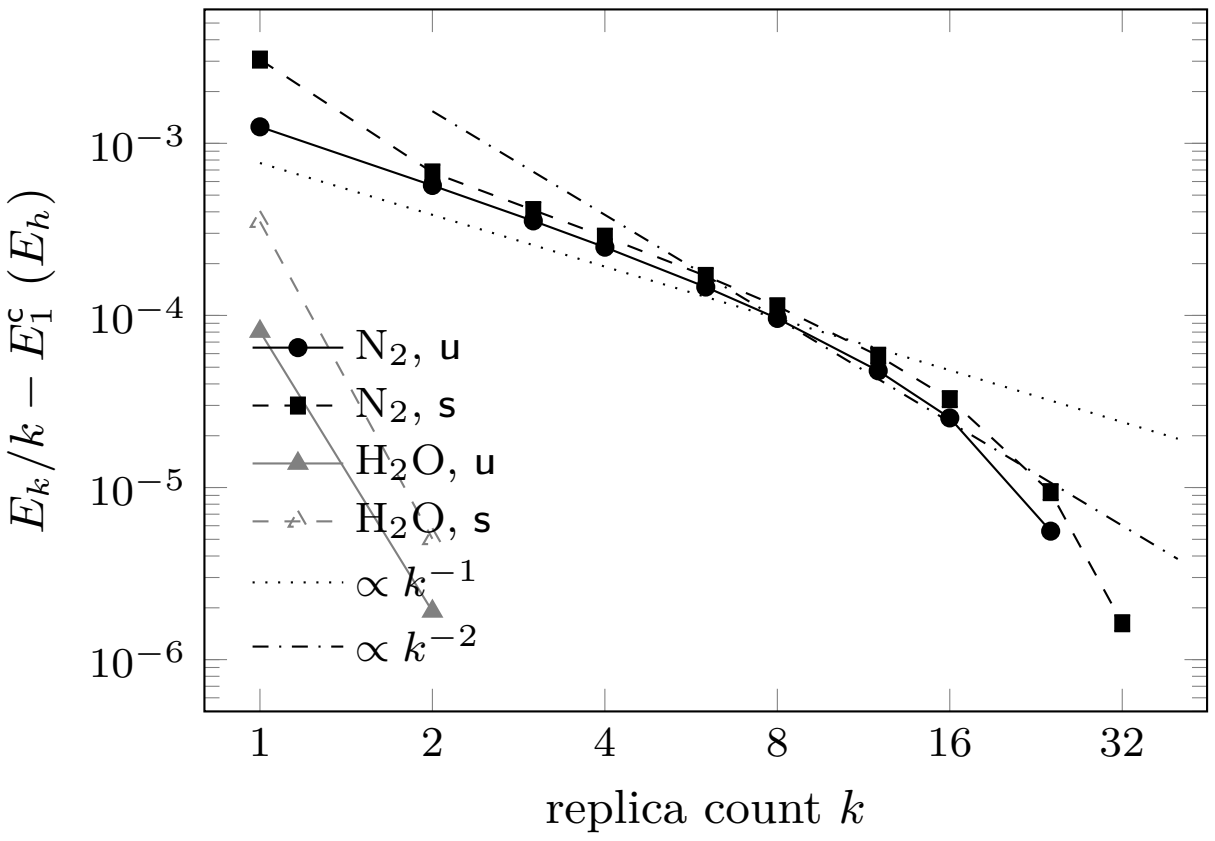

%% file: tables/molecules_reduced_h2o.tex
\begin{table}[t]
\caption{Energy per fragment $E_k^{\mathsf x}/k$ of the H$_2$O fragment from the reduced problem (hartree; double-precision SDPA 7.4.3): primal / dual objective values, which are not certified bounds. The $k=1$ row is the isolated problem.}
\label{tab:h2oreduced}
\footnotesize
\begin{ruledtabular}
\begin{tabular}{rll}
$k$ & $\mathsf x={\mathsf u}$ & $\mathsf x={\mathsf s}$ \\
\hline
1 & $-75.7310975/-75.7310975$ & $-75.7308271/-75.7308271$ \\
2 & $-75.7311763/-75.7311763$ & $-75.7311726/-75.7311737$ \\
3 & $-75.7311782/-75.7311782$ & $-75.7311782/-75.7311786$ \\
4 & $-75.7311782/-75.7311782$ & $-75.7311782/-75.7311784$ \\
8 & $-75.7311782/-75.7311782$ & $-75.7311782/-75.7311782$ \\
32 & $-75.7311782/-75.7311782$ & $-75.7311781/-75.7311786$ \\
\end{tabular}
\end{ruledtabular}
\end{table}

%% file: tables/molecules_reduced_n2.tex
\begin{table}[t]
\caption{Energy per fragment $E_k^{\mathsf x}/k$ of the N$_2$ fragment from the reduced problem (hartree; double-precision SDPA 7.4.3): primal / dual objective values, which are not certified bounds. The $k=1$ row is the isolated problem.}
\label{tab:n2reduced}
\footnotesize
\begin{ruledtabular}
\begin{tabular}{rll}
$k$ & $\mathsf x={\mathsf u}$ & $\mathsf x={\mathsf s}$ \\
\hline
1 & $-108.7143086/-108.7143086$ & $-108.7124885/-108.7124886$ \\
2 & $-108.7149865/-108.7149869$ & $-108.7148781/-108.7148785$ \\
3 & $-108.7152015/-108.7152016$ & $-108.7151463/-108.7151464$ \\
4 & $-108.7153066/-108.7153067$ & $-108.7152690/-108.7152691$ \\
6 & $-108.7154096/-108.7154096$ & $-108.7153862/-108.7153862$ \\
8 & $-108.7154598/-108.7154598$ & $-108.7154428/-108.7154428$ \\
12 & $-108.7155082/-108.7155082$ & $-108.7154975/-108.7154975$ \\
16 & $-108.7155304/-108.7155306$ & $-108.7155232/-108.7155233$ \\
24 & $-108.7155500/-108.7155505$ & $-108.7155463/-108.7155466$ \\
32 & $-108.7155548/-108.7155555$ & $-108.7155543/-108.7155541$ \\
\end{tabular}
\end{ruledtabular}
\end{table}

%% file: tables/ny_compare.tex
\begin{table*}[t]
\caption{Comparison with Ref.~\onlinecite{NakataYasuda2009}, Sec.~III and Figs.~1--2, at its geometries (STO-6G, hartree). Present values: double-precision SDPA 7.4.3, primal (upper) and dual (lower) objective values; defects from midpoints. The reported N$_2$ limit is the constant term of the empirical fit $E_M/M=-108.71553+0.00302\,M^{-2}$ given there, not a directly computed infinite-replica value. For CH$_4$ no limiting value was reported; the reported decrease from $M=1$ to $2$ is listed as the singlet defect.}
\label{tab:NYcompare}
\begin{ruledtabular}
\begin{tabular}{llll}
 & & CH$_4$ & N$_2$ \\
\hline
Isolated singlet $E_1^{\mathsf s}$ & present & \begin{tabular}[t]{@{}l@{}}$-40.2105266$\\$-40.2105266$\end{tabular} & \begin{tabular}[t]{@{}l@{}}$-108.7124885$\\$-108.7124886$\end{tabular} \\
 & Ref.~\onlinecite{NakataYasuda2009} & $-40.2105267$ & $-108.712489$ \\
Replication limit & present, $E_1^{\mathsf c}$ & \begin{tabular}[t]{@{}l@{}}$-40.2108331$\\$-40.2108332$\end{tabular} & \begin{tabular}[t]{@{}l@{}}$-108.7155558$\\$-108.7155559$\end{tabular} \\
 & Ref.~\onlinecite{NakataYasuda2009} & --- & $-108.71553$ (fit) \\
Singlet defect $E_1^{\mathsf s}-E_1^{\mathsf c}$ & present & $3.07\times10^{-4}$ & $3.07\times10^{-3}$ \\
 & Ref.~\onlinecite{NakataYasuda2009} & $3\times10^{-4}$ & $3.04\times10^{-3}$ (from fit) \\
Unrestricted defect $E_1^{\mathsf u}-E_1^{\mathsf c}$ & present & $4.02\times10^{-5}$ & $1.25\times10^{-3}$ \\
\end{tabular}
\end{ruledtabular}
\end{table*}

%% file: tables/molecules_region_cuts.tex
\begin{table}[t]
\caption{Real-space atomic-region integer cuts ($\ell=\lfloor\mu_\Omega\rfloor$ in every region, imposed simultaneously; double-precision SDPA 7.4.3, primal values, hartree). All imposed cuts are inactive at these solutions, and the deficit $E_1^{\mathsf c}-E_1^{\mathsf u}$ is unchanged within the solver spread.}
\label{tab:regioncuts}
\begin{ruledtabular}
\begin{tabular}{lllr}
System & $E_1^{\mathsf c}-E_1^{\mathsf u}$ (cutless) & $E_1^{\mathsf c}-E_1^{\mathsf u}$ (cuts) & change \\
\hline
N$_2$ & $-1.247\times10^{-3}$ & $-1.247\times10^{-3}$ & $-2.7\times10^{-9}$ \\
H$_2$O & $-8.068\times10^{-5}$ & $-8.065\times10^{-5}$ & $3.2\times10^{-8}$ \\
CH$_4$ & $-4.018\times10^{-5}$ & $-4.018\times10^{-5}$ & $-4.4\times10^{-10}$ \\
\end{tabular}
\end{ruledtabular}
\end{table}

%% file: tables/hubbard_archive.tex
\begin{table*}[t]
\caption{Dimensionless energy per four-site Hubbard fragment ($t=1$, $U=4$). The $k$ columns are global-singlet values obtained with SDPA-GMP (512-bit working arithmetic; data described in the supplementary material). The PQG rows agree with the double-precision values (the same problems rebuilt with the input generators of this repository and re-solved with SDPA 7.4.3) within the sum of the primal-dual width of the double-precision solution and the SDPA-GMP objective gap ($\HubGmpGapKone\,t$ at $k=1$; at most $\HubGmpGapMaxKtwoThree\,t$ at $k=2,3$; Table~\ref{tab:saturation}). The last column gives local SDP values $E_1^{\mathsf c}/t$ or $E_1^{\mathsf c,T}/t$, computed in this work with double-precision SDPA 7.4.3 and only the $S_z=0$ spin condition. Entries are primal-dual midpoints, accepted with phase pdOPT, pdFEAS, or pFEAS, primal feasibility error below $10^{-8}$, and dual feasibility error below $10^{-6}$, with primal-dual width at most $10^{-6}\,t$. These are convergence diagnostics, not error guarantees. An unaccepted local solve is shown as ---; the FCI local entry is not applicable. Equality with the replication limit is proved for PQG only, not for the three-index conditions. The $k$ columns of the three-index rows are archived SDPA-GMP solutions; the $k=3$ runs with $T_2$ or $T_2'$ did not finish (Sec.~\ref{sec:threeindex}). Rounded digits are not certified error bounds.}
\label{tab:energies}
\begin{ruledtabular}
\begin{tabular}{lrrrr}
Conditions & $k=1$ & $k=2$ & $k=3$ & Local SDP \\
\hline
FCI & $-1.9531453$ & $-1.9531453$ & $-1.9531453$ & --- \\
PQG & $-2.0017922$ & $-2.0090322$ & $-2.0090322$ & $\HubLone$ \\
PQG+$T_1$ & $-1.9548934$ & $-1.9550140$ & $-1.9550667$ & $-1.9551123$ \\
PQG+$T_1$+$T_2$ & $-1.9531707$ & $-1.9531707$ & --- & $-1.9532178$ \\
PQG+$T_1$+$T_2'$ & $-1.9531707$ & $-1.9531707$ & --- & $-1.9531707$ \\
\end{tabular}
\end{ruledtabular}
\end{table*}

%% file: tables/hubbard_reduced.tex
\begin{table}[t]
\caption{Energy per fragment $E_k^{\mathsf x}/(kt)$ of the four-site Hubbard fragment from the reduced problem (double-precision SDPA 7.4.3): primal / dual objective values, in units of $t$. Primal and dual values are not certified bounds. The local SDP gives $E_1^{\mathsf c}/t=\HubLone$.}
\label{tab:reduced}
\begin{ruledtabular}
\begin{tabular}{rll}
$k$ & $\mathsf x={\mathsf u}$ & $\mathsf x={\mathsf s}$ \\
\hline
1 & $-2.0087310/-2.0087310$ & $-2.0017921/-2.0017921$ \\
2 & $-2.0090322/-2.0090322$ & $-2.0090321/-2.0090326$ \\
3 & $-2.0090322/-2.0090322$ & $-2.0090322/-2.0090322$ \\
4 & $-2.0090322/-2.0090322$ & $-2.0090322/-2.0090322$ \\
5 & $-2.0090322/-2.0090322$ & $-2.0090322/-2.0090322$ \\
6 & $-2.0090322/-2.0090323$ & $-2.0090322/-2.0090323$ \\
8 & $-2.0090322/-2.0090323$ & $-2.0090322/-2.0090322$ \\
12 & $-2.0090322/-2.0090322$ & $-2.0090322/-2.0090324$ \\
16 & $-2.0090322/-2.0090322$ & $-2.0090321/-2.0090326$ \\
24 & $-2.0090322/-2.0090322$ & $-2.0090322/-2.0090322$ \\
32 & $-2.0090322/-2.0090322$ & $-2.0090322/-2.0090322$ \\
64 & $-2.0090322/-2.0090323$ & $-2.0090322/-2.0090327$ \\
1000 & $-2.0090322/-2.0090322$ & --- \\
\end{tabular}
\end{ruledtabular}
\end{table}

%% file: tables/hubbard_saturation.tex
\begin{table*}[t]
\caption{Replication error of the four-site Hubbard fragment in units of $t$. Upper block: midpoints of the primal-dual pairs of the full (unreduced) double-precision SDPA runs ($k\le4$). Lower block: 512-bit SDPA-GMP global-singlet solutions ($\mathsf x={\mathsf s}$). $E_k-kE_1^{\mathsf x}$ is the total nonadditivity; $E_k/k-E_1^{\mathsf c}$ uses the double-precision midpoint of $E_1^{\mathsf c}$. All entries are numerical observations, not certified bounds; exact saturation at $k=2$ is not concluded.}
\label{tab:saturation}
\begin{ruledtabular}
\begin{tabular}{lrrrrrr}
$\mathsf x$ & $k$ & $E_k/k$ & width$/k$ & $E_k-kE_1^{\mathsf x}$ & $E_k/k-E_1^{\mathsf x}$ & $E_k/k-E_1^{\mathsf c}$ \\
\hline
s & 1 & $-2.0017921$ & $1.2\times10^{-8}$ & $0$ & $0$ & $7.24\times10^{-3}$ \\
s & 2 & $-2.0090322$ & $4.4\times10^{-8}$ & $-1.448\times10^{-2}$ & $-7.24\times10^{-3}$ & $-1.19\times10^{-8}$ \\
s & 3 & $-2.0090322$ & $1.5\times10^{-7}$ & $-2.172\times10^{-2}$ & $-7.24\times10^{-3}$ & $-5.84\times10^{-8}$ \\
s & 4 & $-2.0090325$ & $8\times10^{-7}$ & $-2.896\times10^{-2}$ & $-7.24\times10^{-3}$ & $-3.42\times10^{-7}$ \\
u & 1 & $-2.0087310$ & $5\times10^{-10}$ & $0$ & $0$ & $3.01\times10^{-4}$ \\
u & 2 & $-2.0090322$ & $1.1\times10^{-7}$ & $-6.025\times10^{-4}$ & $-3.013\times10^{-4}$ & $-4.15\times10^{-8}$ \\
u & 3 & $-2.0090322$ & $8.1\times10^{-8}$ & $-9.038\times10^{-4}$ & $-3.013\times10^{-4}$ & $-3.52\times10^{-8}$ \\
u & 4 & $-2.0090324$ & $4.7\times10^{-7}$ & $-1.206\times10^{-3}$ & $-3.014\times10^{-4}$ & $-2.11\times10^{-7}$ \\
\hline
${\mathsf s}$ (GMP) & 1 & $-2.00179217$ & --- & $0$ & $0$ & $7.2\times10^{-3}$ \\
${\mathsf s}$ (GMP) & 2 & $-2.00903218$ & --- & $-1.448\times10^{-2}$ & $-7.24\times10^{-3}$ & $2\times10^{-10}$ \\
${\mathsf s}$ (GMP) & 3 & $-2.00903218$ & --- & $-2.172\times10^{-2}$ & $-7.24\times10^{-3}$ & $2\times10^{-10}$ \\
\end{tabular}
\end{ruledtabular}
\end{table*}

%% file: tables/hubbard_lift.tex
\begin{table}[t]
\caption{Numerically evaluated thresholds of the projected-covariance lift at the numerical local PQG minimizer of the four-site Hubbard fragment. The candidate count is $K_{\rm lift}=\max\{2,\lceil1+1/\tau_c\rceil\}$, a sufficient condition for this construction, not a certified saturation onset.}
\label{tab:liftthresholds}
\begin{ruledtabular}
\begin{tabular}{lcc}
Controlled operators & $\tau_c$ & $K_{\rm lift}$ \\
\hline
$\hat N_F$ (unrestricted spin) & 8.8640 & 2 \\
$\hat N_F,\hat{\bm S}_F$ (singlet) & 0.5063 & 3 \\
\end{tabular}
\end{ruledtabular}
\end{table}

%% file: tables/hubbard_deficit.tex
\begin{table}[t]
\caption{Decomposition of the isolated singlet defect $E_1^{\mathsf s}-E_1^{\mathsf c}$ of the four-site Hubbard fragment into the ordered contributions $E_1^{\mathsf s}-E_1^{\mathsf u}$ (spin) and $E_1^{\mathsf u}-E_1^{\mathsf c}$ (local electron number), in units of $t$, evaluated at the midpoints of the primal-dual pairs. The shares are percentages of the total; the two parts are canceled by different mechanisms (see text).}
\label{tab:deficit}
\begin{ruledtabular}
\begin{tabular}{lrr}
Contribution & Value & Share (\%) \\
\hline
$E_1^{\mathsf s}-E_1^{\mathsf u}$ (spin) & $6.939\times10^{-3}$ & 95.8 \\
$E_1^{\mathsf u}-E_1^{\mathsf c}$ (number) & $3.012\times10^{-4}$ & 4.2 \\
$E_1^{\mathsf s}-E_1^{\mathsf c}$ (total) & $7.24\times10^{-3}$ & $100$ \\
\end{tabular}
\end{ruledtabular}
\end{table}

%% file: sections/conclusions.tex
\section{Conclusions}
\label{sec:conclusions}
For identical, strictly noninteracting fragments, the PQG energy per
fragment converges to the minimum $E_1^{\mathsf c}$ of a single-fragment SDP that
fixes only the mean electron number and keeps the particle--hole
covariance condition. The limit is the same with unrestricted spin and
with a global singlet. Restriction gives the lower bound at every replica
count; an explicit lift with negative interfragment covariances of order
$1/(k-1)$ gives the matching upper bound. The energy per fragment is
nonincreasing in the replica count, replication additivity holds exactly
when $E_1^{\mathsf c}=E_1^{\mathsf x}$, and in a fixed finite basis the energy per fragment reaches $E_1^{\mathsf c}$ exactly at a finite replica count. Every finite-replica problem
reduces exactly to an SDP whose size does not depend on the replica count.

The same limit governs weakly coupled chains of identical units: their
PQG energy per unit exists and lies within $\lambda B$ of $E_1^{\mathsf c}$
for coupling strength $\lambda$ (Theorem~\ref{thm:chainlimit}), so the
defect persists in weakly coupled polymers.

The limiting defect $E_1^{\mathsf x}-E_1^{\mathsf c}$ is tied to the local number
variance and, in the singlet setting, also to local spin fluctuations; it
is an artifact of PQG when the exact problem is stable against charge
disproportionation and, in the singlet setting, has a singlet ground state. Integer electron-count cuts that use only the mean of a
fragment's count are implied by PQG and leave it unchanged; cuts on atomic
regions also leave it unchanged in the molecules studied here. The equality that makes
each fragment's count sharp restores additivity with unrestricted spin. In the singlet setting the replicated energy per fragment at sharp local
number reaches $E_1^{\mathsf u}$ for sufficiently large replica counts: the
lowering by number fluctuations is blocked, while the lowering by the spin
part, through interfragment spin correlation, remains; in the Hubbard
example this happens already at $k=2$.

Numerically, the energy per fragment reaches $E_1^{\mathsf c}$ within the
resolution at $k=3$ for H$_2$O, whereas for N$_2$ the energy per fragment
still approaches $E_1^{\mathsf c}$ at $k=32$, where the solver residuals do
not decide whether it has been reached; for the four-site Hubbard fragment it does
so at $k=2$, where SDPA-GMP solutions agree. These are numerical
observations; they do not establish exact saturation counts.

The analysis is restricted to PQG. With the three-index conditions the
restriction bound still holds, so that $E_k/k$ stays bounded below. With
$T_1$ alone, products remain feasible, which gives the upper bound by the
isolated value and the existence of the limit
(Proposition~\ref{prop:t1products}); with $T_1$ and $T_2$ these rest on
Conjecture~\ref{conj:threeindex} (Sec.~\ref{sec:threeindex}). Whether a
lift with the same properties exists is not addressed here. For full $p$-positivity products remain
feasible and the limit exists (Proposition~\ref{prop:pproducts}), and we
conjecture that it is again the single-fragment relaxation
(Conjecture~\ref{conj:ppos}), as it is for $p=2$ and for exact
$N$-representability. The results neither
restore size consistency nor imply $N$-representability of the replicated
solutions.

%% file: sections/backmatter.tex
\section*{Supplementary Material}
The supplementary material contains the complete proofs of the results
stated in Secs.~\ref{sec:problems}--\ref{sec:cuts}, the details of the
numerical protocol and solver diagnostics, and the comparison with
SDPA-GMP solutions in 512-bit arithmetic.

\section*{Author Declarations}
\subsection*{Conflict of Interest}
The author has no conflicts to disclose.

\subsection*{Use of AI Tools}
The author used AI tools in this work: Claude Opus 5.5 (Anthropic) through
Claude Code, and ChatGPT 6.1-Sol and ChatGPT 6.2-Sol (OpenAI) and
GLM-5.3-Flash (Z.ai) through
the pi coding agent, to write and run scripts, to review the code and the
manuscript, and to draft text; ChatGPT 6 Astra Max (OpenAI) was used for
an adversarial review of the manuscript. Findings of the reviews were
checked against the code, the data, and the cited sources before any
change was made. All numbers in the manuscript are
generated by scripts in the accompanying repository and are checked
automatically against the stored data. The author designed the study,
checked all results and statements, and takes full responsibility for the
content of this article.

\section*{Data Availability}
The data and scripts that support the findings of this study are openly
available in Zenodo at \url{https://doi.org/10.5281/zenodo.24293501}.
They include the scripts that rebuild every SDPA input
file, the derived results from which the tables and figures are generated,
the verification programs, and the inputs, outputs, and RDMs of the
archived SDPA-GMP calculations, with a manifest of their origin and
checksums. Because of their size, the double-precision SDPA outputs are
not included; they can be regenerated by solving the rebuilt inputs with
SDPA 7.4.3 using the scripts provided.

\bibliography{references}

%% file: manuscript.bbl
\begin{thebibliography}{50}%
\makeatletter
\providecommand \@ifxundefined [1]{%
 \@ifx{#1\undefined}
}%
\providecommand \@ifnum [1]{%
 \ifnum #1\expandafter \@firstoftwo
 \else \expandafter \@secondoftwo
 \fi
}%
\providecommand \@ifx [1]{%
 \ifx #1\expandafter \@firstoftwo
 \else \expandafter \@secondoftwo
 \fi
}%
\providecommand \natexlab [1]{#1}%
\providecommand \enquote  [1]{``#1''}%
\providecommand \bibnamefont  [1]{#1}%
\providecommand \bibfnamefont [1]{#1}%
\providecommand \citenamefont [1]{#1}%
\providecommand \href@noop [0]{\@secondoftwo}%
\providecommand \href [0]{\begingroup \@sanitize@url \@href}%
\providecommand \@href[1]{\@@startlink{#1}\@@href}%
\providecommand \@@href[1]{\endgroup#1\@@endlink}%
\providecommand \@sanitize@url [0]{\catcode `\\12\catcode `\$12\catcode
  `\&12\catcode `\#12\catcode `\^12\catcode `\_12\catcode `\%12\relax}%
\providecommand \@@startlink[1]{}%
\providecommand \@@endlink[0]{}%
\providecommand \url  [0]{\begingroup\@sanitize@url \@url }%
\providecommand \@url [1]{\endgroup\@href {#1}{\urlprefix }}%
\providecommand \urlprefix  [0]{URL }%
\providecommand \Eprint [0]{\href }%
\providecommand \doibase [0]{https://doi.org/}%
\providecommand \selectlanguage [0]{\@gobble}%
\providecommand \bibinfo  [0]{\@secondoftwo}%
\providecommand \bibfield  [0]{\@secondoftwo}%
\providecommand \translation [1]{[#1]}%
\providecommand \BibitemOpen [0]{}%
\providecommand \bibitemStop [0]{}%
\providecommand \bibitemNoStop [0]{.\EOS\space}%
\providecommand \EOS [0]{\spacefactor3000\relax}%
\providecommand \BibitemShut  [1]{\csname bibitem#1\endcsname}%
\let\auto@bib@innerbib\@empty
\bibitem [{\citenamefont {Coleman}(1963)}]{Coleman1963}%
  \BibitemOpen
  \bibfield  {author} {\bibinfo {author} {\bibfnamefont {A.~J.}\ \bibnamefont
  {Coleman}},\ }\bibfield  {title} {\enquote {\bibinfo {title} {Structure of
  fermion density matrices},}\ }\href
  {https://doi.org/10.1103/RevModPhys.35.668} {\bibfield  {journal} {\bibinfo
  {journal} {Rev. Mod. Phys.}\ }\textbf {\bibinfo {volume} {35}},\ \bibinfo
  {pages} {668--686} (\bibinfo {year} {1963})}\BibitemShut {NoStop}%
\bibitem [{\citenamefont {Garrod}\ and\ \citenamefont
  {Percus}(1964)}]{GarrodPercus1964}%
  \BibitemOpen
  \bibfield  {author} {\bibinfo {author} {\bibfnamefont {C.}~\bibnamefont
  {Garrod}}\ and\ \bibinfo {author} {\bibfnamefont {J.~K.}\ \bibnamefont
  {Percus}},\ }\bibfield  {title} {\enquote {\bibinfo {title} {Reduction of the
  {$N$}-particle variational problem},}\ }\href
  {https://doi.org/10.1063/1.1704098} {\bibfield  {journal} {\bibinfo
  {journal} {J. Math. Phys.}\ }\textbf {\bibinfo {volume} {5}},\ \bibinfo
  {pages} {1756--1776} (\bibinfo {year} {1964})}\BibitemShut {NoStop}%
\bibitem [{\citenamefont {Garrod}, \citenamefont {Mihailovi{\'c}},\ and\
  \citenamefont {Rosina}(1975)}]{GarrodMihailovicRosina1975}%
  \BibitemOpen
  \bibfield  {author} {\bibinfo {author} {\bibfnamefont {C.}~\bibnamefont
  {Garrod}}, \bibinfo {author} {\bibfnamefont {M.~V.}\ \bibnamefont
  {Mihailovi{\'c}}},\ and\ \bibinfo {author} {\bibfnamefont {M.}~\bibnamefont
  {Rosina}},\ }\bibfield  {title} {\enquote {\bibinfo {title} {The variational
  approach to the two-body density matrix},}\ }\href
  {https://doi.org/10.1063/1.522634} {\bibfield  {journal} {\bibinfo  {journal}
  {J. Math. Phys.}\ }\textbf {\bibinfo {volume} {16}},\ \bibinfo {pages}
  {868--874} (\bibinfo {year} {1975})}\BibitemShut {NoStop}%
\bibitem [{\citenamefont {Rosina}\ and\ \citenamefont
  {Garrod}(1975)}]{RosinaGarrod1975}%
  \BibitemOpen
  \bibfield  {author} {\bibinfo {author} {\bibfnamefont {M.}~\bibnamefont
  {Rosina}}\ and\ \bibinfo {author} {\bibfnamefont {C.}~\bibnamefont
  {Garrod}},\ }\bibfield  {title} {\enquote {\bibinfo {title} {The variational
  calculation of reduced density matrices},}\ }\href
  {https://doi.org/10.1016/0021-9991(75)90004-2} {\bibfield  {journal}
  {\bibinfo  {journal} {J. Comput. Phys.}\ }\textbf {\bibinfo {volume} {18}},\
  \bibinfo {pages} {300--310} (\bibinfo {year} {1975})}\BibitemShut {NoStop}%
\bibitem [{\citenamefont {Mihailovi{\'c}}\ and\ \citenamefont
  {Rosina}(1975)}]{MihailovicRosina1975}%
  \BibitemOpen
  \bibfield  {author} {\bibinfo {author} {\bibfnamefont {M.~V.}\ \bibnamefont
  {Mihailovi{\'c}}}\ and\ \bibinfo {author} {\bibfnamefont {M.}~\bibnamefont
  {Rosina}},\ }\bibfield  {title} {\enquote {\bibinfo {title} {The variational
  approach to the density matrix for light nuclei},}\ }\href
  {https://doi.org/10.1016/0375-9474(75)90420-0} {\bibfield  {journal}
  {\bibinfo  {journal} {Nucl. Phys. A}\ }\textbf {\bibinfo {volume} {237}},\
  \bibinfo {pages} {221--228} (\bibinfo {year} {1975})}\BibitemShut {NoStop}%
\bibitem [{\citenamefont {Garrod}\ and\ \citenamefont
  {Fusco}(1976)}]{GarrodFusco1976}%
  \BibitemOpen
  \bibfield  {author} {\bibinfo {author} {\bibfnamefont {C.}~\bibnamefont
  {Garrod}}\ and\ \bibinfo {author} {\bibfnamefont {M.~A.}\ \bibnamefont
  {Fusco}},\ }\bibfield  {title} {\enquote {\bibinfo {title} {A density matrix
  variational calculation for atomic {Be}},}\ }\href
  {https://doi.org/10.1002/qua.560100311} {\bibfield  {journal} {\bibinfo
  {journal} {Int. J. Quantum Chem.}\ }\textbf {\bibinfo {volume} {10}},\
  \bibinfo {pages} {495--510} (\bibinfo {year} {1976})}\BibitemShut {NoStop}%
\bibitem [{\citenamefont {Nakata}\ \emph {et~al.}(2001)\citenamefont {Nakata},
  \citenamefont {Nakatsuji}, \citenamefont {Ehara}, \citenamefont {Fukuda},
  \citenamefont {Nakata},\ and\ \citenamefont {Fujisawa}}]{Nakata2001}%
  \BibitemOpen
  \bibfield  {author} {\bibinfo {author} {\bibfnamefont {M.}~\bibnamefont
  {Nakata}}, \bibinfo {author} {\bibfnamefont {H.}~\bibnamefont {Nakatsuji}},
  \bibinfo {author} {\bibfnamefont {M.}~\bibnamefont {Ehara}}, \bibinfo
  {author} {\bibfnamefont {M.}~\bibnamefont {Fukuda}}, \bibinfo {author}
  {\bibfnamefont {K.}~\bibnamefont {Nakata}},\ and\ \bibinfo {author}
  {\bibfnamefont {K.}~\bibnamefont {Fujisawa}},\ }\bibfield  {title} {\enquote
  {\bibinfo {title} {Variational calculations of fermion second-order reduced
  density matrices by semidefinite programming algorithm},}\ }\href
  {https://doi.org/10.1063/1.1360199} {\bibfield  {journal} {\bibinfo
  {journal} {J. Chem. Phys.}\ }\textbf {\bibinfo {volume} {114}},\ \bibinfo
  {pages} {8282--8292} (\bibinfo {year} {2001})}\BibitemShut {NoStop}%
\bibitem [{\citenamefont {Zhao}\ \emph {et~al.}(2004)\citenamefont {Zhao},
  \citenamefont {Braams}, \citenamefont {Fukuda}, \citenamefont {Overton},\
  and\ \citenamefont {Percus}}]{Zhao2004}%
  \BibitemOpen
  \bibfield  {author} {\bibinfo {author} {\bibfnamefont {Z.}~\bibnamefont
  {Zhao}}, \bibinfo {author} {\bibfnamefont {B.~J.}\ \bibnamefont {Braams}},
  \bibinfo {author} {\bibfnamefont {M.}~\bibnamefont {Fukuda}}, \bibinfo
  {author} {\bibfnamefont {M.~L.}\ \bibnamefont {Overton}},\ and\ \bibinfo
  {author} {\bibfnamefont {J.~K.}\ \bibnamefont {Percus}},\ }\bibfield  {title}
  {\enquote {\bibinfo {title} {The reduced density matrix method for electronic
  structure calculations and the role of three-index representability
  conditions},}\ }\href {https://doi.org/10.1063/1.1636721} {\bibfield
  {journal} {\bibinfo  {journal} {J. Chem. Phys.}\ }\textbf {\bibinfo {volume}
  {120}},\ \bibinfo {pages} {2095--2104} (\bibinfo {year} {2004})}\BibitemShut
  {NoStop}%
\bibitem [{\citenamefont {Erdahl}(1978)}]{Erdahl1978}%
  \BibitemOpen
  \bibfield  {author} {\bibinfo {author} {\bibfnamefont {R.~M.}\ \bibnamefont
  {Erdahl}},\ }\bibfield  {title} {\enquote {\bibinfo {title}
  {Representability},}\ }\href {https://doi.org/10.1002/qua.560130603}
  {\bibfield  {journal} {\bibinfo  {journal} {Int. J. Quantum Chem.}\ }\textbf
  {\bibinfo {volume} {13}},\ \bibinfo {pages} {697--718} (\bibinfo {year}
  {1978})}\BibitemShut {NoStop}%
\bibitem [{\citenamefont {Fukuda}\ \emph {et~al.}(2007)\citenamefont {Fukuda},
  \citenamefont {Braams}, \citenamefont {Nakata}, \citenamefont {Overton},
  \citenamefont {Percus}, \citenamefont {Yamashita},\ and\ \citenamefont
  {Zhao}}]{Fukuda2007}%
  \BibitemOpen
  \bibfield  {author} {\bibinfo {author} {\bibfnamefont {M.}~\bibnamefont
  {Fukuda}}, \bibinfo {author} {\bibfnamefont {B.~J.}\ \bibnamefont {Braams}},
  \bibinfo {author} {\bibfnamefont {M.}~\bibnamefont {Nakata}}, \bibinfo
  {author} {\bibfnamefont {M.~L.}\ \bibnamefont {Overton}}, \bibinfo {author}
  {\bibfnamefont {J.~K.}\ \bibnamefont {Percus}}, \bibinfo {author}
  {\bibfnamefont {M.}~\bibnamefont {Yamashita}},\ and\ \bibinfo {author}
  {\bibfnamefont {Z.}~\bibnamefont {Zhao}},\ }\bibfield  {title} {\enquote
  {\bibinfo {title} {Large-scale semidefinite programs in electronic structure
  calculation},}\ }\href {https://doi.org/10.1007/s10107-006-0027-y} {\bibfield
   {journal} {\bibinfo  {journal} {Math. Program.}\ }\textbf {\bibinfo {volume}
  {109}},\ \bibinfo {pages} {553--580} (\bibinfo {year} {2007})}\BibitemShut
  {NoStop}%
\bibitem [{\citenamefont {Nakata}\ \emph {et~al.}(2008)\citenamefont {Nakata},
  \citenamefont {Braams}, \citenamefont {Fujisawa}, \citenamefont {Fukuda},
  \citenamefont {Percus}, \citenamefont {Yamashita},\ and\ \citenamefont
  {Zhao}}]{Nakata2008}%
  \BibitemOpen
  \bibfield  {author} {\bibinfo {author} {\bibfnamefont {M.}~\bibnamefont
  {Nakata}}, \bibinfo {author} {\bibfnamefont {B.~J.}\ \bibnamefont {Braams}},
  \bibinfo {author} {\bibfnamefont {K.}~\bibnamefont {Fujisawa}}, \bibinfo
  {author} {\bibfnamefont {M.}~\bibnamefont {Fukuda}}, \bibinfo {author}
  {\bibfnamefont {J.~K.}\ \bibnamefont {Percus}}, \bibinfo {author}
  {\bibfnamefont {M.}~\bibnamefont {Yamashita}},\ and\ \bibinfo {author}
  {\bibfnamefont {Z.}~\bibnamefont {Zhao}},\ }\bibfield  {title} {\enquote
  {\bibinfo {title} {Variational calculation of second-order reduced density
  matrices by strong {$N$}-representability conditions and an accurate
  semidefinite programming solver},}\ }\href
  {https://doi.org/10.1063/1.2911696} {\bibfield  {journal} {\bibinfo
  {journal} {J. Chem. Phys.}\ }\textbf {\bibinfo {volume} {128}},\ \bibinfo
  {pages} {164113} (\bibinfo {year} {2008})}\BibitemShut {NoStop}%
\bibitem [{\citenamefont {Mazziotti}(2004)}]{Mazziotti2004}%
  \BibitemOpen
  \bibfield  {author} {\bibinfo {author} {\bibfnamefont {D.~A.}\ \bibnamefont
  {Mazziotti}},\ }\bibfield  {title} {\enquote {\bibinfo {title} {Realization
  of quantum chemistry without wave functions through first-order semidefinite
  programming},}\ }\href {https://doi.org/10.1103/PhysRevLett.93.213001}
  {\bibfield  {journal} {\bibinfo  {journal} {Phys. Rev. Lett.}\ }\textbf
  {\bibinfo {volume} {93}},\ \bibinfo {pages} {213001} (\bibinfo {year}
  {2004})}\BibitemShut {NoStop}%
\bibitem [{\citenamefont {Gidofalvi}\ and\ \citenamefont
  {Mazziotti}(2008)}]{GidofalviMazziotti2008}%
  \BibitemOpen
  \bibfield  {author} {\bibinfo {author} {\bibfnamefont {G.}~\bibnamefont
  {Gidofalvi}}\ and\ \bibinfo {author} {\bibfnamefont {D.~A.}\ \bibnamefont
  {Mazziotti}},\ }\bibfield  {title} {\enquote {\bibinfo {title} {Active-space
  two-electron reduced-density-matrix method: Complete active-space
  calculations without diagonalization of the {$N$}-electron {Hamiltonian}},}\
  }\href {https://doi.org/10.1063/1.2983652} {\bibfield  {journal} {\bibinfo
  {journal} {J. Chem. Phys.}\ }\textbf {\bibinfo {volume} {129}},\ \bibinfo
  {pages} {134108} (\bibinfo {year} {2008})}\BibitemShut {NoStop}%
\bibitem [{\citenamefont {Mazziotti}(2007)}]{Mazziotti2007Book}%
  \BibitemOpen
  \bibinfo {editor} {\bibfnamefont {D.~A.}\ \bibnamefont {Mazziotti}},\ ed.,\
  \href {https://doi.org/10.1002/0470106603} {\emph {\bibinfo {title}
  {Reduced-Density-Matrix Mechanics: With Application to Many-Electron Atoms
  and Molecules}}},\ \bibinfo {series} {Advances in Chemical Physics}, Vol.\
  \bibinfo {volume} {134}\ (\bibinfo  {publisher} {Wiley},\ \bibinfo {year}
  {2007})\BibitemShut {NoStop}%
\bibitem [{\citenamefont {DePrince}(2024)}]{DePrince2024}%
  \BibitemOpen
  \bibfield  {author} {\bibinfo {author} {\bibfnamefont {A.~E.}\ \bibnamefont
  {DePrince}, \bibfnamefont {III}},\ }\bibfield  {title} {\enquote {\bibinfo
  {title} {Variational determination of the two-electron reduced density
  matrix: A tutorial review},}\ }\href {https://doi.org/10.1002/wcms.1702}
  {\bibfield  {journal} {\bibinfo  {journal} {WIREs Comput. Mol. Sci.}\
  }\textbf {\bibinfo {volume} {14}},\ \bibinfo {pages} {e1702} (\bibinfo {year}
  {2024})},\ \Eprint {https://arxiv.org/abs/2310.10746} {arXiv:2310.10746
  [physics.chem-ph]} \BibitemShut {NoStop}%
\bibitem [{\citenamefont {Nakata}, \citenamefont {Ehara},\ and\ \citenamefont
  {Nakatsuji}(2002)}]{Nakata2002}%
  \BibitemOpen
  \bibfield  {author} {\bibinfo {author} {\bibfnamefont {M.}~\bibnamefont
  {Nakata}}, \bibinfo {author} {\bibfnamefont {M.}~\bibnamefont {Ehara}},\ and\
  \bibinfo {author} {\bibfnamefont {H.}~\bibnamefont {Nakatsuji}},\ }\bibfield
  {title} {\enquote {\bibinfo {title} {Density matrix variational theory:
  Application to the potential energy surfaces and strongly correlated
  systems},}\ }\href {https://doi.org/10.1063/1.1453961} {\bibfield  {journal}
  {\bibinfo  {journal} {J. Chem. Phys.}\ }\textbf {\bibinfo {volume} {116}},\
  \bibinfo {pages} {5432--5439} (\bibinfo {year} {2002})}\BibitemShut {NoStop}%
\bibitem [{\citenamefont {Gidofalvi}\ and\ \citenamefont
  {Mazziotti}(2005)}]{GidofalviMazziotti2005}%
  \BibitemOpen
  \bibfield  {author} {\bibinfo {author} {\bibfnamefont {G.}~\bibnamefont
  {Gidofalvi}}\ and\ \bibinfo {author} {\bibfnamefont {D.~A.}\ \bibnamefont
  {Mazziotti}},\ }\bibfield  {title} {\enquote {\bibinfo {title} {Application
  of variational reduced-density-matrix theory to organic molecules},}\ }\href
  {https://doi.org/10.1063/1.1855885} {\bibfield  {journal} {\bibinfo
  {journal} {J. Chem. Phys.}\ }\textbf {\bibinfo {volume} {122}},\ \bibinfo
  {pages} {094107} (\bibinfo {year} {2005})}\BibitemShut {NoStop}%
\bibitem [{\citenamefont {van Aggelen}\ \emph {et~al.}(2009)\citenamefont {van
  Aggelen}, \citenamefont {Bultinck}, \citenamefont {Verstichel}, \citenamefont
  {Van~Neck},\ and\ \citenamefont {Ayers}}]{VanAggelen2009}%
  \BibitemOpen
  \bibfield  {author} {\bibinfo {author} {\bibfnamefont {H.}~\bibnamefont {van
  Aggelen}}, \bibinfo {author} {\bibfnamefont {P.}~\bibnamefont {Bultinck}},
  \bibinfo {author} {\bibfnamefont {B.}~\bibnamefont {Verstichel}}, \bibinfo
  {author} {\bibfnamefont {D.}~\bibnamefont {Van~Neck}},\ and\ \bibinfo
  {author} {\bibfnamefont {P.~W.}\ \bibnamefont {Ayers}},\ }\bibfield  {title}
  {\enquote {\bibinfo {title} {Incorrect diatomic dissociation in variational
  reduced density matrix theory arises from the flawed description of
  fractionally charged atoms},}\ }\href {https://doi.org/10.1039/B907624G}
  {\bibfield  {journal} {\bibinfo  {journal} {Phys. Chem. Chem. Phys.}\
  }\textbf {\bibinfo {volume} {11}},\ \bibinfo {pages} {5558--5560} (\bibinfo
  {year} {2009})}\BibitemShut {NoStop}%
\bibitem [{\citenamefont {van Aggelen}\ \emph {et~al.}(2010)\citenamefont {van
  Aggelen}, \citenamefont {Verstichel}, \citenamefont {Bultinck}, \citenamefont
  {Van~Neck}, \citenamefont {Ayers},\ and\ \citenamefont
  {Cooper}}]{VanAggelen2010}%
  \BibitemOpen
  \bibfield  {author} {\bibinfo {author} {\bibfnamefont {H.}~\bibnamefont {van
  Aggelen}}, \bibinfo {author} {\bibfnamefont {B.}~\bibnamefont {Verstichel}},
  \bibinfo {author} {\bibfnamefont {P.}~\bibnamefont {Bultinck}}, \bibinfo
  {author} {\bibfnamefont {D.}~\bibnamefont {Van~Neck}}, \bibinfo {author}
  {\bibfnamefont {P.~W.}\ \bibnamefont {Ayers}},\ and\ \bibinfo {author}
  {\bibfnamefont {D.~L.}\ \bibnamefont {Cooper}},\ }\bibfield  {title}
  {\enquote {\bibinfo {title} {Chemical verification of variational
  second-order density matrix based potential energy surfaces for the {N}$_2$
  isoelectronic series},}\ }\href {https://doi.org/10.1063/1.3354910}
  {\bibfield  {journal} {\bibinfo  {journal} {J. Chem. Phys.}\ }\textbf
  {\bibinfo {volume} {132}},\ \bibinfo {pages} {114112} (\bibinfo {year}
  {2010})}\BibitemShut {NoStop}%
\bibitem [{\citenamefont {Verstichel}\ \emph {et~al.}(2010)\citenamefont
  {Verstichel}, \citenamefont {van Aggelen}, \citenamefont {Van~Neck},
  \citenamefont {Ayers},\ and\ \citenamefont {Bultinck}}]{Verstichel2010}%
  \BibitemOpen
  \bibfield  {author} {\bibinfo {author} {\bibfnamefont {B.}~\bibnamefont
  {Verstichel}}, \bibinfo {author} {\bibfnamefont {H.}~\bibnamefont {van
  Aggelen}}, \bibinfo {author} {\bibfnamefont {D.}~\bibnamefont {Van~Neck}},
  \bibinfo {author} {\bibfnamefont {P.~W.}\ \bibnamefont {Ayers}},\ and\
  \bibinfo {author} {\bibfnamefont {P.}~\bibnamefont {Bultinck}},\ }\bibfield
  {title} {\enquote {\bibinfo {title} {Subsystem constraints in variational
  second order density matrix optimization: Curing the dissociative
  behavior},}\ }\href {https://doi.org/10.1063/1.3354911} {\bibfield  {journal}
  {\bibinfo  {journal} {J. Chem. Phys.}\ }\textbf {\bibinfo {volume} {132}},\
  \bibinfo {pages} {114113} (\bibinfo {year} {2010})}\BibitemShut {NoStop}%
\bibitem [{\citenamefont {van Aggelen}\ \emph {et~al.}(2011)\citenamefont {van
  Aggelen}, \citenamefont {Verstichel}, \citenamefont {Bultinck}, \citenamefont
  {Van~Neck}, \citenamefont {Ayers},\ and\ \citenamefont
  {Cooper}}]{VanAggelen2011}%
  \BibitemOpen
  \bibfield  {author} {\bibinfo {author} {\bibfnamefont {H.}~\bibnamefont {van
  Aggelen}}, \bibinfo {author} {\bibfnamefont {B.}~\bibnamefont {Verstichel}},
  \bibinfo {author} {\bibfnamefont {P.}~\bibnamefont {Bultinck}}, \bibinfo
  {author} {\bibfnamefont {D.}~\bibnamefont {Van~Neck}}, \bibinfo {author}
  {\bibfnamefont {P.~W.}\ \bibnamefont {Ayers}},\ and\ \bibinfo {author}
  {\bibfnamefont {D.~L.}\ \bibnamefont {Cooper}},\ }\bibfield  {title}
  {\enquote {\bibinfo {title} {Variational second order density matrix study of
  {$\mathrm{F}_3^-$}: Importance of subspace constraints for
  size-consistency},}\ }\href {https://doi.org/10.1063/1.3532409} {\bibfield
  {journal} {\bibinfo  {journal} {J. Chem. Phys.}\ }\textbf {\bibinfo {volume}
  {134}},\ \bibinfo {pages} {054115} (\bibinfo {year} {2011})}\BibitemShut
  {NoStop}%
\bibitem [{\citenamefont {Nakata}\ and\ \citenamefont
  {Anderson}(2012)}]{NakataAnderson2012}%
  \BibitemOpen
  \bibfield  {author} {\bibinfo {author} {\bibfnamefont {M.}~\bibnamefont
  {Nakata}}\ and\ \bibinfo {author} {\bibfnamefont {J.~S.~M.}\ \bibnamefont
  {Anderson}},\ }\bibfield  {title} {\enquote {\bibinfo {title} {On the
  size-consistency of the reduced-density-matrix method and the unitary
  invariant diagonal {$N$}-representability conditions},}\ }\href
  {https://doi.org/10.1063/1.4740246} {\bibfield  {journal} {\bibinfo
  {journal} {AIP Adv.}\ }\textbf {\bibinfo {volume} {2}},\ \bibinfo {pages}
  {032125} (\bibinfo {year} {2012})}\BibitemShut {NoStop}%
\bibitem [{\citenamefont {Nakata}(2026{\natexlab{a}})}]{PartI}%
  \BibitemOpen
  \bibfield  {author} {\bibinfo {author} {\bibfnamefont {M.}~\bibnamefont
  {Nakata}},\ }\bibfield  {title} {\enquote {\bibinfo {title} {Integer
  electron-count cuts in variational second-order reduced-density-matrix
  theory. {I}. {Molecular} dissociation with fixed real-space and orbital
  partitions},}\ }\href@noop {} {\bibfield  {journal} {\bibinfo  {journal}
  {arXiv:2610.36514}\ } (\bibinfo {year} {2026}{\natexlab{a}})},\ \bibinfo
  {note} {submitted to J. Chem. Phys.},\ \Eprint
  {https://arxiv.org/abs/2610.36514} {arXiv:2610.36514 [physics.chem-ph]}
  \BibitemShut {NoStop}%
\bibitem [{\citenamefont {Nakata}(2026{\natexlab{b}})}]{PartII}%
  \BibitemOpen
  \bibfield  {author} {\bibinfo {author} {\bibfnamefont {M.}~\bibnamefont
  {Nakata}},\ }\bibfield  {title} {\enquote {\bibinfo {title} {Integer
  electron-count cuts in variational second-order reduced-density-matrix
  theory. {II}. {A} convergent hierarchy of rigorous relaxations for real-space
  pair-density representability},}\ }\href@noop {} {\bibfield  {journal}
  {\bibinfo  {journal} {arXiv:2611.11707}\ } (\bibinfo {year}
  {2026}{\natexlab{b}})},\ \bibinfo {note} {code, verification programs, and
  data: https://doi.org/10.5281/zenodo.24063922},\ \Eprint
  {https://arxiv.org/abs/2611.11707} {arXiv:2611.11707 [physics.chem-ph]}
  \BibitemShut {NoStop}%
\bibitem [{\citenamefont {Boyn}(2026)}]{Boyn2026}%
  \BibitemOpen
  \bibfield  {author} {\bibinfo {author} {\bibfnamefont {J.-N.}\ \bibnamefont
  {Boyn}},\ }\href {https://doi.org/10.26434/chemrxiv.15011952/v1} {\enquote
  {\bibinfo {title} {Mixed-valence chemistry without fractional electrons in
  variational 2-electron reduced density matrix theory via local integer-charge
  constraints},}\ }\bibinfo {howpublished} {ChemRxiv preprint, version 1,
  posted 10 November 2026} (\bibinfo {year} {2026})\BibitemShut {NoStop}%
\bibitem [{\citenamefont {Nakata}\ and\ \citenamefont
  {Yasuda}(2009)}]{NakataYasuda2009}%
  \BibitemOpen
  \bibfield  {author} {\bibinfo {author} {\bibfnamefont {M.}~\bibnamefont
  {Nakata}}\ and\ \bibinfo {author} {\bibfnamefont {K.}~\bibnamefont
  {Yasuda}},\ }\bibfield  {title} {\enquote {\bibinfo {title} {Size extensivity
  of the variational reduced-density-matrix method},}\ }\href
  {https://doi.org/10.1103/PhysRevA.80.042109} {\bibfield  {journal} {\bibinfo
  {journal} {Phys. Rev. A}\ }\textbf {\bibinfo {volume} {80}},\ \bibinfo
  {pages} {042109} (\bibinfo {year} {2009})}\BibitemShut {NoStop}%
\bibitem [{\citenamefont {Braams}, \citenamefont {Percus},\ and\ \citenamefont
  {Zhao}(2007)}]{BraamsPercusZhao2007}%
  \BibitemOpen
  \bibfield  {author} {\bibinfo {author} {\bibfnamefont {B.~J.}\ \bibnamefont
  {Braams}}, \bibinfo {author} {\bibfnamefont {J.~K.}\ \bibnamefont {Percus}},\
  and\ \bibinfo {author} {\bibfnamefont {Z.}~\bibnamefont {Zhao}},\ }\bibfield
  {title} {\enquote {\bibinfo {title} {The {$T_1$} and {$T_2$} representability
  conditions},}\ }in\ \href {https://doi.org/10.1002/9780470106600.ch5} {\emph
  {\bibinfo {booktitle} {Reduced-Density-Matrix Mechanics: With Application to
  Many-Electron Atoms and Molecules}}},\ \bibinfo {series} {Advances in
  Chemical Physics}, Vol.\ \bibinfo {volume} {134},\ \bibinfo {editor} {edited
  by\ \bibinfo {editor} {\bibfnamefont {D.~A.}\ \bibnamefont {Mazziotti}}}\
  (\bibinfo  {publisher} {Wiley},\ \bibinfo {year} {2007})\ pp.\ \bibinfo
  {pages} {93--101}\BibitemShut {NoStop}%
\bibitem [{\citenamefont {Mazziotti}(2006)}]{Mazziotti2006}%
  \BibitemOpen
  \bibfield  {author} {\bibinfo {author} {\bibfnamefont {D.~A.}\ \bibnamefont
  {Mazziotti}},\ }\bibfield  {title} {\enquote {\bibinfo {title} {Variational
  reduced-density-matrix method using three-particle {$N$}-representability
  conditions with application to many-electron molecules},}\ }\href
  {https://doi.org/10.1103/PhysRevA.74.032501} {\bibfield  {journal} {\bibinfo
  {journal} {Phys. Rev. A}\ }\textbf {\bibinfo {volume} {74}},\ \bibinfo
  {pages} {032501} (\bibinfo {year} {2006})}\BibitemShut {NoStop}%
\bibitem [{\citenamefont {Erdahl}\ and\ \citenamefont
  {Garrod}(1974)}]{ErdahlGarrod1974}%
  \BibitemOpen
  \bibfield  {author} {\bibinfo {author} {\bibfnamefont {R.~M.}\ \bibnamefont
  {Erdahl}}\ and\ \bibinfo {author} {\bibfnamefont {C.}~\bibnamefont
  {Garrod}},\ }\bibfield  {title} {\enquote {\bibinfo {title} {Trace relations,
  one-particle symmetries and the convex structure of reduced density
  operators},}\ }in\ \href@noop {} {\emph {\bibinfo {booktitle} {Reduced
  Density Operators with Applications to Physical and Chemical
  Systems---{II}}}},\ \bibinfo {series and number} {\bibinfo {series} {Queen's
  Papers in Pure and Applied Mathematics}\ No.~\bibinfo {number} {40}},\
  \bibinfo {editor} {edited by\ \bibinfo {editor} {\bibfnamefont {R.~M.}\
  \bibnamefont {Erdahl}}}\ (\bibinfo  {publisher} {Queen's University},\
  \bibinfo {address} {Kingston, Ontario},\ \bibinfo {year} {1974})\ pp.\
  \bibinfo {pages} {22--27}\BibitemShut {NoStop}%
\bibitem [{\citenamefont {Mazziotti}(2026)}]{Mazziotti2026}%
  \BibitemOpen
  \bibfield  {author} {\bibinfo {author} {\bibfnamefont {D.~A.}\ \bibnamefont
  {Mazziotti}},\ }\bibfield  {title} {\enquote {\bibinfo {title}
  {Representability for quantum theory beyond particle-number conservation},}\
  }\href {https://doi.org/10.1103/yt4s-dlwr} {\bibfield  {journal} {\bibinfo
  {journal} {Phys. Rev. Lett.}\ }\textbf {\bibinfo {volume} {136}},\ \bibinfo
  {pages} {120202} (\bibinfo {year} {2026})}\BibitemShut {NoStop}%
\bibitem [{\citenamefont {Fekete}(1923)}]{Fekete1923}%
  \BibitemOpen
  \bibfield  {author} {\bibinfo {author} {\bibfnamefont {M.}~\bibnamefont
  {Fekete}},\ }\bibfield  {title} {\enquote {\bibinfo {title} {{{\"U}ber die
  Verteilung der Wurzeln bei gewissen algebraischen Gleichungen mit
  ganzzahligen Koeffizienten}},}\ }\href {https://doi.org/10.1007/BF01504345}
  {\bibfield  {journal} {\bibinfo  {journal} {Math. Z.}\ }\textbf {\bibinfo
  {volume} {17}},\ \bibinfo {pages} {228--249} (\bibinfo {year}
  {1923})}\BibitemShut {NoStop}%
\bibitem [{\citenamefont {de~Finetti}(1937)}]{deFinetti1937}%
  \BibitemOpen
  \bibfield  {author} {\bibinfo {author} {\bibfnamefont {B.}~\bibnamefont
  {de~Finetti}},\ }\bibfield  {title} {\enquote {\bibinfo {title} {La
  pr{\'e}vision : ses lois logiques, ses sources subjectives},}\ }\href
  {http://www.numdam.org/item/AIHP_1937__7_1_1_0/} {\bibfield  {journal}
  {\bibinfo  {journal} {Ann. Inst. Henri Poincar{\'e}}\ }\textbf {\bibinfo
  {volume} {7}},\ \bibinfo {pages} {1--68} (\bibinfo {year}
  {1937})}\BibitemShut {NoStop}%
\bibitem [{\citenamefont {Diaconis}\ and\ \citenamefont
  {Freedman}(1980)}]{DiaconisFreedman1980}%
  \BibitemOpen
  \bibfield  {author} {\bibinfo {author} {\bibfnamefont {P.}~\bibnamefont
  {Diaconis}}\ and\ \bibinfo {author} {\bibfnamefont {D.}~\bibnamefont
  {Freedman}},\ }\bibfield  {title} {\enquote {\bibinfo {title} {Finite
  exchangeable sequences},}\ }\href {https://doi.org/10.1214/aop/1176994663}
  {\bibfield  {journal} {\bibinfo  {journal} {Ann. Probab.}\ }\textbf {\bibinfo
  {volume} {8}},\ \bibinfo {pages} {745--764} (\bibinfo {year}
  {1980})}\BibitemShut {NoStop}%
\bibitem [{\citenamefont {K{\"o}nig}\ and\ \citenamefont
  {Renner}(2005)}]{KoenigRenner2005}%
  \BibitemOpen
  \bibfield  {author} {\bibinfo {author} {\bibfnamefont {R.}~\bibnamefont
  {K{\"o}nig}}\ and\ \bibinfo {author} {\bibfnamefont {R.}~\bibnamefont
  {Renner}},\ }\bibfield  {title} {\enquote {\bibinfo {title} {A de {F}inetti
  representation for finite symmetric quantum states},}\ }\href
  {https://doi.org/10.1063/1.2146188} {\bibfield  {journal} {\bibinfo
  {journal} {J. Math. Phys.}\ }\textbf {\bibinfo {volume} {46}},\ \bibinfo
  {pages} {122108} (\bibinfo {year} {2005})}\BibitemShut {NoStop}%
\bibitem [{\citenamefont {Christandl}\ \emph {et~al.}(2007)\citenamefont
  {Christandl}, \citenamefont {K{\"o}nig}, \citenamefont {Mitchison},\ and\
  \citenamefont {Renner}}]{CKMR2007}%
  \BibitemOpen
  \bibfield  {author} {\bibinfo {author} {\bibfnamefont {M.}~\bibnamefont
  {Christandl}}, \bibinfo {author} {\bibfnamefont {R.}~\bibnamefont
  {K{\"o}nig}}, \bibinfo {author} {\bibfnamefont {G.}~\bibnamefont
  {Mitchison}},\ and\ \bibinfo {author} {\bibfnamefont {R.}~\bibnamefont
  {Renner}},\ }\bibfield  {title} {\enquote {\bibinfo {title} {One-and-a-half
  quantum de {F}inetti theorems},}\ }\href
  {https://doi.org/10.1007/s00220-007-0189-3} {\bibfield  {journal} {\bibinfo
  {journal} {Commun. Math. Phys.}\ }\textbf {\bibinfo {volume} {273}},\
  \bibinfo {pages} {473--498} (\bibinfo {year} {2007})}\BibitemShut {NoStop}%
\bibitem [{\citenamefont {Krumnow}, \citenamefont {Zimbor{\'a}s},\ and\
  \citenamefont {Eisert}(2017)}]{KZE2017}%
  \BibitemOpen
  \bibfield  {author} {\bibinfo {author} {\bibfnamefont {C.}~\bibnamefont
  {Krumnow}}, \bibinfo {author} {\bibfnamefont {Z.}~\bibnamefont
  {Zimbor{\'a}s}},\ and\ \bibinfo {author} {\bibfnamefont {J.}~\bibnamefont
  {Eisert}},\ }\bibfield  {title} {\enquote {\bibinfo {title} {A fermionic de
  {F}inetti theorem},}\ }\href {https://doi.org/10.1063/1.4998944} {\bibfield
  {journal} {\bibinfo  {journal} {J. Math. Phys.}\ }\textbf {\bibinfo {volume}
  {58}},\ \bibinfo {pages} {122204} (\bibinfo {year} {2017})}\BibitemShut
  {NoStop}%
\bibitem [{\citenamefont {Lewin}, \citenamefont {Nam},\ and\ \citenamefont
  {Rougerie}(2014)}]{LNR2014}%
  \BibitemOpen
  \bibfield  {author} {\bibinfo {author} {\bibfnamefont {M.}~\bibnamefont
  {Lewin}}, \bibinfo {author} {\bibfnamefont {P.~T.}\ \bibnamefont {Nam}},\
  and\ \bibinfo {author} {\bibfnamefont {N.}~\bibnamefont {Rougerie}},\
  }\bibfield  {title} {\enquote {\bibinfo {title} {Derivation of {H}artree's
  theory for generic mean-field {B}ose systems},}\ }\href
  {https://doi.org/10.1016/j.aim.2013.12.010} {\bibfield  {journal} {\bibinfo
  {journal} {Adv. Math.}\ }\textbf {\bibinfo {volume} {254}},\ \bibinfo {pages}
  {570--621} (\bibinfo {year} {2014})}\BibitemShut {NoStop}%
\bibitem [{\citenamefont {Doherty}, \citenamefont {Parrilo},\ and\
  \citenamefont {Spedalieri}(2004)}]{DPS2004}%
  \BibitemOpen
  \bibfield  {author} {\bibinfo {author} {\bibfnamefont {A.~C.}\ \bibnamefont
  {Doherty}}, \bibinfo {author} {\bibfnamefont {P.~A.}\ \bibnamefont
  {Parrilo}},\ and\ \bibinfo {author} {\bibfnamefont {F.~M.}\ \bibnamefont
  {Spedalieri}},\ }\bibfield  {title} {\enquote {\bibinfo {title} {Complete
  family of separability criteria},}\ }\href
  {https://doi.org/10.1103/PhysRevA.69.022308} {\bibfield  {journal} {\bibinfo
  {journal} {Phys. Rev. A}\ }\textbf {\bibinfo {volume} {69}},\ \bibinfo
  {pages} {022308} (\bibinfo {year} {2004})}\BibitemShut {NoStop}%
\bibitem [{\citenamefont {Perdew}\ \emph {et~al.}(1982)\citenamefont {Perdew},
  \citenamefont {Parr}, \citenamefont {Levy},\ and\ \citenamefont
  {Balduz}}]{PPLB1982}%
  \BibitemOpen
  \bibfield  {author} {\bibinfo {author} {\bibfnamefont {J.~P.}\ \bibnamefont
  {Perdew}}, \bibinfo {author} {\bibfnamefont {R.~G.}\ \bibnamefont {Parr}},
  \bibinfo {author} {\bibfnamefont {M.}~\bibnamefont {Levy}},\ and\ \bibinfo
  {author} {\bibfnamefont {J.~L.}\ \bibnamefont {Balduz}, \bibfnamefont
  {Jr.}},\ }\bibfield  {title} {\enquote {\bibinfo {title} {Density-functional
  theory for fractional particle number: Derivative discontinuities of the
  energy},}\ }\href {https://doi.org/10.1103/PhysRevLett.49.1691} {\bibfield
  {journal} {\bibinfo  {journal} {Phys. Rev. Lett.}\ }\textbf {\bibinfo
  {volume} {49}},\ \bibinfo {pages} {1691--1694} (\bibinfo {year}
  {1982})}\BibitemShut {NoStop}%
\bibitem [{\citenamefont {Yang}, \citenamefont {Zhang},\ and\ \citenamefont
  {Ayers}(2000)}]{YZA2000}%
  \BibitemOpen
  \bibfield  {author} {\bibinfo {author} {\bibfnamefont {W.}~\bibnamefont
  {Yang}}, \bibinfo {author} {\bibfnamefont {Y.}~\bibnamefont {Zhang}},\ and\
  \bibinfo {author} {\bibfnamefont {P.~W.}\ \bibnamefont {Ayers}},\ }\bibfield
  {title} {\enquote {\bibinfo {title} {Degenerate ground states and a
  fractional number of electrons in density and reduced density matrix
  functional theory},}\ }\href {https://doi.org/10.1103/PhysRevLett.84.5172}
  {\bibfield  {journal} {\bibinfo  {journal} {Phys. Rev. Lett.}\ }\textbf
  {\bibinfo {volume} {84}},\ \bibinfo {pages} {5172--5175} (\bibinfo {year}
  {2000})}\BibitemShut {NoStop}%
\bibitem [{\citenamefont {Gatermann}\ and\ \citenamefont
  {Parrilo}(2004)}]{GatermannParrilo2004}%
  \BibitemOpen
  \bibfield  {author} {\bibinfo {author} {\bibfnamefont {K.}~\bibnamefont
  {Gatermann}}\ and\ \bibinfo {author} {\bibfnamefont {P.~A.}\ \bibnamefont
  {Parrilo}},\ }\bibfield  {title} {\enquote {\bibinfo {title} {Symmetry
  groups, semidefinite programs, and sums of squares},}\ }\href
  {https://doi.org/10.1016/j.jpaa.2003.12.011} {\bibfield  {journal} {\bibinfo
  {journal} {J. Pure Appl. Algebra}\ }\textbf {\bibinfo {volume} {192}},\
  \bibinfo {pages} {95--128} (\bibinfo {year} {2004})}\BibitemShut {NoStop}%
\bibitem [{\citenamefont {Permenter}\ and\ \citenamefont
  {Parrilo}(2018)}]{PermenterParrilo2018}%
  \BibitemOpen
  \bibfield  {author} {\bibinfo {author} {\bibfnamefont {F.}~\bibnamefont
  {Permenter}}\ and\ \bibinfo {author} {\bibfnamefont {P.}~\bibnamefont
  {Parrilo}},\ }\bibfield  {title} {\enquote {\bibinfo {title} {Partial facial
  reduction: simplified, equivalent {SDPs} via approximations of the {PSD}
  cone},}\ }\href {https://doi.org/10.1007/s10107-017-1169-9} {\bibfield
  {journal} {\bibinfo  {journal} {Math. Program.}\ }\textbf {\bibinfo {volume}
  {171}},\ \bibinfo {pages} {1--54} (\bibinfo {year} {2018})}\BibitemShut
  {NoStop}%
\bibitem [{\citenamefont {Erdahl}\ and\ \citenamefont
  {Jin}(2000{\natexlab{a}})}]{ErdahlJin2000}%
  \BibitemOpen
  \bibfield  {author} {\bibinfo {author} {\bibfnamefont {R.}~\bibnamefont
  {Erdahl}}\ and\ \bibinfo {author} {\bibfnamefont {B.}~\bibnamefont {Jin}},\
  }\bibfield  {title} {\enquote {\bibinfo {title} {On calculating approximate
  and exact density matrices},}\ }in\ \href
  {https://doi.org/10.1007/978-1-4615-4211-7_4} {\emph {\bibinfo {booktitle}
  {Many-Electron Densities and Reduced Density Matrices}}},\ \bibinfo {series
  and number} {Mathematical and Computational Chemistry}\ (\bibinfo
  {publisher} {Springer US},\ \bibinfo {year} {2000})\ pp.\ \bibinfo {pages}
  {57--84}\BibitemShut {NoStop}%
\bibitem [{\citenamefont {Erdahl}\ and\ \citenamefont
  {Jin}(2000{\natexlab{b}})}]{ErdahlJin2000THEOCHEM}%
  \BibitemOpen
  \bibfield  {author} {\bibinfo {author} {\bibfnamefont {R.~M.}\ \bibnamefont
  {Erdahl}}\ and\ \bibinfo {author} {\bibfnamefont {B.}~\bibnamefont {Jin}},\
  }\bibfield  {title} {\enquote {\bibinfo {title} {The lower bound method for
  reduced density matrices},}\ }\href
  {https://doi.org/10.1016/S0166-1280(00)00494-2} {\bibfield  {journal}
  {\bibinfo  {journal} {J. Mol. Struct. THEOCHEM}\ }\textbf {\bibinfo {volume}
  {527}},\ \bibinfo {pages} {207--220} (\bibinfo {year}
  {2000}{\natexlab{b}})}\BibitemShut {NoStop}%
\bibitem [{\citenamefont {Mazziotti}\ and\ \citenamefont
  {Erdahl}(2001)}]{MazziottiErdahl2001}%
  \BibitemOpen
  \bibfield  {author} {\bibinfo {author} {\bibfnamefont {D.~A.}\ \bibnamefont
  {Mazziotti}}\ and\ \bibinfo {author} {\bibfnamefont {R.~M.}\ \bibnamefont
  {Erdahl}},\ }\bibfield  {title} {\enquote {\bibinfo {title} {Uncertainty
  relations and reduced density matrices: Mapping many-body quantum mechanics
  onto four particles},}\ }\href {https://doi.org/10.1103/PhysRevA.63.042113}
  {\bibfield  {journal} {\bibinfo  {journal} {Phys. Rev. A}\ }\textbf {\bibinfo
  {volume} {63}},\ \bibinfo {pages} {042113} (\bibinfo {year}
  {2001})}\BibitemShut {NoStop}%
\bibitem [{\citenamefont {Yamashita}\ \emph {et~al.}(2012)\citenamefont
  {Yamashita}, \citenamefont {Fujisawa}, \citenamefont {Fukuda}, \citenamefont
  {Kobayashi}, \citenamefont {Nakata},\ and\ \citenamefont {Nakata}}]{SDPA7}%
  \BibitemOpen
  \bibfield  {author} {\bibinfo {author} {\bibfnamefont {M.}~\bibnamefont
  {Yamashita}}, \bibinfo {author} {\bibfnamefont {K.}~\bibnamefont {Fujisawa}},
  \bibinfo {author} {\bibfnamefont {M.}~\bibnamefont {Fukuda}}, \bibinfo
  {author} {\bibfnamefont {K.}~\bibnamefont {Kobayashi}}, \bibinfo {author}
  {\bibfnamefont {K.}~\bibnamefont {Nakata}},\ and\ \bibinfo {author}
  {\bibfnamefont {M.}~\bibnamefont {Nakata}},\ }\bibfield  {title} {\enquote
  {\bibinfo {title} {Latest developments in the {SDPA} family for solving
  large-scale {SDPs}},}\ }in\ \href
  {https://doi.org/10.1007/978-1-4614-0769-0_24} {\emph {\bibinfo {booktitle}
  {Handbook on Semidefinite, Conic and Polynomial Optimization}}},\ \bibinfo
  {editor} {edited by\ \bibinfo {editor} {\bibfnamefont {M.~F.}\ \bibnamefont
  {Anjos}}\ and\ \bibinfo {editor} {\bibfnamefont {J.~B.}\ \bibnamefont
  {Lasserre}}}\ (\bibinfo  {publisher} {Springer},\ \bibinfo {address} {New
  York},\ \bibinfo {year} {2012})\ pp.\ \bibinfo {pages} {687--713}\BibitemShut
  {NoStop}%
\bibitem [{\citenamefont {Sun}\ \emph {et~al.}(2018)\citenamefont {Sun},
  \citenamefont {Berkelbach}, \citenamefont {Blunt}, \citenamefont {Booth},
  \citenamefont {Guo}, \citenamefont {Li}, \citenamefont {Liu}, \citenamefont
  {McClain}, \citenamefont {Sayfutyarova}, \citenamefont {Sharma},
  \citenamefont {Wouters},\ and\ \citenamefont {Chan}}]{PySCF}%
  \BibitemOpen
  \bibfield  {author} {\bibinfo {author} {\bibfnamefont {Q.}~\bibnamefont
  {Sun}}, \bibinfo {author} {\bibfnamefont {T.~C.}\ \bibnamefont {Berkelbach}},
  \bibinfo {author} {\bibfnamefont {N.~S.}\ \bibnamefont {Blunt}}, \bibinfo
  {author} {\bibfnamefont {G.~H.}\ \bibnamefont {Booth}}, \bibinfo {author}
  {\bibfnamefont {S.}~\bibnamefont {Guo}}, \bibinfo {author} {\bibfnamefont
  {Z.}~\bibnamefont {Li}}, \bibinfo {author} {\bibfnamefont {J.}~\bibnamefont
  {Liu}}, \bibinfo {author} {\bibfnamefont {J.~D.}\ \bibnamefont {McClain}},
  \bibinfo {author} {\bibfnamefont {E.~R.}\ \bibnamefont {Sayfutyarova}},
  \bibinfo {author} {\bibfnamefont {S.}~\bibnamefont {Sharma}}, \bibinfo
  {author} {\bibfnamefont {S.}~\bibnamefont {Wouters}},\ and\ \bibinfo {author}
  {\bibfnamefont {G.~K.-L.}\ \bibnamefont {Chan}},\ }\bibfield  {title}
  {\enquote {\bibinfo {title} {{PySCF}: The {Python}-based simulations of
  chemistry framework},}\ }\href {https://doi.org/10.1002/wcms.1340} {\bibfield
   {journal} {\bibinfo  {journal} {WIREs Comput. Mol. Sci.}\ }\textbf {\bibinfo
  {volume} {8}},\ \bibinfo {pages} {e1340} (\bibinfo {year}
  {2018})}\BibitemShut {NoStop}%
\bibitem [{\citenamefont {Hehre}, \citenamefont {Stewart},\ and\ \citenamefont
  {Pople}(1969)}]{Hehre1969}%
  \BibitemOpen
  \bibfield  {author} {\bibinfo {author} {\bibfnamefont {W.~J.}\ \bibnamefont
  {Hehre}}, \bibinfo {author} {\bibfnamefont {R.~F.}\ \bibnamefont {Stewart}},\
  and\ \bibinfo {author} {\bibfnamefont {J.~A.}\ \bibnamefont {Pople}},\
  }\bibfield  {title} {\enquote {\bibinfo {title} {Self-consistent
  molecular-orbital methods. {I}. use of {Gaussian} expansions of {Slater}-type
  atomic orbitals},}\ }\href {https://doi.org/10.1063/1.1672392} {\bibfield
  {journal} {\bibinfo  {journal} {J. Chem. Phys.}\ }\textbf {\bibinfo {volume}
  {51}},\ \bibinfo {pages} {2657--2664} (\bibinfo {year} {1969})}\BibitemShut
  {NoStop}%
\bibitem [{\citenamefont {Hubbard}(1963)}]{Hubbard1963}%
  \BibitemOpen
  \bibfield  {author} {\bibinfo {author} {\bibfnamefont {J.}~\bibnamefont
  {Hubbard}},\ }\bibfield  {title} {\enquote {\bibinfo {title} {Electron
  correlations in narrow energy bands},}\ }\href
  {https://doi.org/10.1098/rspa.1963.0204} {\bibfield  {journal} {\bibinfo
  {journal} {Proc. R. Soc. London, Ser. A}\ }\textbf {\bibinfo {volume}
  {276}},\ \bibinfo {pages} {238--257} (\bibinfo {year} {1963})}\BibitemShut
  {NoStop}%
\bibitem [{\citenamefont {Nakata}(2010)}]{Nakata2010SDPAGMP}%
  \BibitemOpen
  \bibfield  {author} {\bibinfo {author} {\bibfnamefont {M.}~\bibnamefont
  {Nakata}},\ }\bibfield  {title} {\enquote {\bibinfo {title} {A numerical
  evaluation of highly accurate multiple-precision arithmetic version of
  semidefinite programming solver: {SDPA-GMP}, {-QD} and {-DD}},}\ }in\ \href
  {https://doi.org/10.1109/CACSD.2010.5612693} {\emph {\bibinfo {booktitle}
  {2010 IEEE International Symposium on Computer-Aided Control System
  Design}}}\ (\bibinfo  {publisher} {IEEE},\ \bibinfo {year} {2010})\ pp.\
  \bibinfo {pages} {29--34}\BibitemShut {NoStop}%
\end{thebibliography}%
